\documentclass[%
notitlepage,
twocolumn,
superscriptaddress,
nobibnotes, 
noeprint, 
 amsmath,amssymb,
prb,
floatfix
]{revtex4-2}

\usepackage{graphicx}
\usepackage{dcolumn}
\usepackage{bm}
\usepackage{units}

\usepackage{placeins}
\usepackage{mleftright}

\usepackage{subfigure}
\usepackage{xcolor}
\usepackage{mathtools}
\usepackage{bbold}

\usepackage{cancel}

\makeatletter
\newcommand{\hstrike}[1]{\mathpalette\@hstrike{#1}}
\newcommand{\@hstrike}[2]{\mathord{\setbox\z@\hbox{$\m@th#1#2$}%
  \ooalign{\copy\z@\cr$\m@th#1\vcenter{\hrule height 0.4pt width \wd\z@}$\cr}}}

\makeatother  

\usepackage{empheq}

\usepackage[most]{tcolorbox}

\tcbset{colback=yellow!10!white, colframe=black!50!black, 
        highlight math style= {enhanced, 
            colframe=black,colback=red!10!white,boxsep=0pt}
        }

\newcommand*{\efun}[1]{\mathrm e^{#1}}

\newcommand*{\rpar}{\mathbf r_{\parallel}}
\newcommand*{\rparp}{\mathbf r_{\parallel}^{\prime}}
\newcommand*{\rparpp}{\mathbf r_{\parallel}^{\prime\prime}}

\newcommand*{\Rpar}{\mathbf R_{\parallel}}
\newcommand*{\Rparp}{\mathbf R_{\parallel}^{\prime}}

\newcommand*{\kpar}[1]{\mathbf k_{\parallel}\vphantom{\mathbf k}_{#1}}
\newcommand*{\kparp}[1]{\mathbf k_{\parallel}^{\prime}\vphantom{\mathbf k}_{#1}}

\newcommand*{\qpar}{ \mathbf q_{\parallel}}
\newcommand*{\qparp}{\mathbf q_{\parallel}^{\prime}}

\newcommand*{\Qpar}{ \mathbf Q_{\parallel}}

\newcommand*{\Gpar}{\mathbf G_{\parallel}}
\newcommand*{\Gparp}{\mathbf G_{\parallel}^{\prime}}
\newcommand*{\Gparpp}{\mathbf G_{\parallel}^{\prime\prime}}
\newcommand*{\Gparppp}{\mathbf G_{\parallel}^{\prime\prime\prime}}
\newcommand*{\Gpari}[1]{\mathbf G_{\parallel}^{\vphantom{\prime}}\vphantom{\mathbf G}_{#1}}

\newcommand*{\Poloptwo}[2]{{{}\hat{P}^{\vphantom{\dagger}}_{}}{\vphantom{P}}_{#1}^{#2}}

\newcommand*{\Poldagtwo}[2]{{{}\hat{P}^{\dagger}}{\vphantom{P}}_{#1}^{#2}}
\newcommand*{\Pol}[1]{P_{#1}^{}}
\newcommand*{\Poltwo}[2]{P{\vphantom{P}}_{#1}^{#2}}

\newcommand*{\adag}[1]{\hat{a}^{\dagger}_{#1}}
\newcommand*{\andag}[1]{\hat{a}^{\vphantom{\dagger}}_{#1}}
\newcommand*{\adagtwo}[2]{{\hat{a}^{\dagger}_{}}{\vphantom{a}}_{#1}^{#2}}
\newcommand*{\andagtwo}[2]{{\hat{a}^{\vphantom{\dagger}}_{}}{\vphantom{a}}_{#1}^{#2}}
\newcommand*{\vdag}[1]{\hat{v}^{\dagger}_{#1}}
\newcommand*{\vndag}[1]{\hat{v}_{#1}^{\vphantom{\dagger}}}
\newcommand*{\cdag}[1]{\hat{c}^{\dagger}_{#1}}
\newcommand*{\cndag}[1]{\hat{c}_{#1}^{\vphantom{\dagger}}}

\newcommand*{\vdagtwo}[2]{{\hat{v}^{\dagger}_{}}{\vphantom{\hat{v}}}_{#1}^{#2}}
\newcommand*{\vndagtwo}[2]{{\hat{v}_{}^{\vphantom{\dagger}}}{\vphantom{\hat{v}}}_{#1}^{#2}}
\newcommand*{\cdagtwo}[2]{{\hat{c}_{}^{\dagger}}{\vphantom{\hat{c}}}_{#1}^{#2}}
\newcommand*{\cndagtwo}[2]{{\hat{c}_{}^{\vphantom{\dagger}}}{\vphantom{\hat{c}}}_{#1}^{#2}}

\newcommand*{\Pndagtwo}[2]{{{}\hat{P}^{\vphantom{\dagger}}_{}}{\vphantom{P}}_{#1}^{#2}}

\newcommand*{\Pdagtwo}[2]{{{}\hat{P}^{\dagger}}{\vphantom{P}}_{#1}^{#2}}

\newcommand{\ExWFtwo}[2]{\varphi_{#1}^{#2}}
\newcommand{\ExWFstartwo}[2]{\varphi^*{\vphantom{\varphi}}_{#1}^{#2}}

\renewcommand*{\bf}[1]{\textbf{#1}}
\renewcommand*{\eqref}[1]{Eq.~(\ref{#1})}

\usepackage[ left = 1.5cm, right=1.5cm, bottom=1.5cm, top=1.5cm]{geometry}

\definecolor{custom}{RGB}{15,45,150}
\usepackage[urlcolor=blue]{hyperref}
\hypersetup{colorlinks=true,linktoc=page,citecolor = blue,linkcolor=blue}

\usepackage{framed}
\colorlet{shadecolor}{gray!15}

\setcitestyle{numbers,square}

\begin{document}
\makeatletter
\renewcommand\@biblabel[1]{[#1]}
\makeatletter


\title{Coulomb Interaction and
Density-Independent Exciton-Scattering Processes in Atomically Thin Semiconductors
}

\author{Henry Mittenzwey}
\email{henry.mittenzwey@uni-giessen.de}
\affiliation{Institut für Theoretische Physik und Zentrum für Materialforschung, Justus-Liebig-Universität Gie{\ss}en, 
35392 Gie{\ss}en, Germany}
\affiliation{Nichtlineare Optik und Quantenelektronik, Institut für Physik und Astronomie (IFPA), Technische Universität Berlin, 10623 Berlin, Germany}

\author{Andreas Knorr}
\affiliation{Nichtlineare Optik und Quantenelektronik, Institut für Physik und Astronomie (IFPA), Technische Universität Berlin, 10623 Berlin, Germany}

\author{Thorsten Deilmann}
\affiliation{Institut für Festkörpertheorie, Universität Münster, 48149 Münster, Germany}


\begin{abstract}
In quantum-kinetic approaches to the dynamics of Coulomb-bound many-body correlations such as excitons, trions, biexcitons or higher-order correlations, a detailed knowledge of the many-body Coulomb Hamiltonian serving as a starting point is important. 
In this manuscript, the second-quantized description of the Coulomb interaction between Bloch electrons in a Heisenberg-equation-of-motion approach in atomically thin semiconductors is derived and reviewed. Emphasis is put on a discussion of Umklapp processes and the dielectric screening including all local-field effects. A link between \textit{ab initio} methods of screening and few-band models in effective-mass approximations for the quantum kinetics is established and important Coulomb scattering processes contributing to the exciton energy landscape and density-independent exciton scattering are discussed.
\end{abstract}
\date{\today}
\maketitle
\tableofcontents

\section{Introduction}
\label{sec:introduction}

In this manuscript, we review and discuss a Heisenberg-equations-of-motion approach to the second-quantized Coulomb interaction between Bloch electrons in thin semiconductors \cite{mittenzwey2026diss}. Our theory serves as a starting point for the calculation of many-body complexes such as excitons \cite{wang2018colloquium,katsch2018theory,chernikov2014exciton}, trions \cite{perea2024trion,van2023excitons,katsch2022excitonic,katsch2022doping,efimkin2018exciton,combescot2017three,efimkin2017many,esser2001theory,druppel2017diversity,arora2019excited,deilmann2018interlayer,mak2013tightly,ross2013electrical}, biexcitons \cite{deckert2025coherent,katsch2020exciton,katsch2019theory,steinhoff2018biexciton,takayama2002t,tang2025theoretical,barbone2018charge} and higher correlations in intense optical fields in undoped semiconductors \cite{schafer2025distinct,mittenzwey2026excitonic,axt2001evidence,bolton2000demonstration,meier1999excitons,bartels1997coherent} or in small optical fields in heavily doped semiconductors \cite{dijkstra2025ten,choi2024emergence,van2022six} and their quantum kinetics. Within a projection technique on electron--hole pair operators \cite{katsch2018theory,ivanov1993self} up to second order dynamics-controlled truncation \cite{axt1994dynamics,victor1995hierarchy}, the exciton dynamics in the undoped limit within an effective-mass approximation are dominated by incoherent exciton scattering via exciton--phonon interaction \cite{thranhardt2000quantum,herbst2000temporally,axt1996influence,selig2018dark,brem2020phonon} and density-independent incoherent Coulomb scattering such as intervalley exchange \cite{selig2020suppression,selig2019ultrafast,yu2014valley,combescot2023ab,glazov2014exciton}, Dexter interaction \cite{berghauser2018inverted,bernal2018exciton,dogadov2026diss} or disorder-mediated coherent intervalley exchange \cite{schmidt2016ultrafast}. Recent advances go beyond an effective-mass approximation by marrying \textit{ab initio} methods with quantum-kinetic approaches \cite{amit2025ab,steinhoff2025wannier,lee2024phonon,chan2025exciton,hu2023excitonic,amit2023ultrafast,zhang2022ab,perfetto2022real,stefanucci2023and,molina2017ab,attaccalite2011real}.

While the topic of second-quantized many-body Coulomb interaction is covered by various textbooks \cite{czycholl2007theoretische,cohen2016fundamentals,ziman2001electrons,mahan2000many,haug2009quantum,book:schaefer_semiconductor_optics}, 
its derivation starting from the classical Hamiltonian 
and a detailed discussion of Umklapp processes and the origin of local-field effects and microscopic screening in atomically thin semiconductors is often not provided. Our approach shall serve as a starting point to combine state-of-the-art \textit{ab initio} methods with many-body quantum kinetics.

The manuscript is organized as follows: In Sec.~\ref{app:CoulombInteractionHamiltonian}, we derive the general many-body Coulomb Hamiltonian in second quantization, cf.~\eqref{eq:Coulomb_Hamiltonian_SecondQuantized_Formfactors_Evaluated}:
\begin{multline}
\label{eq:introduction:Coulomb_Hamiltonian_SecondQuantized_Formfactors_Evaluated}
\hat H_{\text{Coul}}=  \frac{1}{2}\sum_{\substack{\lambda_1,\dots\lambda_4,\kpar{},\kparp{},\\\qpar,\Gpar,\Gparp,\\
    s,s^{\prime},n_1,\dots n_4}} V^{n_1,n_2,n_3,n_4}_{\qpar,\Gpar,\Gparp}(\omega=0)\\
    \times
    \overline \Upsilon_{\kpar{}+\qpar+\Gpar,\kpar{}}^{\lambda_1,\lambda_4,s}\overline \Upsilon_{\kparp{}-\qpar-\Gparp,\kparp{}}^{\lambda_2,\lambda_3,s^{\prime}}\\
    \times
    \bigg(\adagtwo{\lambda_1,\kpar{}+\qpar}{s,n_1}(t)\adagtwo{\lambda_2,\kparp{}-\qpar}{s^{\prime},n_2}(t)\andagtwo{\lambda_3,\kparp{}}{s^{\prime},n_3}(t)\andagtwo{\lambda_4,\kpar{}}{s,n_4}(t)\\
     + \,\delta_{\lambda_2,\lambda_4}^{\kparp{},\kpar{}+\qpar}\delta_{s,s^{\prime}}^{n_2,n_4} \adagtwo{\lambda_1,\kpar{}+\qpar}{s,n_1}(t)\andagtwo{\lambda_3,\kparp{}}{s^{\prime},n_3}(t)\bigg)
    ,
\end{multline}
in its most general form including all possible Coulomb-scattering processes ($\adag{1}\adag{2}\andag{3}\andag{4}$) and local-field effects ($\Gpar$, $\Gparp$) with the ground-state screened quantum-confined Coulomb potential $V^{n_1,n_2,n_3,n_4}_{\qpar,\Gpar,\Gparp}(\omega=0)$, cf.~\eqref{eq:introduction:coulomb_potential_quantum_confined_epsilon_abinitio} in an adiabatic approximation, cf.\ \eqref{eq:Coulomb_Hamiltonian_SecondQuantized_ValleyExpanded_adiabatic_approx}, (reduced) Bloch form factors $\overline \Upsilon_{\kparp{},\kpar{}}^{\lambda,\lambda^{\prime},s}$, cf.~\eqref{eq:reduced_form_factor}, and annihilation (creation) operators $\hat a^{(\dagger)}\vphantom{a}_{\lambda,\kpar{}}^{s,n}$ in band $\lambda$, in-plane quasi-momentum $\kpar{}$, spin $s$ and confinement quantum number $n$.

In Sec.~\ref{sec:screening}, we discuss the concepts of screening, where we, first, provide a microscopic model (Sec.~\ref{sec:microscopic_screening}) for the quantum-confined Coulomb potential $V_{\qpar,\Gpar,\Gparp}^{n_1,n_2,n_3,n_4}(\omega,\omega^{\prime})$, cf.~\eqref{eq:coulomb_potential_quantum_confined_epsilon_abinitio}:
\begin{multline}
    V_{\qpar,\Gpar,\Gparp}^{n_1,n_2,n_3,n_4}(\omega,\omega^{\prime})\\
    = \sum_{G_z,G_z^{\prime}}
    \left(\varepsilon^{-1}_{\text{mic}}\right)_{\qpar,\Gpar,\Gparp,G_z,G_z^{\prime}}(\omega,\omega^{\prime})F_{G_z}^{n_1,n_4} F_{-G_z^{\prime}}^{n_2,n_3}\\
    \times V_{0,\qpar+\Gparp,G_z^{\prime}}^{\text{trunc}},
    \label{eq:introduction:coulomb_potential_quantum_confined_epsilon_abinitio}
\end{multline}
with the three-dimensional microscopic dielectric function $\varepsilon_{\text{mic},\mathbf q,\mathbf G,\mathbf G^{\prime}}(\omega,-\omega^{\prime})$ from \textit{ab initio} supercell calculations in random-phase approximation, cf.~\eqref{eq:dielectric_function_microscopic}:
\begin{multline}
    \varepsilon_{\text{mic},\mathbf q,\mathbf G,\mathbf G^{\prime}}(\omega,-\omega^{\prime}) \\
    = 2\pi\delta(\omega-\omega^{\prime})\delta_{\mathbf G,\mathbf G^{\prime}}
    - 2\pi\delta(\omega-\omega^{\prime})V_{0,\mathbf q+\mathbf G}\\
    \times\sum_{\lambda,\lambda^{\prime},\mathbf k}
    \frac{\overline \Upsilon_{\mathbf k-\mathbf q-\mathbf G,\mathbf k}^{\lambda,\lambda^{\prime},s}
    \overline \Upsilon_{\mathbf k,\mathbf k-\mathbf q-\mathbf G^{\prime}}^{\lambda^{\prime},\lambda,s}\left(f_{\lambda,\mathbf k-\mathbf q}^s - f_{\lambda^{\prime},\mathbf k}^s\right)}{ \tilde E_{\lambda,\mathbf k-\mathbf q}^s - \tilde E_{\lambda^{\prime},\mathbf k}^s +\hbar\omega + \mathrm i\hbar\gamma  },
    \label{eq:introduction:dielectric_function_microscopic}
\end{multline}
with renormalized single-particle dispersions $\tilde E_{\lambda,\mathbf k}^s$ and Fermi distributions $f_{\lambda,\mathbf k}^s$, 
and, second, review an analytical macroscopic approach (Sec.~\ref{sec:macroscopic_screening}), where the quantum-confined Coulomb potential in a layered dielectric environment with substrate and superstrate without local fields ($\Gpar=\Gparp=\mathbf 0$) in a static, i.e., memory-free limit ($\omega=\omega^{\prime}=0$) with $V_{\qpar,\Gpar,\Gparp}^{n_1,n_2,n_3,n_4}(\omega,\omega^{\prime}) \rightarrow V_{\qpar}$ can be expressed as, cf.~\eqref{eq:Coulomb_potential_no_vacuum_gap_delta_confinement}:
\begin{align}
\begin{split}
    V_{\qpar}    =
    \frac{e^2}{2\mathcal A\epsilon_0\epsilon_{\text{M}}|\qpar|}\frac{\gamma_- + \delta_+ -\gamma_+\mathrm e^{|\qpar|d} - \delta_-\mathrm e^{-|\qpar|d}}{\delta_-\mathrm e^{-|\qpar|d}-\gamma_+\mathrm e^{|\qpar|d}}.
    \end{split}
    \label{eq:introduction:Coulomb_potential_no_vacuum_gap_delta_confinement}
\end{align}

In Sec.~\ref{sec:excitons}, we derive an effective Bethe--Salpeter equation in a static approximation, cf.~\eqref{eq:BSE}:
\begin{multline}
    \left(E_{c,\mathbf q}-E_{v,\mathbf q} + \Sigma_{c,v,\mathbf q}^{\text{H}} + \Sigma_{c,v,\mathbf q}^{\text{SEX}} + \Sigma_{c,v,\mathbf q}^{\text{COH}}\right)\Phi_{\mu,v,c,\mathbf q}\\
    + \sum_{\mathbf q^{\prime},v^{\prime},c^{\prime}}\left(K_{\text{eh}}^{\text{dir}}\,\vphantom{K}_{v,c,\mathbf q}^{v^{\prime},c^{\prime},\mathbf q^{\prime}} + K_{\text{eh}}^{\text{exch}}\,\vphantom{K}_{v,c,\mathbf q}^{v^{\prime},c^{\prime},\mathbf q^{\prime}}\right)\Phi_{\mu,v^{\prime},c^{\prime},\mathbf q^{\prime}}\\
    = E_{\mu}\Phi_{\mu,v,c,\mathbf q},
    \label{eq:introduction:BSE}
\end{multline}
for the excitonic energies $E_{\mu}$ and excitonic wave functions $\Phi_{\mu,v,c,\mathbf q}$ defined over the entire Brillouin zone, 
and its few-band effective-mass version, the Wannier equation, cf.~\eqref{eq:WannierEquation}:
\begin{multline}
    \left(\tilde E_{\text{gap}}^{\xi,\xi^{\prime},s,s^{\prime}} + \frac{\hbar^2\mathbf q^2}{2m_{\text{r},s,s^{\prime}}^{\xi,\xi^{\prime}}}\right)\ExWFtwo{\mu,\mathbf q}{\xi,\xi^{\prime},s,s^{\prime}}\\
    - \sum_{\mathbf q^{\prime}} K_{\text{eh}}^{\text{dir}}\,\vphantom{K}_{\mathbf q^{\prime}}^{\xi,\xi^{\prime},s,s^{\prime}}  \ExWFtwo{\mu,\mathbf q-\mathbf q^{\prime}}{\xi,\xi^{\prime},s,s^{\prime}} 
    = E_{\mu}^{\xi,\xi^{\prime},s,s^{\prime}}\ExWFtwo{\mu,\mathbf q}{\xi,\xi^{\prime},s,s^{\prime}},
    \label{eq:introduction:WannierEquation}
\end{multline}
for excitonic energies $E_{\mu}^{\xi,\xi^{\prime},s,s^{\prime}}$ and excitonic wave functions $\ExWFtwo{\mu,\mathbf q}{\xi,\xi^{\prime},s,s^{\prime}}$ defined in the vicinity of the high-symmetry points $\xi$ ($\xi^{\prime}$) with respect to holes (electrons), 
which describe Coulomb-bound electron--hole pairs, the excitons, and discuss their signatures in optical absorption spectra. 

In Sec.~\ref{sec:direct_interaction} (direct electron--hole interaction $\adagtwo{\lambda,\kpar{}+\qpar}{s}\adagtwo{\lambda^{\prime},\kparp{}-\qpar}{s^{\prime}}\andagtwo{\lambda^{\prime},\kparp{}}{s^{\prime}}\andagtwo{\lambda,\kpar{}}{s}$ with \mbox{$\lambda\neq\lambda^{\prime}$}) and Sec.~\ref{sec:exchange_interaction} (electron--hole exchange interaction $\adagtwo{\lambda^{\prime},\kpar{}+\qpar}{s}\adagtwo{\lambda,\kparp{}-\qpar}{s^{\prime}}\andagtwo{\lambda^{\prime},\kparp{}}{s^{\prime}}\andagtwo{\lambda,\kpar{}}{s}$ with \mbox{$\lambda\neq\lambda^{\prime}$}), we dissect many-body Coulomb scattering processes described by \eqref{eq:introduction:Coulomb_Hamiltonian_SecondQuantized_Formfactors_Evaluated} beyond the Wannier equation in an effective-mass two-band model, which are important for exciton quantum kinetics in the density-independent limit.

\section{Many-Body Coulomb Interaction Hamiltonian}
\label{app:CoulombInteractionHamiltonian}
The classical electron--electron Coulomb interaction Hamiltonian in Coulomb gauge reads \cite{book:photons_atoms_Tannoudji1989}:
\begin{align}
    H_{\text{Coul}} = \frac{1}{2}\int\mathrm d^3r\,\rho(\mathbf r,t)\phi(\mathbf r,t),
\end{align}
where \mbox{$\rho(\mathbf r,t) = \sum_i q_i\delta(\mathbf r-\mathbf r_i(t))$} is the electronic charge density with charge $q_i = -e$ and location $\mathbf r_i(t)$ of the $i$-th electron and $\phi(\mathbf r,t)$ is the electron--electron Coulomb potential, which solves the generalized Poisson equation:
\begin{align}
\label{eq:sec_derivation_coulomb_poissons_equation}
    \nabla_{\mathbf r}^{}\cdot \int\mathrm d^3r^{\prime}\,\mathrm dt^{\prime}\,\epsilon(\mathbf r,\mathbf r^{\prime},t,t^{\prime})\nabla_{\mathbf r^{\prime}}^{}\phi(\mathbf r^{\prime},t^{\prime})=-\frac{1}{\epsilon_0}\rho(\mathbf r,t).
\end{align}
Here, $\epsilon(\mathbf r,\mathbf r^{\prime},t,t^{\prime})$ is the relative spatially and temporally nonlocal background permittivity covering the dielectric environment and the thin semiconductor itself. The solution of \eqref{eq:sec_derivation_coulomb_poissons_equation} can be written in terms of a Green's function $G(\mathbf r,\mathbf r^{\prime},t,t^{\prime})$:
\begin{align}
    \phi(\mathbf r,t) = \int\mathrm d^3r^{\prime}\,\mathrm dt^{\prime}\,G(\mathbf r,\mathbf r^{\prime},t,t^{\prime})\rho(\mathbf r^{\prime},t^{\prime}).
    \label{eq:coulomb_potential_greens_function}
\end{align}
Plugging this into the Hamiltonian, we obtain:
\begin{align}
    H_{\text{Coul}} = \frac{1}{2}\int\mathrm d^3r\,\mathrm d^3r^{\prime}\,\mathrm dt^{\prime}\,\rho(\mathbf r,t)G(\mathbf r,\mathbf r^{\prime},t,t^{\prime})\rho(\mathbf r^{\prime},t^{\prime}).
    \label{eq:coul_ham_greensfunction}
\end{align}
The quantized electron charge density \mbox{$\rho(\mathbf r,t)\rightarrow \hat \rho(\mathbf r,t)$} reads \cite{hedin1970effects}:
\begin{align}
\begin{split}
    \hat \rho(\mathbf r,t)=&\,-e\int\mathrm d^3r^{\prime}\,\hat \Psi^{\dagger}_{}(\mathbf r^{\prime},t)\delta(\mathbf r-\mathbf r^{\prime})\hat \Psi(\mathbf r^{\prime},t)\\
    =&\,-e\hat \Psi^{\dagger}_{}(\mathbf r,t)\hat \Psi(\mathbf r,t),
    \end{split}
\end{align}
where $\hat \Psi(\mathbf r,t)$ is the corresponding fermionic Schrödinger field operator with fermionic equal-time commutation relations for a constrained system \cite{gergely2002hamiltonian,Schiff1968,gitman1990quantization,sundermeyer1982constrained}:
\begin{align}
    \left[\hat{\Psi}(\mathbf r,t),\hat{\Psi}^{\dagger}(\mathbf r^{\prime},t)\right]_+ = &\, \hat{\Psi}(\mathbf r,t)\hat{\Psi}^{\dagger}(\mathbf r^{\prime},t) + \hat{\Psi}^{\dagger}(\mathbf r^{\prime},t)\hat{\Psi}(\mathbf r,t)\notag\\
    = &\, \delta(\mathbf r-\mathbf r^{\prime}).
    \label{eq:SchrödingerFieldCommutator}
\end{align}
In the following, we set $\mathbf r = \begin{pmatrix}\rpar & z\end{pmatrix}^{\top}$, where $\rpar$ ($z$) represent the in-plane (out-of-plane) coordinates. The field operator is expanded in single-particle states:
\begin{align}
    \hat \Psi(\rpar,z,t) = \sum_{\lambda,\kpar{},s,n}\psi_{\lambda,\kpar{}}^{s}(\rpar)\chi_s\zeta_n(z)\andagtwo{\lambda,\kpar{}}{s,n}(t),
    \label{eq:field_operator}
\end{align}
where:
\begin{align}
    \psi_{\lambda,\mathbf k_{\parallel}}^{s}(\mathbf r_{\parallel}) = \frac{1}{\sqrt{\mathcal A}}\mathrm e^{\mathrm i\mathbf k_{\parallel}\cdot\mathbf r_{\parallel}}u_{\lambda,\mathbf k_{\parallel}}^{s}(\mathbf r_{\parallel})
\end{align}
are the Bloch functions with lattice-periodic Bloch factors $u_{\lambda,\mathbf k_{\parallel}}^{s}(\mathbf r_{\parallel})$ describing electrons in a two-dimensional periodic lattice in band $\lambda$ and in-plane momentum $\mathbf k_{\parallel}$ from within the first Brillouin zone (BZ) normalization area $\mathcal A$ within Born--von~K\'arm\'an boundary conditions \cite{vonkarman1912schwingungen}. 
By Fourier expanding $u_{\lambda,\mathbf k_{\parallel}}^{s}(\mathbf r_{\parallel}) = \sum_{\mathbf G_{\parallel}}\mathrm e^{\mathrm i\mathbf G_{\parallel}\cdot\mathbf r_{\parallel}}u_{\lambda,\mathbf k_{\parallel}+\mathbf G_{\parallel}}^{s}$ within the periodic-gauge condition satisfying
$\psi_{\lambda,\mathbf k_{\parallel}+\mathbf G_{\parallel}}^{s}=\psi_{\lambda,\mathbf k_{\parallel}}^{s}$ \cite{resta2000manifestations,martin2020electronic,Kittel1963}, we can also obtain the expression:
\begin{align}
\psi_{\lambda,\mathbf k_{\parallel}}^{s}(\mathbf r_{\parallel}) = \frac{1}{\sqrt{\mathcal A}}\sum_{\mathbf G_{\parallel}}\mathrm e^{\mathrm i(\mathbf k_{\parallel}+\mathbf G_{\parallel})\cdot\mathbf r_{\parallel}}u_{\lambda,\mathbf k_{\parallel}+\mathbf G_{\parallel}}^{s}.
\end{align}
Hence, a Bloch electron---in contrast to a free electron---does not possess a single momentum but is rather described as a wave packet, i.e., a superposition of plane waves with momenta $\mathbf k_{\parallel}+\mathbf G_{\parallel}$ weighted by the corresponding Bloch factor in reciprocal space $u_{\lambda,\mathbf k_{\parallel}+\mathbf G_{\parallel}}^{s}$. Moreover, in \eqref{eq:field_operator}, 
$\chi_s^{}$ are the eigenfunctions of the spin angular momentum operator in $z$ direction $\hat S_z = \frac{\hbar}{2}\sigma_z$ with Pauli matrix $\sigma_z$, which are normalized according to $\bar \chi_{s}^{}\cdot\chi_{s^{\prime}}^{}=\delta_{s,s^{\prime}}$, and $\zeta_n^{}(z)$ are the confinement functions in $z$ direction 
with confinement quantum number $n$. $a^{(\dagger)}\vphantom{a}_{\lambda,\mathbf k_{\parallel}}^{s,n}(t)$ are the annihilation (creation) operators, whose commutation relations follow from \eqref{eq:SchrödingerFieldCommutator}:
\begin{align}
    \left[\andag{i}(t),\adag{j}(t)\right] = \delta_{i,j}.
    \label{eq:commutation_relation_creation_annihilation_operators}
\end{align}
The factorization in \eqref{eq:field_operator} is a quasi-mesoscopic ansatz, which assumes in-plane, out-of-plane and spin contributions to be separable. We note that this is an approximation to simplify the notation. In App.~\ref{app:supercell_hamiltonian}, we show an exact approach, which is able to incorporate Bloch functions obtained from \textit{ab initio} calculations using supercells. We also note that we work in the reduced-zone scheme throughout the manuscript, i.e., the quasi- or crystal momenta $\mathbf k_{\parallel}$ are always restricted to the first Brillouin zone. 
In second quantization $H_{\text{Coul}}\rightarrow \hat H_{\text{Coul}}$, \eqref{eq:coul_ham_greensfunction} translates to:
\begin{multline}
    \hat H_{\text{Coul}} = \frac{1}{2}\sum_{\substack{\lambda_1,\dots\lambda_4,\\\mathbf k_{\parallel,1},\dots\mathbf k_{\parallel,4},\\s_1,\dots s_4,\\n_1,\dots n_4}}\int\mathrm dt^{\prime}\,\frac{e^2}{\mathcal A^2}\int\mathrm d^2r_{\parallel}^{}\,\mathrm d^2r^{\prime}_{\parallel}\,\mathrm dz\,\mathrm dz^{\prime}\,\\
    \times G(\mathbf r_{\parallel}^{},\mathbf r^{\prime}_{\parallel},z,z^{\prime},t,t^{\prime})\left(\bar{\chi}^{}_{s_1}\cdot\chi^{}_{s_4}\right)\left(\bar{\chi}^{}_{s_2}\cdot\chi^{}_{s_3}\right)\\
    \times \mathrm e^{-\mathrm i(\mathbf k_{\parallel,1}-\mathbf k_{\parallel,4})\cdot\mathbf r_{\parallel}^{}}
    u^*\vphantom{u}_{\lambda_1,\mathbf k_{\parallel,1}}^{s_1}(\mathbf r_{\parallel}^{})
    u^{s_4}_{\lambda_4,\mathbf k_{\parallel,4}}(\mathbf r_{\parallel})\zeta^*_{n_1}(z)\zeta^{}_{n_4}(z)\\
    \times \mathrm e^{-\mathrm i(\mathbf k_{\parallel,2}-\mathbf k_{\parallel,3})\cdot\mathbf r^{\prime}_{\parallel}}
    u^*\vphantom{u}_{\lambda_2,\mathbf k_{\parallel,2}}^{s_2}(\mathbf r^{\prime}_{\parallel})
    u^{s_3}_{\lambda_3,\mathbf k_{\parallel,3}}(\mathbf r^{\prime}_{\parallel})\zeta^*_{n_2}(z^{\prime})\zeta^{}_{n_3}(z^{\prime})
    \\
    \times \adagtwo{\lambda_1,\mathbf k_{\parallel,1}}{s_1,n_1}(t)\andagtwo{\lambda_4,\mathbf k_{\parallel,4}}{s_4,n_4}(t)\adagtwo{\lambda_2,\mathbf k_{\parallel,2}}{s_2,n_2}(t^{\prime})\andagtwo{\lambda_3,\mathbf k_{\parallel,3}}{s_3,n_3}(t^{\prime}).
\end{multline}
Note that we do not perform a normal ordering yet, since we cannot simply apply the equal-time commutation relation from \eqref{eq:commutation_relation_creation_annihilation_operators} in this effective two-time Hamiltonian here. 
We expand the Coulomb Green's function $G(\rpar,\rparp,z,z^{\prime},t,t^{\prime})$ in terms of its Fourier components \cite{wiser1963dielectric,adler1962quantum,rohlfing2000electron}:
\begin{multline}
    G(\rpar,\rparp,z,z^{\prime},t,t^{\prime}) 
    = 
    \frac{1}{\mathcal A}\sum_{\qpar{},\Gpar{},\Gparp{}}\mathrm e^{\mathrm i(\qpar{}+\Gpar{})\cdot\rpar}\\
    \times G_{\qpar{},\Gpar{},\Gparp{}}(z,z^{\prime},t,t^{\prime})\mathrm e^{-\mathrm i(\qpar{}+\Gparp{})\cdot\rparp},
    \label{eq:coulomb_greens_function_fourier_expansion_GGprime}
\end{multline}
where $\qpar{}$ is a momentum within the first Brillouin zone and $\Gpar{}$ and $\Gparp{}$ are in-plane reciprocal lattice vectors. Here, the summation runs over all reciprocal lattice vectors $\Gpar{}$ and $\Gparp{}$ to cover the total reciprocal lattice. The restriction to $\qpar{}$ results from demanding lattice periodicity: \mbox{$G(\rpar+\Rpar,\rparp+\Rpar,z,z^{\prime},t,t^{\prime}) \overset{!}{=} G(\rpar,\rparp,z,z^{\prime},t,t^{\prime})$}. In Fourier space, it also holds: $G_{\qpar{}+\Gparpp{},\Gpar{},\Gparp{}}^{}(z,z^{\prime},t,t^{\prime}) = G_{\qpar{},\Gpar{}+\Gparpp{},\Gparp{}+\Gparpp{}}^{}(z,z^{\prime},t,t^{\prime})$. The Fourier expansion in \eqref{eq:coulomb_greens_function_fourier_expansion_GGprime} accounts for all microscopic local-field effects with $\Gpar{}\neq\Gparp{}$ \cite{wiser1963dielectric,hybertsen1987ab}, whose explicit calculation requires \textit{ab initio} methods \cite{louie1975local,hybertsen1987ab,rohlfing2000electron,qiu2013optical,qiu2015nonanalyticity,qiu2016screening,andersen2015dielectric,sondersted2024improved,latini2015excitons,thygesen2017calculating}. 
These local-field effects originate from spatial variations of the charge distributions within the unit cell, which are on the scale of the lattice constant of approximately $0.3$\,nm for the common TMDC monolayers \cite{kormanyos2015k} or below, much smaller than the exciton Bohr radius of approximately 1\,nm. From a rigorous theoretical viewpoint, they are the result of an elimination of unwanted degrees of freedom \cite{ismail2001coupling,rosner2015wannier,erben2018excitation}, i.e., deeper valence/core bands or specific quantum-mechanical hierarchy strains, which will be discussed later on.

To illustrate the distinction between macroscopic and microscopic, i.e., local fields, we perform an averaging on the scale of the lattice vectors $\Rpar$ and $\Rparp$ over the unit-cell area $\mathcal A_{\text{UC}}$ by:
\begin{align}
\begin{split}
    &G(\Rpar,\Rparp,z,z^{\prime},t,t^{\prime}) \\
    &= \frac{1}{\mathcal A_{\text{UC}}}\int\mathrm d^2r_{\parallel}^{}\,\frac{1}{\mathcal A_{\text{UC}}}\int\mathrm d^2r^{\prime}_{\parallel}\,G(\rpar+\Rpar,\rpar+\Rparp,z,z^{\prime},t,t^{\prime})\\
    &=\frac{1}{\mathcal A}\sum_{\qpar,\Gpar,\Gparp}G_{\qpar{},\Gpar,\Gparp}^{}(z,z^{\prime},t,t^{\prime})\\
    &\quad\quad\times
    \frac{1}{\mathcal A_{\text{UC}}}\int_{\mathcal A_{\text{UC}}}\mathrm d^2r_{\parallel}^{}\, \mathrm e^{\mathrm i(\qpar+\Gpar)\cdot(\rpar+\Rpar)} \\
    &\quad\quad\times\frac{1}{\mathcal A_{\text{UC}}}\int_{\mathcal A_{\text{UC}}}\mathrm d^2r^{\prime}_{\parallel}\,
   \mathrm e^{-\mathrm i(\qpar+\Gparp)\cdot(\rparp+\Rparp)}\\
   &= \frac{1}{\mathcal A}\sum_{\qpar,\Gpar,\Gparp}G_{\qpar,\Gpar,\Gparp}^{}(z,z^{\prime},t,t^{\prime})
   \mathrm e^{\mathrm i\qpar\cdot (\Rpar-\Rparp)}\\
    &\quad\quad\times\frac{1}{\mathcal A_{\text{UC}}}\int_{\mathcal A_{\text{UC}}}\mathrm d^2r_{\parallel}^{}\, \mathrm e^{\mathrm i(\qpar+\Gpar)\cdot\rpar} \\
    &\quad\quad \times \frac{1}{\mathcal A_{\text{UC}}}\int_{\mathcal A_{\text{UC}}}\mathrm d^2r^{\prime}_{\parallel}\,
   \mathrm e^{-\mathrm i(\qpar+\Gparp)\cdot\rparp},
   \end{split}
\end{align}
where we used \mbox{$\mathrm e^{\mathrm i\Gpar\cdot\Rpar} = 1$} in the last step. The spatial integrals can be evaluated by assuming \mbox{$|\qpar|\ll |\Gpar|$}:
\begin{align}
\begin{split}
    \frac{1}{\mathcal A_{\text{UC}}}\int_{\mathcal A_{\text{UC}}}\mathrm d^2r_{\parallel}^{}\, \mathrm e^{\mathrm i(\qpar+\Gpar)\cdot\rpar} &\approx \frac{1}{\mathcal A_{\text{UC}}}\int_{\mathcal A_{\text{UC}}}\mathrm d^2r_{\parallel}^{}\, \mathrm e^{\mathrm i\Gpar\cdot\rpar}\\
    &= \delta_{\Gpar,\mathbf 0},
    \end{split}
\end{align}
as $\mathrm e^{\mathrm i\Gpar\cdot\rpar}$ varies rapidly within the unit cell. Altogether, we obtain:
\begin{align}
\begin{split}
    G(\Rpar,\Rparp,z,z^{\prime},t,t^{\prime}) = &\, 
    \frac{1}{\mathcal A}\sum_{\mathbf q}G_{\qpar,\mathbf 0,\mathbf 0}^{}(z,z^{\prime},t,t^{\prime})
   \mathrm e^{\mathrm i\qpar\cdot (\Rpar-\Rparp)}\\
   \equiv &\, G(\Rpar-\Rparp,z,z^{\prime},t,t^{\prime}).
   \label{eq:greens_function_macroscopically_averaged}
   \end{split}
\end{align}
Hence, the macroscopic Coulomb Green's function $G(\Rpar,\Rparp,z,z^{\prime},t,t^{\prime})$, which is the average over the unit cell of the microscopic Green's function $G(\rpar,\rparp,z,z^{\prime},t,t^{\prime})$, does not contain any local fields anymore and 
depends only on spatial distances. 
From \eqref{eq:greens_function_macroscopically_averaged}, the macroscopic limit---not to be confused with the limit $\qpar\rightarrow 0$---in Fourier space can be identified as $G_{\qpar,\mathbf 0,\mathbf 0}^{}(z,z^{\prime})$.

With the microscopic Coulomb Green's function in \eqref{eq:coulomb_greens_function_fourier_expansion_GGprime}, we obtain:
\begin{multline}
    \label{eq:Coulomb_Hamiltonian_SecondQuantized}
    \hat H_{\text{Coul}} = \frac{1}{2}\sum_{\substack{\lambda_1,\dots\lambda_4,\\\kpar{1},\dots\kpar{4},\\\qpar,\Gpar,\Gparp,s, s^{\prime},\\n_1,\dots n_4}}\int\mathrm dt^{\prime}\,V^{n_1,n_2,n_3,n_4}_{\qpar,\Gpar,\Gparp}(t,t^{\prime})\\
    \times \Upsilon_{\kpar{1},\kpar{4},\qpar+\Gpar}^{\lambda_1,\lambda_4,s}\Upsilon_{\kpar{2},\kpar{3},-\qpar-\Gparp}^{\lambda_2,\lambda_3,s^{\prime}} \\
    \times\adagtwo{\lambda_1,\kpar{1}}{s,n_1}(t)\andagtwo{\lambda_4,\kpar{4}}{s,n_4}(t)\adagtwo{\lambda_2,\kpar{2}}{s^{\prime},n_2}(t^{\prime})\andagtwo{\lambda_3,\kpar{3}}{s^{\prime},n_3}(t^{\prime}),
\end{multline}
where we defined the form factors as:
\begin{align}
    \Upsilon_{\kpar{},\kparp{},\qpar}^{\lambda,\lambda^{\prime},s} = & \,\frac{1}{\mathcal A}\int\mathrm d^2r_{\parallel}^{}\, \mathrm e^{-\mathrm i(\kpar-\kparp-\qpar)\cdot\mathbf r_{\parallel}^{}}u^*\vphantom{u}_{\lambda,\kpar{}}^s(\mathbf r_{\parallel}^{})u^{s}_{\lambda^{\prime},\kparp{}}(\mathbf r_{\parallel}^{}),
    \label{eq:formfactor_full}
\end{align}
and identified the time-nonlocal quantum-confined Coulomb potential with all local-field contributions $\Gpar$ and $\Gparp$ as:
\begin{multline}
\label{eq:Coulomb_Potential_Energy_SecondQuantized}
    V^{n_1,n_2,n_3,n_4}_{\qpar,\Gpar,\Gparp}(t,t^{\prime}) \\
     = \frac{e^2}{\mathcal A} \int \mathrm dz\,\mathrm dz^{\prime}\,\zeta^*_{n_1}(z)\zeta^{}_{n_4}(z)\zeta^*_{n_2}(z^{\prime})\zeta^{}_{n_3}(z^{\prime})\\
    \times 
    G_{\qpar,\Gpar,\Gparp}^{}(z,z^{\prime},t,t^{\prime}).
\end{multline}

Note that the Coulomb Hamiltonian in \eqref{eq:Coulomb_Hamiltonian_SecondQuantized} equally assumes a screened Coulomb interaction for direct electron--hole processes ($\adagtwo{\lambda,\kpar{1}}{}\adagtwo{\lambda^{\prime},\kpar{2}}{}\andagtwo{\lambda^{\prime},\kpar{3}}{}\andagtwo{\lambda,\kpar{4}}{}$, $\lambda\neq\lambda^{\prime}$) and electron--hole exchange processes ($\adagtwo{\lambda,\kpar{1}}{}\adagtwo{\lambda^{\prime},\kpar{2}}{}\andagtwo{\lambda,\kpar{3}}{}\andagtwo{\lambda^{\prime},\kpar{4}}{}$, $\lambda\neq\lambda^{\prime}$). In a fully microscopic treatment, the exchange contributions are unscreened, while the direct contributions are screened \cite{qiu2015nonanalyticity,rohlfing2000electron}. However, within a reduced few-band model, it turns out that screening the exchange interaction via the dielectric function of the material itself is necessary \cite{benedict2002screening}, since the interaction with all other bands is neglected. Similarly, if any substrate/superstrate material is present, the exchange contributions also have to take into account the screening by the dielectric environment \cite{deilmann2019important,qiu2017environmental}. 
Hence, within a few-band model, exchange and direct interaction can be treated on the same footing with respect to the screening. 
This way, we avoid an overestimation of the exchange interaction strength at the expense of a minor underestimation, since, if we use the full dielectric function, we double-count the influence of those bands, which we consider in the few-band model as active. We note that the exchange effects in, e.g., TMDCs are comparably small.

\subsection{Evaluating the Form Factors}
\label{sec:formfactors}

\begin{figure}
    \centering
    \subfigure[]{
    \label{fig:normal_process_square_lattice}
    \includegraphics[width=0.48\linewidth]{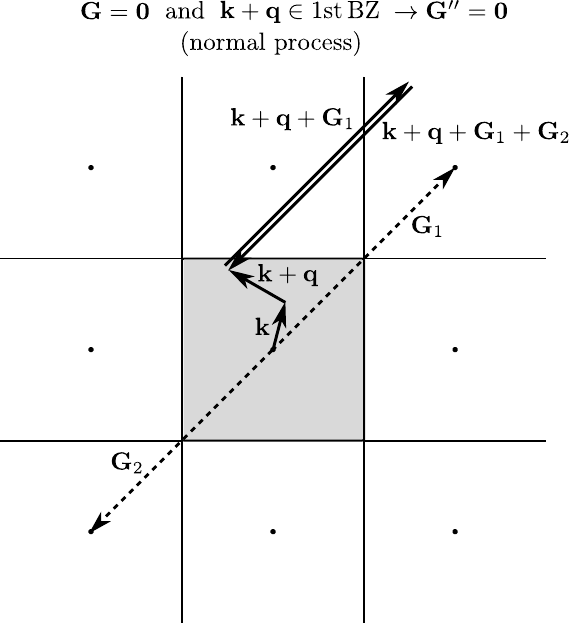}}
    \subfigure[]{
    \label{fig:umklapp_process_square_lattice}
    \includegraphics[width=0.465\linewidth]{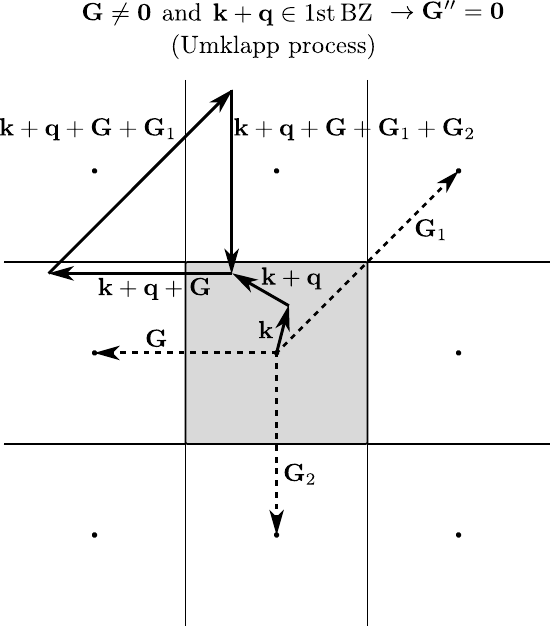}}
    \caption{Normal (a) and Umklapp (b) processes for a vanishing reciprocal lattice vector $\mathbf G^{\prime\prime}$ in an example square lattice.}
    \label{fig:square_lattice}
\end{figure}

Expanding the spatial integrals in \eqref{eq:formfactor_full} over the unit cells in a periodic lattice \cite{ziman2001electrons}:
\begin{multline}
    \Upsilon_{\kpar{},\kparp{},\qpar+\Gpar}^{\lambda,\lambda^{\prime},s}
    =\frac{1}{\mathcal A}\sum_{\Rpar}\mathrm e^{-\mathrm i(\kpar{}-\kparp{}-\qpar)\cdot\Rpar}\\
    \times\int_{\mathcal A_{\text{UC}}}\mathrm d^2r_{\parallel}^{}\, \mathrm e^{-\mathrm i(\kpar{}-\kparp{}-\qpar-\Gpar)\cdot\mathbf r}u^*\vphantom{u}^s_{\lambda,\kpar{}}(\rpar)u^{s}_{\lambda^{\prime},\kparp{}}(\rpar),
\end{multline}
evaluating the lattice sum:
\begin{align}
    \sum_{\Rpar}\mathrm e^{-\mathrm i(\kpar{}-\kparp{}-\qpar)\cdot\Rpar}
    = \mathcal N\sum_{\Gparpp
    }\delta_{\Gparpp,\kpar{}-\kparp{}-\qpar},
    \label{eq:lattice_sum}
\end{align}
where $\mathcal N$ is the total number of unit cells with area $\mathcal A_{\text{UC}}$, and Fourier expanding the Bloch factors as follows:
\begin{align}
    \label{eq:u_fourier_expansion_rG}
    u_{\lambda,\kpar{}}^s(\rpar) = \sum_{\Gpar}u_{\lambda,\kpar{},\Gpar}^s\mathrm e^{\mathrm i\Gpar\cdot\rpar},
\end{align}
which satisfy the periodic-gauge condition of the Bloch functions: $\psi_{\lambda,\kpar{}+\Gpar}^{s}=\psi_{\lambda,\kpar{}}^{s}$ \cite{resta2000manifestations,martin2020electronic,Kittel1963}, 
the form factors can be expressed as:
\begin{multline}
    \Upsilon_{\kpar{},\kparp{},\qpar+\Gpar}^{\lambda,\lambda^{\prime},s}\\
    = 
    \sum_{\Gparpp,\Gpari{1},\Gpari{2}}u^*\vphantom{u}_{\lambda,\kparp{}+\qpar+\Gparpp,-\Gpari{1}}^s u_{\lambda^{\prime},\kparp{},\Gpari{2}}^{s}\\
    \times\delta_{\Gparpp,\kpar{}-\kparp{}-\qpar}
    \delta_{\Gparpp,\Gpar+ \Gpari{1} + \Gpari{2}}.
    \label{eq:form_factor_evaluated}
\end{multline}
This expression is often used in explicit DFT/\textit{GW} methods \cite{hybertsen1986electron,martin2020electronic,kaasbjerg2013acoustic}, as it relies on a full reciprocal-space formulation. 
Here, the first Kronecker delta ensures that $\kparp{}+\qpar+\Gparpp\in \text{1st BZ}$ after evaluating the $\kpar{}$ sum in the Hamiltonian, while the second Kronecker delta ensures momentum conservation of the total scattering process. 
In an extended classification, we call processes with $\Gpar=\Gparpp=\mathbf 0$ \textit{normal} processes, cf.~Fig.~\ref{fig:normal_process_square_lattice}, and processes with $\Gpar\neq\mathbf 0$ \textit{or} $\Gparpp\neq\mathbf 0$ \textit{Umklapp} processes, cf.~Fig.~\ref{fig:umklapp_process_square_lattice}. This implies that even a process with $\kparp{}+\qpar\in \text{1st BZ}$ is classified as an \textit{Umklapp} process, as long as a nonzero reciprocal lattice vector $\Gpar$ entering the argument of the Coulomb potential $V_{\qpar,\Gpar,\Gparp}$ is involved, cf.~Fig.~\ref{fig:umklapp_process_square_lattice}.
Our definition of Umklapp processes closely resembles the one in Refs.~\cite{kaasbjerg2013acoustic,vogl1976microscopic,cohen2016fundamentals} and is not equal to the usual narrower definition in textbooks \cite{pines2018elementary,ziman2001electrons,mahan2000many,czycholl2007theoretische}, which only relates processes with $\kpar{}+\qpar\notin\text{1st BZ}$ to Umklapp processes, i.e., processes with $\Gparpp\neq\mathbf 0$. However, even the classification of a process being Umklapp in the standard and narrower sense contains some arbitrariness, i.e., the same process $\kpar{}+\qpar$ can be classified as Umklapp or not depending on the exact choice of the Brillouin zone. As a consequence, the magnitude of the scattering matrix elements, i.e., the Bloch form factors $\Upsilon_{\kpar{},\kparp{},\qpar+\Gpar}^{\lambda,\lambda^{\prime},s}$, \textit{do not} depend on the reciprocal lattice vector $\Gparpp$ originating from the lattice sum in \eqref{eq:lattice_sum}---in contrast to the reciprocal lattice vector $\Gpar$ originating from the Fourier transformation in \eqref{eq:coulomb_greens_function_fourier_expansion_GGprime}. In the end, it is just a classification. What matters, is that the description captures all possible processes, and not, how they are labeled.

\begin{figure}
    \centering
    \includegraphics[width=1\linewidth]{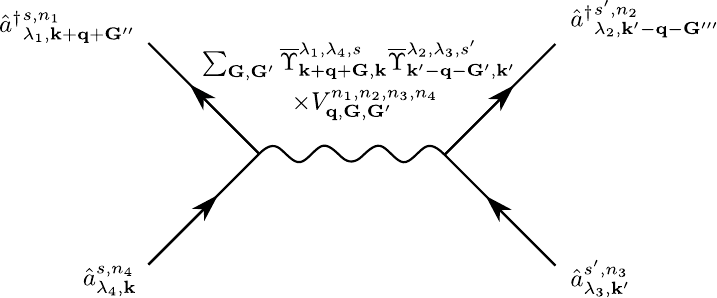}
    \caption{Scheme of the Coulomb scattering processes described by \eqref{eq:Coulomb_Hamiltonian_SecondQuantized_Formfactors_Evaluated}.}
    \label{fig:coulomb_scattering_processes_general}
\end{figure}

The resulting Hamiltonian then reads:
\begin{multline}
    \label{eq:Coulomb_Hamiltonian_SecondQuantized_Formfactors_Evaluated}
    \hat H_{\text{Coul}} 
    = \frac{1}{2}\sum_{\substack{\lambda_1,\dots\lambda_4,\kpar{}\kparp{},\qpar,\\\Gpar,\Gparp,s, s^{\prime},n_1,\dots n_4,\\
    \Gparpp|_{\kpar{}+\qpar+\Gparpp\in\text{1st BZ}},\\
    \Gparppp|_{\kparp{}-\qpar-\Gparppp\in\text{1st BZ}}}}
    \int\mathrm dt^{\prime}\,V^{n_1,n_2,n_3,n_4}_{\qpar,\Gpar,\Gparp}(t,t^{\prime})\\
    \times
    \overline \Upsilon_{\kpar{}+\qpar+\Gpar,\kpar{}}^{\lambda_1,\lambda_4,s}
    \overline \Upsilon_{\kparp{}-\qpar-\Gparp,\kparp{}}^{\lambda_2,\lambda_3,s^{\prime}}\\
    \times\adagtwo{\lambda_1,\kpar{}+\qpar+\Gparpp}{s,n_1}(t)\andagtwo{\lambda_4,\kpar{}}{s,n_4}(t)\adagtwo{\lambda_2,\kparp{}-\qpar-\Gparppp}{s^{\prime},n_2}(t^{\prime})\andagtwo{\lambda_3,\kparp{}}{s^{\prime},n_3}(t^{\prime}),
\end{multline}
where:
\begin{align}
    \overline \Upsilon_{\kpar{},\kparp{}}^{\lambda,\lambda^{\prime},s} = \sum_{\Gpar}u^*\vphantom{u}_{\lambda,\kpar{},\Gpar}^s u^{s}_{\lambda^{\prime},\kparp{},\Gpar},
    \label{eq:reduced_form_factor}
\end{align}
is the reduced form factor. In Fig.~\ref{fig:coulomb_scattering_processes_general}, we schematically depict the scattering processes described by \eqref{eq:Coulomb_Hamiltonian_SecondQuantized_Formfactors_Evaluated}.

Note that in the calculation of standard BSE kernels \cite{qiu2015nonanalyticity,deslippe2012berkeleygw}, only the reciprocal lattice vectors $\Gpar$ and $\Gparp$ originating from the Fourier transformation in \eqref{eq:coulomb_greens_function_fourier_expansion_GGprime} appear, as $\Gparpp$ and $\Gparppp$ originate from explicitly evaluating the lattice sum in \eqref{eq:lattice_sum}, which is often only done implicitly.

Also, we note that, contrary to their outer appearance in \eqref{eq:Coulomb_Hamiltonian_SecondQuantized_Formfactors_Evaluated} and Fig.~\ref{fig:coulomb_scattering_processes_general}, many-body scattering processes with $\Gparpp\neq\Gparppp$ do not violate actual momentum conservation, since the quasi-momentum of a Bloch electron is not equal to the kinetic momentum of a bare electron.

From now on, we treat the reciprocal lattice vectors $\Gparpp$ and $\Gparppp$ describing Umklapp processes in the narrower sense implicitly, which is often done in the literature, and drop their summation. 
This can be safely done, since the matrix elements do not depend on these specific reciprocal lattice vectors $\Gparpp$ and $\Gparppp$, as discussed earlier.

\subsection{Expansion Over the Band Extrema}
\label{sec:expansion_band_extrema}

We expand the momenta around the band extrema denoted by $\xi$, i.e., for a TMDC, the high-symmetry points $\mathbf K^{\xi}$:
\begin{align}
    \kpar{}\rightarrow \kpar{}+\mathbf K^{\xi},\quad\quad \kparp{}\rightarrow \kparp{}+\mathbf K^{\xi^{\prime}}.
\end{align}
where $\xi = \Gamma, K, K^{\prime},\Lambda,\Lambda^{\prime}$ with $\mathbf K^{\Gamma} = \mathbf 0$. 
Momentum transfers $\qpar$ within the first Brillouin zone can also expanded around the band extrema:
\begin{align}
    \qpar\rightarrow \qpar + \mathbf K^{\xi^{\prime\prime}}.
\end{align}
The Bloch factors are then written as:
\begin{align}
    u_{\lambda,\kpar{}}^{s} \rightarrow u_{\lambda,\kpar{}+\mathbf K^{\xi}}^{s} \equiv u_{\lambda,\kpar{}}^{\xi,s},
\end{align}
where $u_{\lambda,\kpar{}}^{\xi,s}$ is the Bloch factor valid for small momenta $\kpar{}$ around valley $\xi$. Correspondingly, the valley-expanded Bloch factors read:
\begin{align}
    \overline \Upsilon_{\kpar{},\kparp{}}^{\lambda,\lambda^{\prime},\xi,\xi^{\prime},s} = \sum_{\Gpar}u^*\vphantom{u}_{\lambda,\kpar{},\Gpar}^{\xi,s} u^{\xi^{\prime},s}_{\lambda^{\prime},\kparp{},\Gpar}.
    \label{eq:reduced_form_factor_valleyexpanded}
\end{align}
The resulting Hamiltonian reads:
\begin{multline}
\label{eq:Coulomb_Hamiltonian_SecondQuantized_ValleyExpanded}
    \hat H_{\text{Coul}}=  \frac{1}{2}\sum_{\substack{\lambda_1,\dots\lambda_4,\kpar{},\kparp{},\\\qpar,\Gpar,\Gparp,\\
    s,s^{\prime},\xi,\xi^{\prime},\xi^{\prime\prime},n_1,\dots n_4
    }} \int\mathrm dt^{\prime}\,V^{n_1,n_2,n_3,n_4}_{\qpar + \mathbf K^{\xi^{\prime\prime}},\Gpar,\Gparp}(t,t^{\prime})\\
    \times
    \overline \Upsilon_{\kpar{}+\qpar+\Gpar,\kpar{}}^{\lambda_1,\lambda_4,\xi+\xi^{\prime\prime},\xi,s}\overline \Upsilon_{\kparp{}-\qpar-\Gparp,\kparp{}}^{\lambda_2,\lambda_3,\xi^{\prime}-\xi^{\prime\prime},\xi^{\prime},s^{\prime}}\\
    \times\adagtwo{\lambda_1,\kpar{}+\qpar}{\xi+\xi^{\prime\prime},s,n_1}(t)\andagtwo{\lambda_4,\kpar{}}{\xi,s,n_4}(t)\adagtwo{\lambda_2,\kparp{}-\qpar}{\xi^{\prime}-\xi^{\prime\prime},s^{\prime},n_2}(t^{\prime})\andagtwo{\lambda_3,\kparp{}}{\xi^{\prime},s^{\prime},n_3}(t^{\prime}).
    \end{multline}
Here, as a shorthand notation, we write: $\xi-\xi^{\prime} \equiv \mathbf K^{\xi}-\mathbf K^{\xi^{\prime}}$.

\subsection{Taylor Expansion of the Form Factor in Small Momentum Transfers}
\label{sec:taylor_expansion_form_factors}

A Taylor expansion of the reduced form factors for small momentum transfer $\qpar$ yields:
\begin{multline}
\overline \Upsilon_{\kpar{}+\qpar+\Gpar,\kpar{}}^{\lambda,\lambda^{\prime},\xi,\xi^{\prime},s}\\
= 
    \frac{1}{\mathcal A_{\text{UC}}}\int_{\mathcal A_{\text{UC}}}\mathrm d^2r_{\parallel}^{}\, u^*_{}\vphantom{u}^{\xi,s}_{\lambda,\kpar{}+\qpar}(\rpar)\mathrm e^{\mathrm i\Gpar\cdot\rpar}u^{\xi^{\prime},s}_{\lambda^{\prime},\kpar{}}(\rpar)\\
    = \frac{1}{\mathcal A_{\text{UC}}}\int_{\mathcal A_{\text{UC}}}\mathrm d^2r_{\parallel}^{}\,
    \left(\left(\vphantom{\frac{1}{2}}1+\qpar\cdot\nabla_{\kpar{}}+\frac{1}{2}\mleft(\qpar\cdot\nabla_{\kpar{}}\mright)\right.\right.\\
    \left.\left.\vphantom{\frac{1}{2}}\times\mleft(\qpar\cdot\nabla_{\kpar{}}\mright) + \mathcal O((\qpar)^3)\right)
    u^*_{}\vphantom{u}^{\xi,s}_{\lambda,\kpar{}}(\rpar)\right)\mathrm e^{\mathrm i\Gpar\cdot\rpar}u^{\xi^{\prime},s}_{\lambda^{\prime},\kpar{}}(\rpar).
\end{multline}

We consider terms up to second order:
\begin{multline}
    \overline \Upsilon_{\kpar{}+\qpar+\Gpar,\kpar{}}^{\lambda,\lambda^{\prime},\xi,\xi^{\prime},s}\\
    \approx  \frac{1}{\mathcal A_{\text{UC}}}\int_{\mathcal A_{\text{UC}}}\mathrm d^2r_{\parallel}^{}\, u^*_{}\vphantom{u}^{\xi,s}_{\lambda,\kpar{}}(\rpar)\mathrm e^{\mathrm i\Gpar\cdot\rpar}u^{\xi^{\prime},s}_{\lambda^{\prime},\kpar{}}(\rpar)\\
     + \frac{1}{\mathcal A_{\text{UC}}}\int_{\mathcal A_{\text{UC}}}\mathrm d^2r_{\parallel}^{}\, \left(\left(\qpar\cdot\nabla_{\kpar{}}\right) u^*_{}\vphantom{u}^{\xi,s}_{\lambda,\kpar{}}(\rpar{})\right)\mathrm e^{\mathrm i\Gpar\cdot\rpar}u^{\xi^{\prime},s}_{\lambda^{\prime},\kpar{}}(\rpar)\\
     - \frac{1}{2}\frac{1}{\mathcal A_{\text{UC}}}\int_{\mathcal A_{\text{UC}}}\mathrm d^2r_{\parallel}^{}\, \left(\left(\qpar\cdot\nabla_{\kpar{}}\right) u^*_{}\vphantom{u}^{\xi,s}_{\lambda,\kpar{}}(\rpar)\right)\mathrm e^{\mathrm i\Gpar\cdot\rpar}\\
     \times\left(\qpar\cdot\nabla_{\kpar{}}\right)u^{\xi^{\prime},s}_{\lambda^{\prime},\kpar{}}(\rpar)
    .
\end{multline}
In the last term, we shifted the momentum derivative to the second Bloch factor by assuming vanishing boundary terms in $\kpar{}$ and by neglecting any $\kpar{}$ derivative on the creation/annihilation operators. 
Inserting a unit matrix where appropriate, we obtain:
\begin{multline}
    \overline \Upsilon_{\kpar{}+\qpar+\Gpar,\kpar{}}^{\lambda,\lambda^{\prime},\xi,\xi^{\prime},s} \\
    \approx
    \frac{1}{\mathcal A_{\text{UC}}}\int_{\mathcal A_{\text{UC}}}\mathrm d^2r_{\parallel}^{}\, u^*_{}\vphantom{u}^{\xi,s}_{ \lambda,\kpar{}}(\rpar)
    \mathrm e^{\mathrm i\Gpar\cdot\rpar}u^{\xi^{\prime},s}_{\lambda^{\prime},\kpar{}}(\rpar)\\
    +\sum_{\lambda^{\prime\prime}}\frac{1}{\mathcal A_{\text{UC}}}\int_{\mathcal A_{\text{UC}}}\mathrm d^2r_{\parallel}^{}\, \left(\left(\qpar\cdot\nabla_{\kpar{}}\right)u^*_{}\vphantom{u}^{\xi,s}_{\lambda,\kpar{}}(\rpar)\right)u^{\xi,s}_{\lambda^{\prime\prime},\kpar{}}(\rpar)\\
    \times
    \frac{1}{\mathcal A_{\text{UC}}}\int_{\mathcal A_{\text{UC}}}\mathrm d^2r^{\prime}_{\parallel}\, u^*_{}\vphantom{u}^{\xi,s}_{\lambda^{\prime\prime},\kpar{}}(\rparp)
    \mathrm e^{\mathrm i\Gpar\cdot\rparp}u^{\xi^{\prime},s}_{\lambda^{\prime},\kpar{}}(\rparp)\\
     - \frac{1}{2}\sum_{\lambda^{\prime\prime},\lambda^{\prime\prime\prime}}\frac{1}{\mathcal A_{\text{UC}}}\int_{\mathcal A_{\text{UC}}}\mathrm d^2r_{\parallel}^{}\, \left(\left(\qpar\cdot\nabla_{\kpar{}}\right)u^*_{}\vphantom{u}^{\xi,s}_{\lambda,\kpar{}}(\rpar)\right)\\
     \times u^{\xi,s}_{\lambda^{\prime\prime},\kpar{}}(\rpar)\\
     \times
    \frac{1}{\mathcal A_{\text{UC}}}\int_{\mathcal A_{\text{UC}}}\mathrm d^2r^{\prime}_{\parallel}\, u^*_{}\vphantom{u}^{\xi,s}_{\lambda^{\prime\prime},\kpar{}}(\rparp)
    \mathrm e^{\mathrm i\Gpar\cdot\rparp}
    u^{\xi,s}_{\lambda^{\prime\prime\prime},\kpar{}}(\rparp)\\
     \times\frac{1}{\mathcal A_{\text{UC}}}\int_{\mathcal A_{\text{UC}}}\mathrm d^2r^{\prime\prime}_{\parallel}\,
     u^*_{}\vphantom{u}^{\xi,s}_{\lambda^{\prime\prime\prime},\kpar{}}(\rparpp)
    \left(\qpar\cdot\nabla_{\kpar{}}\right)u^{\xi^{\prime},s}_{\lambda^{\prime},\kpar{}}(\rparpp).
\end{multline}
By identifying:
\begin{align}
    \mathbf d_{\kpar{}}^{\lambda,\lambda^{\prime},\xi,\xi^{\prime},s} = -\mathrm ie\frac{1}{\mathcal A_{\text{UC}}}\int_{\mathcal A_{\text{UC}}}\mathrm d^2r_{\parallel}^{}\,u^*_{}\vphantom{u}_{\lambda,\kpar{}}^{\xi,s}(\rpar)\nabla_{\kpar{}}u_{\lambda^{\prime},\kpar{}}^{\xi^{\prime},s}(\rpar),
    \label{eq:dipole_moment}
\end{align} which constitutes the dipole matrix element for \mbox{$\lambda^{\prime}=\bar\lambda$} and the Berry connection for \mbox{$\lambda^{\prime}=\lambda$} \cite{aversa1995nonlinear,xiao2010berry,mkrtchian2019theory}, we can write:
\begin{align}
    \label{eq:formfactor_zeroth}
    &\overline \Upsilon_{\kpar{}+\qpar+\Gpar,\kpar{}}^{\lambda,\lambda^{\prime},\xi,\xi^{\prime},s} \approx
    \overline \Upsilon_{\kpar{}+\Gpar,\kpar{}}^{\lambda,\lambda^{\prime},\xi,\xi^{\prime},s}\\
    \label{eq:formfactor_first}
    &\quad\quad-\frac{\mathrm i}{e}\sum_{\lambda^{\prime\prime}}
    \left(\qpar\cdot\mathbf d_{\kpar{}}^{\lambda,\lambda^{\prime\prime},\xi,\xi,s}\right)
    \overline \Upsilon_{\kpar{}+\Gpar,\kpar{}}^{\lambda^{\prime\prime},\lambda^{\prime},\xi,\xi^{\prime},s}\\
    \begin{split}
    \label{eq:formfactor_second}
    &\quad\quad - \frac{1}{2e^2}\sum_{\lambda^{\prime\prime},\lambda^{\prime\prime\prime}}
    \left(\qpar\cdot\mathbf d_{\kpar{}}^{\lambda,\lambda^{\prime\prime},\xi,\xi,s}\right)
    \overline \Upsilon_{\kpar{}+\Gpar,\kpar{}}^{\lambda^{\prime\prime},\lambda^{\prime\prime\prime},\xi,\xi,s}\\
    &\quad\quad\quad\quad\times    \left(\qpar\cdot\mathbf d_{\kpar{}}^{\lambda^{\prime\prime\prime},\lambda^{\prime},\xi,\xi^{\prime},s}\right)
    .
    \end{split}
\end{align}

This expression is similar to a $\qpar\cdot\mathbf p_{\parallel}^{}$ expansion up to second order at the band extrema.

\section{Screening}
\label{sec:screening}

In this section, we 
discuss 
the validity of applying the \textit{generalized} Poisson equation in \eqref{eq:sec_derivation_coulomb_poissons_equation} giving rise to the screened quantum-confined Coulomb potential in \eqref{eq:Coulomb_Potential_Energy_SecondQuantized} entering the many-body Coulomb Hamiltonian in \eqref{eq:Coulomb_Hamiltonian_SecondQuantized}.

Starting from scratch, we have to consider the unscreened Poisson equation for the total potential $\phi_{\text{tot}}$ induced by the total charge density of the system $\rho_{\text{tot}}$ derived from Gauss's law $\begin{pmatrix}\nabla_{\rpar}^{} &\partial_z\end{pmatrix}^{\top}\cdot\mathbf E_{\parallel}(\rpar,z,t) = \frac{1}{\epsilon_0}\rho_{\text{tot}}(\rpar,z,t)$ for a longitudinal electric field $\mathbf E_{\parallel}(\rpar,z,t) = -\begin{pmatrix}
    \nabla_{\rpar}&\partial_z
\end{pmatrix}^{\top}\phi_{\text{tot}}(\rpar,z,t)$:
\begin{align}
    \begin{pmatrix}\nabla_{\rpar}\\\partial_z\end{pmatrix}\cdot\begin{pmatrix}\nabla_{\rpar}\\\partial_z\end{pmatrix}\phi_{\text{tot}}(\rpar,z,t) = - \frac{1}{\epsilon_0}\rho_{\text{tot}}(\rpar,z,t).
    \label{eq:poisson_equation_bare}
\end{align}
Expanding the total potential in terms of the bare Green's function $G_{0}(\rpar,\rparp,z,z^{\prime},t,t^{\prime})$:
\begin{multline}
    \phi_{\text{tot}}(\rpar,z,t) \\
    = \int \mathrm d^2r^{\prime}_{\parallel}\,\mathrm dz^{\prime}\,\mathrm dt^{\prime}\,G_{0}(\rpar,\rparp,z,z^{\prime},t,t^{\prime})\rho_{\text{tot}}(\rparp,z^{\prime},t^{\prime}),
    \label{eq:total_potential_total_charge_density}
\end{multline}
and performing a Fourier transformation with respect to the in-plane coordinates $\rpar$, yields the following equation for the bare Green's function $G_{0,\qpar,\Gpar,\Gparp}$:
\begin{multline}
    \left(-\left(\qpar+\Gpar\right)^2 + \partial_z^2\right)G_{0,\qpar,\Gpar,\Gparp}(z,z^{\prime},t,t^{\prime})\\
    = - \frac{1}{\epsilon_0}\delta(z-z^{\prime})\delta(t-t^{\prime})\delta_{\Gpar,\Gparp}.
    \label{eq:poisson_equation_bare_greens_function}
\end{multline}
The solution reads:
\begin{align}
\begin{split}
    G_{0,\qpar,\Gpar,\Gparp}(z,z^{\prime},t,t^{\prime}) = &\, \delta_{\Gpar,\Gparp}\delta(t-t^{\prime})\frac{\mathrm e^{-|\qpar+\Gpar||z-z^{\prime}|}}{2\epsilon_0|\qpar+\Gpar|}\\
    \equiv &\, G_{0,\qpar+\Gpar}(z-z^{\prime})\delta_{\Gpar,\Gparp}\delta(t-t^{\prime}),
    \end{split}
    \label{eq:bare_greens_function_asymmetric}
\end{align}
is diagonal in $\Gpar$ and does not include any memory effects. 
In this case, if we wanted to calculate the optical response of a thin semiconductor, we would 
have to include a large amount of bands in the many-body Coulomb interaction, 
which would result in a large amount of correlations via Heisenberg equations of motion. Hence, we would strongly underestimate the screening if we restricted the considered bands to just a few 
e.g., restricting the phase space to the topmost valence band and the lowest conduction band.

To circumvent the necessity of computing a large amount of correlations, we divide the total charge density $\rho_{\text{tot}}(\rpar,z,t)$ in a background contribution $\rho_{\text{b}}(\rpar,z,t)$ and an active contribution $\rho(\rpar,z,t)$ with:
\begin{align}
    \rho_{\text{tot}}(\rpar,z,t) = \rho_{\text{b}}(\rpar,z,t) + \rho(\rpar,z,t),
    \label{eq:charge_density_total}
\end{align}
where the active contribution $\rho(\rpar,z,t)$ contains the reduced number of active bands and the background contribution $\rho_{\text{b}}(\rpar,z,t)$ contains all other bands including possible sub-/superstrate materials. 
Then, from $\rho_{\text{b}}(\rpar,z,t) = -\begin{pmatrix}\nabla_{\rpar} & \partial_z\end{pmatrix}^{\top}\cdot \mathbf P_{\parallel}(\rpar,z,t)$ 
and using the linear response of a longitudinal polarization density $\mathbf P_{\parallel}(\rpar,z,t)$ due to a spatially and temporally nonlocal background described by the electronic susceptibility $\chi(\rpar,\rparp,z,z^{\prime},t,t^{\prime})$, 
$\mathbf P_{\parallel}(\rpar,z,t) = \epsilon_0\int\mathrm d^2r^{\prime}_{\parallel}\,\mathrm dz^{\prime}\,\mathrm dt^{\prime}\,\chi(\rpar,\rparp,z,z^{\prime},t,t^{\prime})\mathbf E_{\parallel}(\rparp,z^{\prime},t^{\prime})$, we obtain the relation:
\begin{multline}
    \rho_{\text{b}}(\rpar,z,t) = \epsilon_0\begin{pmatrix}\nabla_{\rpar}\\\partial_z\end{pmatrix}\cdot\int\mathrm d^2r^{\prime}_{\parallel}\,\mathrm dz^{\prime}\,\mathrm dt^{\prime}\,\\
    \times \chi(\rpar,\rparp,z,z^{\prime},t,t^{\prime})\begin{pmatrix}\nabla_{\rparp}\\\partial_{z^{\prime}}\end{pmatrix}\phi_{\text{tot}}(\rparp,z^{\prime},t^{\prime}),
    \label{eq:relation_rho_b_phi_tot}
\end{multline}
which, via defining the nonlocal dielectric function:
\begin{multline}
    \epsilon(\rpar,\rparp,z,z^{\prime},t,t^{\prime})\\
    = \delta(\rpar-\rparp)\delta(z-z^{\prime})\delta(t-t^{\prime})
    + \chi(\rpar,\rparp,z,z^{\prime},t,t^{\prime}),
    \label{eq:epsilon_nonlocal_realspace}
\end{multline}
yields the generalized or screened Poisson equation \cite{pasenow2005excitonic,zimmermann2016poisson}, cf.\ also \eqref{eq:sec_derivation_coulomb_poissons_equation}:
\begin{multline}
    \begin{pmatrix}\nabla_{\rpar}\\\partial_z\end{pmatrix}\cdot\int \mathrm d^2r^{\prime}_{\parallel}\,\mathrm dz^{\prime}\,\mathrm dt^{\prime}\,\epsilon(\rpar,\rparp,z,z^{\prime},t,t^{\prime})\\
    \times\begin{pmatrix}\nabla_{\rparp}\\\partial_{z^{\prime}}\end{pmatrix}
    \phi_{\text{tot}}(\rparp,z^{\prime},t^{\prime})
    = -\frac{1}{\epsilon_0}\rho(\rpar,z,t).
    \label{eq:generalized_poisson_equation2}
\end{multline}
Now, the total potential $\phi_{\text{tot}}(\rpar,z,t)$ is induced via the active charge density $\rho(\rpar,z,t)$ only, while the influence of the background is encoded in the nonlocal permittivity $\epsilon(\rpar,\rparp,z,z^{\prime},t,t^{\prime})$. We expand the total potential in terms of a screened Green's function $G(\rpar,\rparp,z,z^{\prime},t,t^{\prime})$ and the active charge density $\rho(\rpar,z,t)$, cf.\ also \eqref{eq:coulomb_potential_greens_function}:
\begin{align}
    \phi_{\text{tot}}(\rpar,z,t) = \int \mathrm d^2r^{\prime}_{\parallel}\,\mathrm dz^{\prime}\,\mathrm dt^{\prime}\,G(\rpar,\rparp,z,z^{\prime},t,t^{\prime})\rho(\rparp,z^{\prime},t^{\prime}),
\end{align}
and, 
after 
an in-plane Fourier transformation, 
we obtain:
\begin{multline}
    \sum_{\Gparp}\int \mathrm dt^{\prime}\,\int\mathrm dz^{\prime}\, \begin{pmatrix}
        \mathrm i(\qpar+\Gpar)\\\partial_z
    \end{pmatrix}^{\top}\epsilon_{\qpar,\Gpar,\Gparp}(z,z^{\prime},t,t^{\prime})\\
    \times\begin{pmatrix}
        \mathrm i(\qpar+\Gparp)\\\partial_{z^{\prime}}
    \end{pmatrix}
    G_{\qpar,\Gparp,\Gparpp}(z^{\prime},z^{\prime\prime},t^{\prime},t^{\prime\prime})\\
    = - \frac{1}{\epsilon_0}\delta_{\Gpar,\Gparpp}\delta(z-z^{\prime\prime})\delta(t-t^{\prime\prime}).
    \label{eq:poisson_equation_screened_greens_function}
\end{multline}
We observe that the screened Green's function $G_{\qpar,\Gparp,\Gparpp}(z,z^{\prime},t,t^{\prime})$ is not diagonal in $\Gpar$ and includes memory effects, as long as the dielectric function $\epsilon_{\qpar,\Gpar,\Gparp}(z,z^{\prime},t,t^{\prime})$ is non-local in space and time. Hence, with the corresponding dielectric function, $G_{\qpar,\Gparp,\Gparpp}(z,z^{\prime},t,t^{\prime})$ is the correct Green's function to use in our few-band many-body Coulomb Hamiltonian in \eqref{eq:Coulomb_Hamiltonian_SecondQuantized_Formfactors_Evaluated} or \eqref{eq:Coulomb_Hamiltonian_SecondQuantized_ValleyExpanded} and the problem of screening is shifted to the evaluation of the microscopic dielectric function $\epsilon_{\qpar,\Gpar,\Gparp}(z,z^{\prime},t,t^{\prime})$. 

In the following, we provide the link to usual literature definitions of the dielectric function in the macroscopic limit in three and two dimensions. 
In a three-dimensional periodic lattice in the static limit ($\omega = 0$) and by redefining:
\begin{align}
    \tilde \epsilon_{\mathbf q,\mathbf G,\mathbf G^{\prime}}
    = \frac{
    \left(\mathbf q+\mathbf G\right)\cdot \left(\mathbf q + \mathbf G^{\prime}\right)
    }{\left| \mathbf q+\mathbf G \right| \left| \mathbf q+\mathbf G^{\prime} \right| 
    }
    \epsilon_{\mathbf q,\mathbf G,\mathbf G^{\prime}},
\end{align}
which is the longitudinal dielectric function \cite{mochan1985electromagnetic_2,johnson1975local} and
$\frac{\left(\mathbf q+\mathbf G\right)\cdot \left(\mathbf q + \mathbf G^{\prime}\right)}{\left| \mathbf q+\mathbf G \right| \left| \mathbf q+\mathbf G^{\prime} \right|}$ is the longitudinal projector. Hence, in a microscopic approach, the longitudinal projection of the dielectric function occurs naturally. We then find for the screened Green's function in the static limit via matrix inversion:
\begin{align}
    G_{\mathbf q,\mathbf G,\mathbf G^{\prime}} = \frac{\left(\tilde \epsilon^{-1}\right)_{\mathbf q,\mathbf G,\mathbf G^{\prime}}}{\epsilon_0\left| \mathbf q+\mathbf G \right| \left| \mathbf q+\mathbf G^{\prime} \right| },
\end{align}
which corresponds exactly to the expression from Rohlfing and Louie in Ref.~\cite{rohlfing2000electron}. Note that 
$\left(\tilde \epsilon^{-1} \right)_{\mathbf q,\mathbf G,\mathbf G^{\prime}}$ is the inverse of the full dielectric matrix 
$\tilde \epsilon_{\mathbf q,\mathbf G,\mathbf G^{\prime}}$.

However, in practical calculations, the dielectric function is defined via relating an external perturbing potential $\phi_{\text{ext}}$ to the total potential $\phi_{\text{tot}}$ in the material. To obtain such a description, we combine \eqref{eq:total_potential_total_charge_density}, \eqref{eq:charge_density_total} and \eqref{eq:relation_rho_b_phi_tot}:
\begin{multline}
    \phi_{\text{tot},\mathbf q+\mathbf G}(\omega)
    = \sum_{\mathbf G^{\prime}} \frac{1}{2\pi}\int\mathrm d\omega^{\prime}\left( \varepsilon^{-1}_{\text{mic}}\right)_{\mathbf q,\mathbf G,\mathbf G^{\prime}}(\omega,- \omega^{\prime}) \\
    \times
    \phi_{\text{ext},\mathbf q+\mathbf G^{\prime}}(\omega^{\prime}),
    \label{eq:relation_total_external_potential_fourierspace}
\end{multline}
with:
\begin{multline}
    \varepsilon_{\text{mic},\mathbf q,\mathbf G,\mathbf G^{\prime}}(\omega,-\omega^{\prime}) = \delta_{\mathbf G,\mathbf G^{\prime}}2\pi\delta(\omega-\omega^{\prime}) \\
    +  \frac{(\mathbf q+\mathbf G)\cdot(\mathbf q+\mathbf G^{\prime})}{(\mathbf q+\mathbf G)\cdot(\mathbf q+\mathbf G)}\chi_{\mathbf q,\mathbf G,\mathbf G^{\prime}}(\omega,-\omega^{\prime}),
    \label{eq:dielectric_function_fourier_space_phi_tot_phi_bare}
\end{multline}
or, analog, in real space:
\begin{align}
    \phi_{\text{tot}}(\mathbf r,t) 
    = \int \mathrm d^3r^{\prime}\,\mathrm dt^{\prime}\,\varepsilon^{-1}_{\text{mic}}(\mathbf r,\mathbf r^{\prime},t,t^{\prime}) \phi_{\text{ext}}(\mathbf r^{\prime},t^{\prime}).
    \label{eq:relation_total_external_potential_realspace}
\end{align}
Note that the dielectric function $\epsilon$ from \eqref{eq:epsilon_nonlocal_realspace} appearing in the generalized Poisson equation in \eqref{eq:generalized_poisson_equation2} is not equal to the dielectric function $\varepsilon_{\text{mic}}$ from \eqref{eq:dielectric_function_fourier_space_phi_tot_phi_bare} appearing in \eqref{eq:relation_total_external_potential_fourierspace} or \eqref{eq:relation_total_external_potential_realspace}, as long as all local-field effects with $\mathbf G\neq\mathbf G^{\prime}$ are included.

In the following, we provide the expressions for the static ($\omega=0$) macroscopic dielectric function in three and two dimensions, which are useful in relating macroscopic screening approaches to fully microscopic calculations.

First, we consider the three-dimensional case. 
We perform a macroscopic averaging over the total potential in \eqref{eq:relation_total_external_potential_realspace}:
\begin{multline}
    \phi_{\text{tot}}(\mathbf R) =  \frac{1}{\mathcal V_{\text{UC}}}\int_{\mathcal V_{\text{UC}}}\mathrm d^3r\,\phi_{\text{tot}}(\mathbf r+ \mathbf R)\\
    =  \int \mathrm d^3r^{\prime}\,\varepsilon^{-1}_{\text{mic}}(\mathbf R,\mathbf r^{\prime})\phi_{\text{ext}}(\mathbf r^{\prime})\\
    =  \sum_{\mathbf R^{\prime}}\int_{\mathcal V_{\text{UC}}} \mathrm d^3r^{\prime}\,\varepsilon^{-1}_{\text{mic}}(\mathbf R,\mathbf r^{\prime}+\mathbf R^{\prime})
     \phi_{\text{ext}}(\mathbf r^{\prime}+\mathbf R^{\prime}),
\end{multline}
where $\mathcal V_{\text{UC}}$ is the volume of the three-dimensional unit cell. 
Next, we assume the external perturbation as macroscopic, i.e., $\phi_{\text{ext}}$ is constant over the size of a unit cell:
\begin{align}
    \phi_{\text{ext}}(\mathbf r^{\prime}+\mathbf R^{\prime}) \approx \phi_{\text{ext}}(\mathbf R^{\prime}) \quad \text{for $ \mathbf r \in \mathcal V_{\text{UC}}$},
\end{align}
which yields:
\begin{align}
    \phi_{\text{tot}}(\mathbf R) 
    \approx \mathcal V_{\text{UC}} \sum_{\mathbf R^{\prime}} \varepsilon^{-1}_{\text{mic}}(\mathbf R,\mathbf R^{\prime})\phi_{\text{ext}}(\mathbf R^{\prime}).
\end{align}
After a Fourier transformation, this expression can be rearranged:
\begin{shaded}
\begin{align}
    \begin{split}
     \frac{\phi_{\text{tot},\mathbf q}}{\phi_{\text{ext},\mathbf q}} & = \left( \varepsilon^{-1}_{\text{mic}}\right)_{\mathbf q,\mathbf 0,\mathbf 0}
     \eqqcolon \frac{1}{\epsilon_{\mathbf q}^{\text{3D}}} \quad  \text{(three dimensions)},
     \end{split}
    \label{eq:macroscopic_dielectric_function_3D_definition}
\end{align}
\end{shaded}
\noindent
and yields the definition of the macroscopic dielectric function $\epsilon_{\qpar,q_z}^{\text{3D}}$ in three dimensions. Note that 
$\epsilon_{\mathbf q}^{\text{3D}} \neq  \varepsilon_{\text{mic},\mathbf q,\mathbf 0,\mathbf 0}$
due to local-field effects.

In two-dimensional confined semiconductors, special care must be taken with respect to the out-of-plane direction. In \textit{ab initio} calculations, an array of periodically arranged layers with supercell distance $L_z$ is usually created. The monolayer case is then obtained by choosing $L_z$ in such a way that the individual layers behave as electrostatically decoupled with the help of a truncated Coulomb potential \cite{rozzi2006exact,ismail2006truncation}. Hence, if we spatially averaged over the supercell distance $L_z$, which can be chosen as arbitrarily large, we would obtain a two-dimensional dielectric function of unity \cite{tian2019electronic,huser2013dielectric}, which is meaningless. Thus, it can be argued that not the dielectric function, but rather the polarizability or susceptibility should be regarded as the fundamental quantity determining the dielectric properties \cite{tian2019electronic}. However, as long as we know, how and where which quantity evaluated via \textit{ab initio} methods enters our description of screening, such reasoning can be avoided. In this work, we take the route of incorporating the results from the Computational Materials Repository established by the workgroup of Thygesen \cite{andersen2015dielectric}. Within this approach, it is possible to obtain a meaningful two-dimensional dielectric function, if we average over the layer thickness $d$ around the center of the material $z_0=0$ and not over the supercell length $L_z$, where $d$ is assumed as the layer distance in the corresponding bulk material \cite{latini2015excitons,huser2013dielectric}:
\begin{align}
\begin{split}
    &\phi_{\text{tot}}(\Rpar,0) \\
    & = \frac{1}{\mathcal A_{\text{UC}}}\int_{\mathcal A_{\text{UC}}}\mathrm d^2r_{\parallel}^{}\,\frac{1}{d}\int_{-\frac{d}{2}}^{\frac{d}{2}}\mathrm dz\,\phi_{\text{tot}}(\rpar+\Rpar,z)\\
    & = \frac{1}{\mathcal A_{\text{UC}}}\int_{\mathcal A_{\text{UC}}}\mathrm d^2r_{\parallel}^{}\,\frac{1}{d}\int_{-\frac{d}{2}}^{\frac{d}{2}}\mathrm dz\,\sum_{\Rparp}\int_{\mathcal A_{\text{UC}}} \mathrm d^2r^{\prime}_{\parallel}\,\int \mathrm dz^{\prime}\,\\
    &\quad\quad\times\varepsilon^{-1}_{\text{mic}}(\rpar+\Rpar,\rparp+\Rparp,z,z^{\prime})\phi_{\text{ext}}(\rparp+\Rparp,z^{\prime}).
\end{split}
    \label{eq:phi_tot_phi_ext_realspace_2d_macr}
\end{align}
Similar to the three-dimensional case, we assume the external perturbation as in-plane macroscopic, but completely independent of the out-of-plane direction: $\phi_{\text{ext}}(\rparp+\Rparp,z^{\prime}) \equiv \phi_{\text{ext}}(\rparp)$ for $\rparp\in \mathcal A_{\text{UC}}$. This choice has to be made, since, otherwise, artificial interactions between different layers in the out-of-plane direction would result. After Fourier expanding:
\begin{multline}
    \varepsilon^{-1}_{\text{mic}}(\rpar,\rparp,z,z^{\prime}) = \frac{1}{\mathcal A}\sum_{\substack{\qpar,\Gpar,\Gparp,\\G_z,G_z^{\prime}}}\mathrm e^{\mathrm i (\qpar+\Gpar)\cdot\rpar}\mathrm e^{\mathrm iG_zz}\\
    \times 
    \left(\varepsilon^{-1}_{\text{mic}}\right)_{\qpar,\Gpar,\Gparp,G_z,G_z^{\prime}}\mathrm e^{-(\qpar+\Gparp)\cdot\rparp}\mathrm e^{-\mathrm iG_z^{\prime} z^{\prime}},
\end{multline}
and rearranging \eqref{eq:phi_tot_phi_ext_realspace_2d_macr}, we obtain:
\begin{shaded}
\begin{align}
\begin{split}
    \frac{\phi_{\text{tot},\qpar}(0)}{\phi_{\text{ext},\qpar}} &=  \sum_{G_z}\frac{1}{d}\int_{-\frac{d}{2}}^{\frac{d}{2}}\mathrm dz\,\mathrm e^{\mathrm iG_zz}\left(\varepsilon^{-1}_{\text{mic}}\right)_{\qpar,\mathbf 0,\mathbf 0,G_z,0} \\
    & \eqqcolon  \frac{1}{\epsilon_{\qpar}^{\text{2D}}}
    \quad\quad \text{(two dimensions)},
    \label{eq:macroscopic_dielectric_function_2D_definition}
\end{split}
\end{align}
\end{shaded}
\noindent
where the spatial integral can be explicitly evaluated as:
\begin{align}
    \frac{1}{d}\int_{-\frac{d}{2}}^{\frac{d}{2}}\mathrm dz\,\mathrm e^{\mathrm iG_zz} =
    \begin{cases} \frac{\sin\mleft(G_z\frac{d}{2}\mright)}{G_z\frac{d}{2}},~G_z\neq 0,\\
    1,~G_z=0.
    \end{cases}
\end{align}
Note that the summation over $G_z$ in \eqref{eq:macroscopic_dielectric_function_2D_definition} is not a computational artifact. This way, a physically meaningful quantum-confined dielectric function for an isolated monolayer can be constructed 
from \textit{ab initio} calculations, which use periodically arranged layers in the out-of-plane direction.

We note that \eqref{eq:macroscopic_dielectric_function_2D_definition} can also be derived from Qiu's and Louie's approach \cite{qiu2016screening,qiu2013optical}. In contrast to H\"user, Latini and Thygesen, Qiu and Louie work directly with the Green's functions giving rise to the total and external potentials induced by a test/active charge density. By rewriting \eqref{eq:relation_total_external_potential_realspace}, we obtain the following relation between the total screened Green's function $G$ and the bare unscreened Green's function $G_0$:
\begin{multline}
    G(\rpar,\rparp,z,z^{\prime})\\
    = \int\mathrm d^2r^{\prime\prime}_{\parallel}\,\mathrm dz^{\prime\prime}\varepsilon^{-1}_{\text{mic}}(\rpar,\rparpp,z,z^{\prime\prime})G_{0}(\rparpp,\rparp,z^{\prime\prime},z^{\prime}).
    \label{eq:total_bare_greens_function_relation}
\end{multline}
We introduce a static, screened $V$  (unscreened $V_{0}$) quantum-confined Coulomb potential, cf.\ \eqref{eq:Coulomb_Potential_Energy_SecondQuantized}:
\begin{align}
    V_{(0)}(\rpar,\rparp) =  \frac{e^2}{\mathcal A}\int\mathrm dz\,\mathrm dz^{\prime}\,|\zeta(z)|^2 G_{(0)}(\rpar,\rparp,z,z^{\prime}) |\zeta(z^{\prime})|^2.
\end{align}
where $G_{(0)}(\rpar,\rparp,z,z^{\prime})$ is the screened (unscreened) Green's function solving \eqref{eq:poisson_equation_screened_greens_function} (\eqref{eq:poisson_equation_bare_greens_function}) and $|\zeta(z)|^2$ describes the charge distribution in the out-of-plane direction. 
Qiu and Louie now define a two-dimensional dielectric function $\epsilon_{\text{2D}}^{-1}$ via:
\begin{align}
    V(\rpar,\rparp) = \int\mathrm d^2r^{\prime\prime}\, \epsilon_{\text{2D}}^{-1}(\rpar,\rparpp) V_0(\rparpp,\rparp).
    \label{eq:dielectric_function_2d_louie_def_realspace}
\end{align}
By Fourier expanding \eqref{eq:dielectric_function_2d_louie_def_realspace}, we obtain the following relation in the in-plane macroscopic limit:
\begin{align}
    \left(\epsilon_{\text{2D}}^{-1}\right)_{\qpar,\mathbf 0,\mathbf 0} = 
    \frac{V_{\qpar,\mathbf 0,\mathbf 0}}{V_{0,\qpar,\mathbf 0,\mathbf 0}},
    \label{eq:dielectric_function_2d_louie_def_fourierspace}
\end{align}
where:
\begin{multline}
    V_{(0),\qpar,\mathbf 0,\mathbf 0}
      = \frac{e^2}{\mathcal A}\int\mathrm dz\,\mathrm dz^{\prime}\, |\zeta(z)|^2 |\zeta(z^{\prime})|^2\\
      \times \sum_{G_z,G_z^{\prime}} \mathrm e^{\mathrm iG_z z}
    G_{(0),\qpar,\mathbf 0,\mathbf 0,G_z,G_z^{\prime}}
    \mathrm e^{-G_z^{\prime} z^{\prime}} .
\end{multline}
If we now assume a step-function confinement around the material center $z_0$: $|\zeta(z)|^2 = \frac{1}{d}\Theta(z-z_0+\frac{d}{2})\Theta(\frac{d}{2}+z_0-z)$, \eqref{eq:dielectric_function_2d_louie_def_fourierspace} can be cast into \eqref{eq:macroscopic_dielectric_function_2D_definition} by using \eqref{eq:total_bare_greens_function_relation}, so that $\left(\epsilon_{\text{2D}}^{-1}\right)_{\qpar,\mathbf 0,\mathbf 0} \equiv \frac{1}{\epsilon_{\qpar}^{\text{2D}}}$. Hence, 
we observe that the definitions of the two-dimensional dielectric functions in Louie's and Thygesen's approaches are equal and the only difference 
stems from the exact choice of the out-of-plane confinement: A Dirac-delta confinement with $|\zeta(z)|^2 = \delta(z-z_0)$ in the former \cite{qiu2013optical,qiu2016screening} compared to a step-function confinement in the latter \cite{latini2015excitons,huser2013dielectric}.

\subsection{Microscopic Screening}
\label{sec:microscopic_screening}
In this section, we derive an explicit expression of the microscopic dielectric function $\varepsilon_{\text{mic}}$. 
To keep the notation as simple as possible, we emphasize that $\mathbf r$, $\mathbf q$ and $\mathbf G$ denote the three-dimensional position, momentum and reciprocal-lattice vector. 
We consider the total Coulomb Hamiltonian:
\begin{align}
\begin{split}
    H_{\text{Coul,tot}} =&\, \frac{1}{2}\int\mathrm d^3r\,\rho_{\text{tot}}(\mathbf r,t)\phi_{\text{tot}}(\mathbf r,t).
    \end{split}
\end{align}
Now, the idea is as follows: We split the total charge density in an active part and a background part: $\rho_{\text{tot}}(\mathbf r,t) = \rho_{\text{b}}(\mathbf r,t) + \rho(\mathbf r,t)$. The active part $\rho(\mathbf r,t)$ is, e.g., a reduced phase space of only one valence and one conduction band, whereas the background part $\rho_{\text{b}}(\mathbf r)$ includes all other bands. Then, we calculate the response of the background due to the ``perturbation'' by the active part.

The total Hamiltonian in Fourier space reads:
\begin{align}
\begin{split}
    \hat H_{\text{Coul,tot}} = &\, 
    \frac{1}{2\mathcal V}\sum_{\mathbf q,\mathbf G}\hat \rho_{\text{tot},-\mathbf q-\mathbf G}(t) G_{0,\mathbf q+\mathbf G}\hat \rho_{\text{tot},\mathbf q+\mathbf G}(t)\\
    = &\, \hat H_{\text{Coul,bb}} + \hat H_{\text{Coul,ba}} + \hat H_{\text{Coul,aa}}.
\end{split}
\end{align}
The background--background interaction $\hat H_{\text{Coul,bb}}$ reads:
\begin{multline}
    \hat H_{\text{Coul,bb}} = \frac{1}{2}\sum_{\substack{\lambda_1\dots\lambda_4,\\\mathbf k,\mathbf k^{\prime},\mathbf q,\mathbf G,s,s^{\prime}}}V_{0,\mathbf q+\mathbf G}\overline \Upsilon_{\mathbf k+\mathbf q+\mathbf G,\mathbf k}^{\lambda_1,\lambda_4,s}\overline \Upsilon_{\mathbf k^{\prime}-\mathbf q-\mathbf G,\mathbf k^{\prime}}^{\lambda_2,\lambda_3,s^{\prime}}\\
    \times\adagtwo{\lambda_1,\mathbf k+\mathbf q}{s}(t)\adagtwo{\lambda_2,\mathbf k^{\prime}-\mathbf q}{s^{\prime}}(t)\andagtwo{\lambda_3,\mathbf k^{\prime}}{s^{\prime}}(t)\andagtwo{\lambda_4,\mathbf k}{s}(t),
    \label{eq:H_background_background}
\end{multline}
and the interaction between the background and active states $\hat H_{\text{Coul,ba}}$ reads:
\begin{multline}
    \hat H_{\text{Coul,ba}} = -e \sum_{\substack{\lambda_1,\lambda_2,\mathbf k,\mathbf q,\mathbf G,s}}\tilde \phi_{\text{ext},\mathbf q+\mathbf G}(t)\overline \Upsilon_{\mathbf k+\mathbf q+\mathbf G,\mathbf k}^{\lambda_1,\lambda_2,s}\\
    \times \adagtwo{\lambda_1,\mathbf k+\mathbf q}{s}(t)\andagtwo{\lambda_2,\mathbf k}{s}(t),
    \label{eq:H_background_active}
\end{multline}
where $\overline \Upsilon_{\mathbf k^{\prime},\mathbf k}^{\lambda,\lambda^{\prime},s}$ is the three-dimensional version of \eqref{eq:reduced_form_factor}. In \eqref{eq:H_background_active}, 
the external potential $\tilde \phi_{\text{ext},\mathbf q+\mathbf G}(t)$ induced by the active charges $\rho_{\mathbf q+\mathbf G}(t)$ reads:
\begin{align}
    \tilde \phi_{\text{ext},\mathbf q+\mathbf G}(t) = \frac{1}{\mathcal V}G_{0,\mathbf q+\mathbf G}\rho_{\mathbf q+\mathbf G}(t).
    \label{eq:active_potential_sec:screening}
\end{align}
The active--active contribution $\hat H_{\text{Coul,aa}}$ is neglected, as it does not couple to the background. In the next step, we calculate the background charge density $\rho_{\text{b},\mathbf q+\mathbf G}$:
\begin{align}
    \rho_{\text{b},\mathbf q+\mathbf G}(t) = -e\sum_{\lambda,\lambda^{\prime},\mathbf k,s}\overline \Upsilon_{\mathbf k-\mathbf q-\mathbf G,\mathbf k}^{\lambda,\lambda^{\prime},s} \langle \adagtwo{\lambda,\mathbf k-\mathbf q}{s}\andagtwo{\lambda^{\prime},\mathbf k}{s}\rangle (t),
\end{align}
where the coherences $\langle \adagtwo{\lambda,\mathbf k-\mathbf q}{s}\andagtwo{\lambda^{\prime},\mathbf k}{s}\rangle (t)$ have to be evaluated via Heisenberg's equations of motion. 
In a Hartree--Fock and a random-phase approximation, the equations of motion for $\langle \adagtwo{\lambda,\mathbf k-\mathbf q}{s}\andagtwo{\lambda^{\prime},\mathbf k}{s}\rangle$ read:
\begin{multline}
    \mathrm i\hbar\partial_t \langle \adagtwo{\lambda,\mathbf k-\mathbf q}{s}\andagtwo{\lambda^{\prime},\mathbf k}{s} \rangle (t)\Big|_{0 + \text{ba} + \text{bb} } \\
    = \left( \tilde E_{\lambda^{\prime},\mathbf k}^s - \tilde E_{\lambda,\mathbf k-\mathbf q}^{s} 
    - \mathrm i\hbar\gamma \right)\langle \adagtwo{\lambda,\mathbf k-\mathbf q}{s}\andagtwo{\lambda^{\prime},\mathbf k}{s} \rangle (t)\\
     - e \sum_{\mathbf G}\left(\tilde \phi_{\text{ext},\mathbf q+\mathbf G}(t) + \tilde \phi_{\text{b},\mathbf q+\mathbf G}(t)\right)\overline \Upsilon_{\mathbf k+\mathbf G,\mathbf k-\mathbf q}^{\lambda^{\prime},\lambda,s}\\
     \times \left(f_{\lambda,\mathbf k-\mathbf q}^s - f_{\lambda^{\prime},\mathbf k}^s\right),
     \label{eq:equations_of_motion_charge_density}
\end{multline}
where $f_{\lambda,\mathbf k}^s$ are static Fermi distributions and $\gamma$ is an additionally included dephasing due to higher-order correlation effects beyond a Hartree--Fock approximation. Note that \eqref{eq:equations_of_motion_charge_density} does not describe excitonic effects as a consequence of the random-phase approximation. These excitonic effects are later calculated within a cluster-expansion approach by using \eqref{eq:Coulomb_Hamiltonian_SecondQuantized_Formfactors_Evaluated} under the influence of the RPA-screening cloud from \eqref{eq:equations_of_motion_charge_density}. $\tilde E_{\lambda,\mathbf k}^{s}$ are the energy dispersions renormalized by electron--electron interaction:
\begin{align}
    \tilde E_{\lambda,\mathbf k}^{s} = E_{\lambda,\mathbf k}^{s} - \sum_{\mathbf q^{\prime},\mathbf G^{\prime},\lambda^{\prime}}V_{0,\mathbf q^{\prime} +\mathbf G^{\prime}} \big|\overline \Upsilon_{\mathbf k,\mathbf k+\mathbf q^{\prime}+\mathbf G^{\prime}}^{\lambda,\lambda^{\prime}}\big|^2 f_{\lambda^{\prime},\mathbf k+\mathbf q^{\prime}}^s,
\end{align}
and $\tilde \phi_{\text{b},\mathbf q+\mathbf G}$ is the background potential:
\begin{align}
    \tilde \phi_{\text{b},\mathbf q+\mathbf G}(t) = \frac{1}{\mathcal V} G_{0,\mathbf q+\mathbf G} \rho_{\text{b},\mathbf q+\mathbf G}(t).
    \label{eq:background_potential_sec:screening}
\end{align}
Note that the definitions for the external and background potentials ($\tilde \phi_{\text{ext},\mathbf q+\mathbf G}(t)$ and $\tilde \phi_{\text{b},\mathbf q+\mathbf G}(t)$) used in \eqref{eq:active_potential_sec:screening} and \eqref{eq:background_potential_sec:screening} are \textit{not} the bare Fourier transforms ($ \phi_{\text{ext},\mathbf q+\mathbf G}(t)$ and $ \phi_{\text{b},\mathbf q+\mathbf G}(t)$), since we additionally included the semiconductor volume $\mathcal V$. This way, the (modified) Fourier transform $\tilde \phi_{\mathbf q+\mathbf G}(t)$ carries equal units compared to the corresponding real-space quantity $\phi(\mathbf r)$. 
By identifying the total potential as $\tilde \phi_{\text{ext},\mathbf q+\mathbf G}(t) + \tilde \phi_{\text{b},\mathbf q+\mathbf G}(t) = \tilde \phi_{\text{tot},\mathbf q+\mathbf G}(t)$, we find the following expression for the background charge density after solving \eqref{eq:equations_of_motion_charge_density} in frequency space:
\begin{multline}
    \rho_{\text{b},\mathbf q+\mathbf G} (\omega) \\
    = e^2\sum_{\lambda,\lambda^{\prime},\mathbf k,\mathbf G^{\prime}}
    \frac{\overline \Upsilon_{\mathbf k-\mathbf q-\mathbf G,\mathbf k}^{\lambda,\lambda^{\prime},s}
    \overline \Upsilon_{\mathbf k,\mathbf k-\mathbf q-\mathbf G^{\prime}}^{\lambda^{\prime},\lambda,s}\left(f_{\lambda,\mathbf k-\mathbf q}^s - f_{\lambda^{\prime},\mathbf k}^s\right)}{ \tilde E_{\lambda,\mathbf k-\mathbf q}^s - \tilde E_{\lambda^{\prime},\mathbf k}^s +\hbar\omega + \mathrm i\hbar\gamma  }\\
    \times \tilde \phi_{\text{tot},\mathbf q+\mathbf G^{\prime}}(\omega).
    \label{eq:rho_b_full_solution}
\end{multline}
Also, from \eqref{eq:relation_rho_b_phi_tot}, we obtain the following expression:
\begin{multline}
    \rho_{\text{b},\mathbf q+\mathbf G} (\omega) = - \mathcal V\sum_{\mathbf G^{\prime}}\frac{1}{2\pi}\int\mathrm d\omega^{\prime}\,\left(\mathbf q+\mathbf G\right)\cdot\left(\mathbf q+\mathbf G^{\prime}\right)\\
    \times\epsilon_0 \chi_{\mathbf q,\mathbf G,\mathbf G^{\prime}}(\omega,-\omega^{\prime}) \tilde \phi_{\text{tot},\mathbf q+\mathbf G^{\prime}}(\omega^{\prime}).
    \label{eq:relation_rho_b_phi_tot_sec:screening}
\end{multline}
By now comparing \eqref{eq:rho_b_full_solution} with \eqref{eq:relation_rho_b_phi_tot_sec:screening}, we can find an explicit expression for the susceptibility $\chi_{\mathbf q,\mathbf G,\mathbf G^{\prime}}(\omega,-\omega^{\prime})$ determining the microscopic dielectric function:
\begin{multline}
    \varepsilon_{\text{mic},\mathbf q,\mathbf G,\mathbf G^{\prime}}(\omega,-\omega^{\prime}) = 2\pi\delta(\omega-\omega^{\prime})\delta_{\mathbf G,\mathbf G^{\prime}}\\
    + \frac{(\mathbf q+\mathbf G)\cdot(\mathbf q+\mathbf G^{\prime})}{(\mathbf q+\mathbf G)\cdot(\mathbf q+\mathbf G)}\chi_{\mathbf q,\mathbf G,\mathbf G^{\prime}}(\omega,-\omega^{\prime}),
    \label{eq:microscopic_dielectric_function_potentialrelation}
\end{multline}
and explicitly obtain:
\begin{shaded}
\begin{multline}
    \varepsilon_{\text{mic},\mathbf q,\mathbf G,\mathbf G^{\prime}}(\omega,-\omega^{\prime}) \\
    = 2\pi\delta(\omega-\omega^{\prime})\delta_{\mathbf G,\mathbf G^{\prime}}
    - 2\pi\delta(\omega-\omega^{\prime})V_{0,\mathbf q+\mathbf G}\\
    \times\sum_{\lambda,\lambda^{\prime},\mathbf k}
    \frac{\overline \Upsilon_{\mathbf k-\mathbf q-\mathbf G,\mathbf k}^{\lambda,\lambda^{\prime},s}
    \overline \Upsilon_{\mathbf k,\mathbf k-\mathbf q-\mathbf G^{\prime}}^{\lambda^{\prime},\lambda,s}\left(f_{\lambda,\mathbf k-\mathbf q}^s - f_{\lambda^{\prime},\mathbf k}^s\right)}{ \tilde E_{\lambda,\mathbf k-\mathbf q}^s - \tilde E_{\lambda^{\prime},\mathbf k}^s +\hbar\omega + \mathrm i\hbar\gamma  }.
    \label{eq:dielectric_function_microscopic}
\end{multline}
\end{shaded}
\eqref{eq:dielectric_function_microscopic} is the usual (renormalized) Lindhard formula from the literature \cite{hybertsen1986electron,adler1962quantum,wiser1963dielectric,ehrenreich1959self,lindhard1954properties}, which describes interband ($\lambda\neq\lambda^{\prime}$) screening, which dominates in charge-neutral semiconductors \cite{walter1970wave}, and intraband ($\lambda=\lambda^{\prime}$) screening of quasi-free carriers, which dominates in metals or doped semiconductors.
By adding the corresponding Hamiltonians within our Heisenberg-equations-of-motion approach, cf.~\eqref{eq:equations_of_motion_charge_density}, the inclusion of other screening mechanisms in \eqref{eq:dielectric_function_microscopic} such as phonon screening \cite{steinhoff2020dynamical,alvertis2024phonon,lee2024phonon,van2024effects,ben2025breakdown} is straightforward.

We emphasize that \eqref{eq:microscopic_dielectric_function_potentialrelation} and \eqref{eq:dielectric_function_microscopic} describe the microscopic dielectric function $\varepsilon_{\text{mic}}$ as it appears in the relation between the total and bare potential or Green's function, \eqref{eq:total_bare_greens_function_relation} or \eqref{eq:relation_total_external_potential_fourierspace}, and \textit{not} the dielectric function $\epsilon$ in the generalized Poisson equation in \eqref{eq:generalized_poisson_equation2}, which reads:
\begin{multline}
    \epsilon_{\mathbf q,\mathbf G,\mathbf G^{\prime}}(\omega,-\omega^{\prime})\\
    = 2\pi\delta(\omega-\omega^{\prime})\delta_{\mathbf G,\mathbf G^{\prime}}
    + \chi_{\mathbf q,\mathbf G,\mathbf G^{\prime}}(\omega,-\omega^{\prime}).
    \label{eq:microscopic_dielectric_function_poissonequation}
\end{multline}
Both expressions in \eqref{eq:microscopic_dielectric_function_potentialrelation} and \eqref{eq:microscopic_dielectric_function_poissonequation} differ slightly, if all local-field effects with $\mathbf G\neq\mathbf G^{\prime}$ are taken into account but coincide in the macroscopic limit with $\mathbf G=\mathbf G^{\prime}=\mathbf 0$. 

We note that the inverse of \eqref{eq:dielectric_function_microscopic}, $\left(\varepsilon^{-1}_{\text{mic}}\right)_{\mathbf q,\mathbf G,\mathbf G^{\prime}}$, is always well-defined, since, in practice, only a finite number of reciprocal lattice vectors is considered, as the matrix elements decay rapidly with increasing $\mathbf G$. 
To numerically calculate the dielectric matrix, a DFT approach with a many-body perturbation theory in \textit{GW} approximation is usually employed \cite{deilmann2023optical,deslippe2012berkeleygw} 
via packages such as, e.g., \textsc{Quantum Espresso} \cite{giannozzi2009quantum}, \textsc{Yambo} \cite{marini2009yambo,sangalli2019many}, \textsc{BerkeleyGW} \cite{hybertsen1986electron,deslippe2012berkeleygw}, \textsc{GPAW} \cite{mortensen2024gpaw,larsen2017atomic} or \textsc{VASP} \cite{hafner2008ab}. 

Usually, we make a static approximation with $\omega = 0$ in \eqref{eq:dielectric_function_microscopic} \cite{rohlfing2000electron}, 
i.e., we assume that the cloud of induced screening charges (virtual charges, i.e., polarizations if $\lambda\neq\lambda^{\prime}$ or real charges if $\lambda=\lambda^{\prime}$) follows the dynamics of a perturbing electron or hole \textit{adiabatically}, which corresponds to a memory-free screening response of the singlet cloud and removes the time integration in the many-body Coulomb Hamiltonian in \eqref{eq:Coulomb_Hamiltonian_SecondQuantized}. This is a reasonable approximation in the interband dielectric function, \eqref{eq:dielectric_function_microscopic} with $\lambda\neq\lambda^{\prime}$, as long as the exciton energy is much smaller than the resonance of the interband plasmon \cite{rohlfing2000electron}. On the other hand, such an assumption fails, if we also take into account screening by a Fermi sea of dopants \cite{glazov2018breakdown,zimmermann1971dynamical} or an optically injected electron--hole/exciton gas, i.e., taking additional intraband contributions in the dielectric function, \eqref{eq:dielectric_function_microscopic} with $\lambda=\lambda^{\prime}$, into account \cite{haug2008quantum,banyai1998ultrafast}. In this regime, the screening cloud of induced charges cannot be assumed to follow the dynamics of the perturbing charges, e,g. optically induced electron--hole pairs, adiabatically, as the intraband plasmon frequency can be of similar magnitude or even less compared to the exciton binding energy. In the case of optically excited electron--hole pairs, full quantum-kinetic calculations in quantum wells show that the impact of time-dependent RPA screening is in the intermediate regime between an unscreened Coulomb potential and a statically screened Coulomb potential \cite{banyai1998ultrafast}, at least within short timescales of coherence decay. 
This behavior might explain the observations that simulations without a density-dependent dielectric function 
already yield good agreement with experiment in out-of-equilibrium ultrafast optics \cite{deckert2025coherent,schafer2025distinct,schafer2024optical,trovatello2022disentangling,katsch2020exciton,ciuti2000strongly,schulzgen1999direct,cundiff1994rabi}.

Now, we again distinguish between in-plane and out-of-plane coordinates $\mathbf r \rightarrow \begin{pmatrix}
    \rpar & z
\end{pmatrix}^{\top}$, crystal momenta $\mathbf q\rightarrow \begin{pmatrix}
    \qpar & q_z
\end{pmatrix}^{\top}$ and reciprocal lattice vectors $\mathbf G \rightarrow \begin{pmatrix}
    \Gpar & G_z
\end{pmatrix}^{\top}$. 
Having determined the dielectric function of the material, cf.~\eqref{eq:dielectric_function_microscopic}, we now construct the quantum-confined Coulomb potential from \eqref{eq:Coulomb_Potential_Energy_SecondQuantized} entering \eqref{eq:Coulomb_Hamiltonian_SecondQuantized_Formfactors_Evaluated} by means of \eqref{eq:total_bare_greens_function_relation} with confinement wave functions $\zeta_n^{}(z)$:
\begin{multline}
    V_{\qpar,\Gpar,\Gparp}^{n_1,n_2,n_3,n_4}(\omega,\omega^{\prime}) = \frac{e^2}{\mathcal A}\int\mathrm dz\,\mathrm dz^{\prime}\,\zeta_{n_1}^*(z)\zeta_{n_4}^{}(z) \\
    \times G_{\qpar,\Gpar,\Gparp}(z,z^{\prime},\omega,\omega^{\prime})  \zeta_{n_2}^*(z^{\prime})\zeta_{n_3}^{}(z^{\prime})\\
    = \frac{e^2}{\mathcal A}\int\mathrm dz\,\mathrm dz^{\prime}\,\zeta_{n_1}^*(z)\zeta_{n_4}^{}(z)\\
   \times \int\mathrm dz^{\prime\prime}\,\left(\varepsilon_{\text{mic}}^{-1}\right)_{\qpar,\Gpar,\Gparp}(z,z^{\prime\prime},\omega,\omega^{\prime})\\
     \times G_{0,\qpar+\Gparp}(z^{\prime\prime}-z^{\prime})\zeta_{n_2}^*(z^{\prime})\zeta_{n_3}^{}(z^{\prime}).
\end{multline}
Fourier expanding the dielectric matrix in the out-of-plane direction and inserting a truncated bare Coulomb Green's function $G_{0,\qpar}(z) \rightarrow G_{0,\qpar}^{\text{trunc}}(z) = \frac{\mathrm e^{-|\qpar| |z|}}{2\epsilon_0 |\qpar|}\Theta (z+\frac{L_{z,\text{SC}}}{2})\Theta (\frac{L_{z,\text{SC}}}{2}-z)$, which eliminates electrostatic interactions between different supercells in the out-of-plane direction with supercell length $L_{z,\text{SC}}$ and total out-of-plane lattice length $L_z$, yields:
\begin{multline}
    V_{\qpar,\Gpar,\Gparp}^{n_1,n_2,n_3,n_4}(\omega,\omega^{\prime})\\
    = \frac{e^2}{\mathcal A L_z}\int\mathrm dz\,\mathrm dz^{\prime}\,
    \zeta_{n_1}^*(z)\zeta_{n_4}^{}(z)\zeta_{n_2}^*(z^{\prime})\zeta_{n_3}^{}(z^{\prime}) \\
    \times\int_{z^{\prime}-\frac{L_{z,\text{SC}}}{2}}^{z^{\prime}+\frac{L_{z,\text{SC}}}{2}}\mathrm dz^{\prime\prime}\,
    \sum_{q_z,G_z,G_z^{\prime}} \mathrm e^{\mathrm i(q_z+G_z)z}\\
    \times\left(\varepsilon_{\text{mic}}^{-1}\right)_{\qpar,\Gpar,\Gparp,q_z,G_z,G_z^{\prime}}(\omega,\omega^{\prime})\mathrm e^{-\mathrm i(q_z+G_z^{\prime})z^{\prime\prime}}
    \\
    \times
    \frac{\mathrm e^{-|\qpar+\Gparp| |z^{\prime\prime}-z^{\prime}|}}{2\epsilon_0 |\qpar+\Gparp|}.
\end{multline}
Shifting $z^{\prime\prime} \rightarrow z^{\prime\prime} + z^{\prime}$, defining confinement form factors:
\begin{align}
    F_{q_z+G_z}^{n,n^{\prime}} = \int\mathrm dz\,\zeta_n^*(z) \zeta_{n^{\prime}}^{}(z) \mathrm e^{\mathrm i(q_z+G_z)z},
    \label{eq:confinement_form_factor}
\end{align}
and evaluating the $z^{\prime\prime}$ integral:
\begin{multline}
    \int_{-\frac{L_{z,\text{SC}}}{2}}^{\frac{L_{z,\text{SC}}}{2}}\mathrm dz^{\prime\prime}\, \mathrm e^{-\mathrm i(q_z+G_z^{\prime})z^{\prime\prime}}\mathrm e^{-|\qpar+\Gparp| |z^{\prime\prime}|}\\
    = \frac{2|\qpar+\Gparp|}{|\qpar+\Gparp|^2 + (q_z+G_z^{\prime})^2}\\
    \times \left( 1 - \mathrm e^{-|\qpar+\Gparp|\frac{L_{z,\text{SC}}}{2}}\cos\mleft((q_z+G_z^{\prime})\frac{L_{z,\text{SC}}}{2}\mright) \right),
\end{multline}
we obtain the following expression:
\begin{shaded}
\begin{multline}
    V_{\qpar,\Gpar,\Gparp}^{n_1,n_2,n_3,n_4}(\omega,\omega^{\prime})\\
    =
    \sum_{q_z,G_z,G_z^{\prime}}
    \left(\varepsilon_{\text{mic}}^{-1}\right)_{\qpar,\Gpar,\Gparp,q_z,G_z,G_z^{\prime}}(\omega,\omega^{\prime})F_{q_z+G_z}^{n_1,n_4} F_{-q_z-G_z^{\prime}}^{n_2,n_3}\\
    \times V_{0,\qpar+\Gparp,q_z+G_z^{\prime}}^{\text{trunc}},
    \label{eq:coulomb_potential_quantum_confined_epsilon_abinitio}
\end{multline}
\end{shaded}
\noindent
where $V_{0,\qpar+\Gpar,q_z+G_z}^{\text{trunc}}$ is the truncated bare Coulomb potential in reciprocal space \cite{qiu2016screening,ismail2006truncation}:
\begin{multline}
    V_{0,\qpar+\Gpar,q_z+G_z}^{\text{trunc}} \\
    = \frac{e^2}{\mathcal A L_{z}} \frac{1 - \mathrm e^{-|\qpar+\Gpar|\frac{L_{z,\text{SC}}}{2}}\cos\mleft((q_z+G_z)\frac{L_{z,\text{SC}}}{2}\mright)}{\epsilon_0 (  |\qpar+\Gpar|^2 + (q_z+G_z)^2)}.
    \label{eq:truncated_coulomb_potential_fourierspace}
\end{multline}
Note that in practical calculations, the three-dimensional unscreened Coulomb potential $V_{0,\qpar+\Gpar,q_z+G_z} = \frac{e^2}{\mathcal V\epsilon_0\left(|\qpar+\Gpar|^2 + (q_z+G_z)^2\right)}$ in \eqref{eq:dielectric_function_microscopic} has also to be replaced by the truncated Coulomb $V_{0,\qpar+\Gpar,q_z+G_z}^{\text{trunc}}$ potential from \eqref{eq:truncated_coulomb_potential_fourierspace} to avoid electrostatic image interactions between layers in different supercells \cite{qiu2016screening,ismail2006truncation}. Also note that the dependence on $q_z$ is usually dropped, since the allowed $q_z$ values are already small if the supercell length is chosen as correspondingly large, i.e., most of the spatial variation in the out-of-plane direction occurs \textit{within} the supercell captured by $G_z$. Moreover, many DFT codes only provide the dominant $q_z=0$ components.

Within a static or adiabatic approximation of the Coulomb potential in \eqref{eq:coulomb_potential_quantum_confined_epsilon_abinitio} (or the dielectric function in \eqref{eq:dielectric_function_microscopic}), at $\omega=0$ \cite{rohlfing2000electron}, the resulting Hamiltonian from 
\eqref{eq:Coulomb_Hamiltonian_SecondQuantized_Formfactors_Evaluated} becomes Hermitian and reads:
\begin{shaded}
\begin{multline}
\label{eq:Coulomb_Hamiltonian_SecondQuantized_ValleyExpanded_adiabatic_approx}
    \hat H_{\text{Coul}}=  \frac{1}{2}\sum_{\substack{\lambda_1,\dots\lambda_4,\kpar{},\kparp{},\\\qpar,\Gpar,\Gparp,\\
    s,s^{\prime},n_1,\dots n_4}} V^{n_1,n_2,n_3,n_4}_{\qpar,\Gpar,\Gparp}(\omega=0)\\
    \times
    \overline \Upsilon_{\kpar{}+\qpar+\Gpar,\kpar{}}^{\lambda_1,\lambda_4,s}\overline \Upsilon_{\kparp{}-\qpar-\Gparp,\kparp{}}^{\lambda_2,\lambda_3,s^{\prime}}\\
    \times
    \bigg(\adagtwo{\lambda_1,\kpar{}+\qpar}{s,n_1}(t)\adagtwo{\lambda_2,\kparp{}-\qpar}{s^{\prime},n_2}(t)\andagtwo{\lambda_3,\kparp{}}{s^{\prime},n_3}(t)\andagtwo{\lambda_4,\kpar{}}{s,n_4}(t)\\
    + \,\delta_{\lambda_2,\lambda_4}^{\kparp{},\kpar{}+\qpar}\delta_{s,s^{\prime}}^{n_2,n_4} \adagtwo{\lambda_1,\kpar{}+\qpar}{s,n_1}(t)\andagtwo{\lambda_3,\kparp{}}{s^{\prime},n_3}(t)\bigg)
    ,
\end{multline}
\end{shaded}
\noindent
where we performed a normal ordering using \eqref{eq:commutation_relation_creation_annihilation_operators}. The single-particle terms correspond to the $V(\mathbf r=\mathbf 0)$ self-energy already occurring in classical electrodynamics \cite{weisskopf1939self,dirac1938classical} and cancel out in dynamical calculations, if we consider a bare unscreened Coulomb potential or a screened Coulomb potential without local fields \cite{haug2009quantum}. However, due to the elimination of the background, the single-particle contribution is now screened and does not vanish in general, if all local fields with $\mathbf G\neq\mathbf G^{\prime}$ are taken into account, i.e., it encodes parts of the screening-inducing higher correlations in this effective description. Since the Coulomb Hamiltonian in an adiabatic approximation in \eqref{eq:Coulomb_Hamiltonian_SecondQuantized_ValleyExpanded_adiabatic_approx} is now Hermitian and time-local, it is a generator of an unitary time evolution via, e.g., Heisenberg equations of motion.

With \eqref{eq:coulomb_potential_quantum_confined_epsilon_abinitio}, we have derived an explicit link between established \textit{ab initio} methods of dielectric screening and our few-band Heisenberg-equations-of-motion approach to many-body exciton dynamics starting from the screened Coulomb Hamiltonian in \eqref{eq:Coulomb_Hamiltonian_SecondQuantized_ValleyExpanded_adiabatic_approx}. 

The model is suited to capture ground-state screening of charge-neutral semiconductors via the interband dielectric function ($\lambda\neq\lambda^{\prime}$) in \eqref{eq:dielectric_function_microscopic}, where an adiabatic approximation is valid \cite{rohlfing2000electron}. We note that this is not fulfilled in every material, in this case, dynamical screening must be already considered on the ground-state level \cite{marini2003dynamical,strinati1984effects}. 
Either way, the calculation of screening effects beyond the ground-state screening (not to be confused with dynamical phase-space filling effects) due to, e.g., doping-induced free charge densities \cite{glazov2018breakdown,zimmermann1971dynamical,scharf2019dynamical}, optically excited electron--hole pairs/excitons \cite{steinhoff2024exciton,banyai1998ultrafast,perfetto2022real,lohof2018prospects,schulz2026optical,erben2022optical} or phonons \cite{steinhoff2020dynamical,alvertis2024phonon,lee2024phonon,van2024effects,ben2025breakdown} cannot be described in an adiabatic approximation in general and requires nonequilibrium Green's function techniques \cite{haug2008quantum,stefanucci2025nonequilibrium} or higher-order correlations within the cluster expansion beyond a Hartree--Fock limit \cite{kira2006many,fricke1996transport}. While the former is more practical if all orders in the quantum hierarchy of a specific interaction need to be considered, the latter is more suited to capture cross interactions, where the ground-state screened Hamiltonian in \eqref{eq:Coulomb_Hamiltonian_SecondQuantized_ValleyExpanded_adiabatic_approx} is a suitable starting point. However, an explicit evaluation of dynamical screening is beyond the scope of this manuscript.

\subsection{Macroscopic Screening}
\label{sec:macroscopic_screening}

In this section, we discuss material and environmental screening, which greatly influences the excitonic properties of layered semiconductor heterostructures, in a macroscopic approach. 

Since substrate screening mainly affects the screening at small momenta \cite{latini2015excitons}, where the macroscopic approach to the dielectric function---both the material dielectric function and the environmental screening---is valid \cite{latini2015excitons,trolle2017model,qiu2016screening}, local-field contributions with $\Gpar\neq\Gparp$ can be calculated reliably with an isolated layer in vacuum. Hence, in a first approximation, it is possible to circumvent microscopic approaches to the screening in layered dielectrics such as the performance-oriented Quantum Electrostatic Heterostructure (QEH) model \cite{andersen2015dielectric,gjerding2020efficient} or more rigorous but computationally more expensive approaches such as explicit multi-layer supercell calculations \cite{hu2023excitonic,qiu2017environmental}.

\begin{figure}[h!]
\centering
		\includegraphics[width=0.7\linewidth]{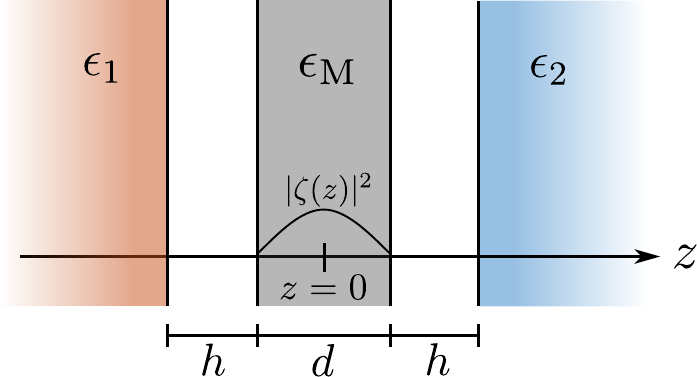}
	\caption[]{Five dielectric volumes: The grey-colored volume represents the semiconductor monolayer $\epsilon_{\text{M}}$ with thickness $d$, red and blue denote superstrate $\epsilon_1$ and substrate $\epsilon_2$, respectively, with a small vacuum gap $h$.}
	\label{fig:five_dielectric_volumes}
\end{figure}

We use the \textit{ansatz} from Ref.~\cite{trolle2017model}, which incorporates material screening via a three-dimensional background dielectric function of the corresponding bulk material and solves the generalized Poisson equation in a layered dielectric environment, cf.~Fig.~\ref{fig:five_dielectric_volumes}, which is local in $z$ direction and time. 
While the inclusion of the dielectric environment is crucial to obtain reasonable exciton binding energies \cite{peimyoo2020engineering,latini2015excitons,chernikov2014exciton}, care should be taken not to overestimate its influence in a macroscopic approach. For that matter, we include a small vacuum gap $h$ between the sample $\epsilon_{\text{M}}$ and the surrounding dielectric materials $\epsilon_1$ and $\epsilon_2$, which has to be adjusted to usual van der Waals distances \cite{florian2018dielectric,druppel2017diversity,deilmann2023optical}. 
Note that the usually applied effective monolayer thickness $d$ of 0.65\,nm \cite{kylanpaa2015binding} actually already includes the van der Waals distance, since the real sample thickness, i.e., the chalcogen--chalcogen distance is on the order of 0.33\,nm \cite{narciso2025interlayer}. Hence, adding a small vacuum gap of $h$ results in slightly overestimated interlayer distances compared to the bulk case, but there is experimental evidence for such an increased gap in monolayers \cite{rooney2017observing} with back-up from theory \cite{florian2018dielectric,druppel2017diversity}. 
Also note that the dielectric constants used in our macroscopic model correspond to the \textit{bulk} limit of the respective material. The confinement, i.e., reduction of dimensionality is then carried out via adjusting the layer thickness $d$. 

The macroscopically screened quantum-confined Coulomb potential $V_{\qpar}$ in the static limit ($\omega=\omega^{\prime}=0$) reads, cf.\ also \eqref{eq:greens_function_macroscopically_averaged} and \eqref{eq:Coulomb_Potential_Energy_SecondQuantized}:
\begin{shaded}
\begin{multline}
    V_{\qpar,\mathbf 0,\mathbf 0}^{1,1,1,1}\mathcolor{blue}{(\omega=0)}\\
    \rightarrow V_{\qpar} =  \frac{e^2}{\mathcal A}\int\mathrm dz\,\mathrm dz^{\prime}\,\left|\zeta(z)\right|^2 G_{\qpar}^{\text{mac}}(z,z^{\prime})\left|\zeta(z^{\prime})\right|^2,
    \label{eq:CoulombPotentialGeneral}
\end{multline}
\end{shaded}
\noindent
where $G_{\qpar}^{\text{mac}}(z,z^{\prime})$ solves the macroscopically screened Poisson equation in a layered dielectric environment with dielectric constants $\epsilon_1$/$\epsilon_2$ of superstrate/substrate and material dielectric constant $\epsilon_{\text{M}}^{}$, cf.~Fig.~\ref{fig:five_dielectric_volumes}:
\begin{align}
\epsilon_{\qpar}(z) = \begin{cases}
    \epsilon_1, \quad z\leq-\frac{d}{2}-h\\
    1, \quad -\frac{d}{2}-h < z \leq  - \frac{d}{2},\\
    \epsilon_{\text{M}}^{}, \quad -\frac{d}{2} < z \leq  \frac{d}{2},\\
    1, \quad \frac{d}{2} < z \leq  \frac{d}{2} + h,\\
    \epsilon_2, \quad \frac{d}{2}+h <  z,\\
\end{cases}
\label{eq:dielectric_environment}
\end{align}
which reads:
\begin{multline}
    -|\qpar|^2\epsilon_{\qpar}(z)G_{\qpar}^{\text{mac}}(z,z^{\prime})+\partial_z\left(\epsilon(z)\partial_zG_{\qpar}^{\text{mac}}(z,z^{\prime})\right)\\
    =-\delta(z-z^{\prime}).
    \label{eq:poisson_equation_layered_dielectric}
\end{multline}
In \eqref{eq:CoulombPotentialGeneral}, we already restricted the confinement wave functions $\zeta_n(z)$ to the respective ground state \mbox{$n=1$} and dropped the confinement quantum number index $\zeta_1(z)\equiv \zeta(z)$.

In the following, we solve the Poisson equation in a layered dielectric environment in \eqref{eq:poisson_equation_layered_dielectric} following a procedure similar to Refs.~\cite{thesis:nanjing_green_function,meckbach2018influence}.

\subsubsection{Poisson's Equation in Layered Media}
\label{eq:poisson_equation_layered_media}
Since the permittivity $\epsilon(z)$ is defined as piecewise constant in each volume, i.e., $\epsilon(z)=\epsilon_{i},~z\in \mathcal V_i$, cf.~\eqref{eq:dielectric_environment}, we can define the Poisson equation from \eqref{eq:poisson_equation_layered_dielectric} piecewise in each volume $\mathcal V_i$:
\begin{equation}
	\label{eq:greens_function_equation_volumes}
    -|\qpar|^2G_{i,j,\qpar}^{\text{mac}}(z,z^{\prime}) + \partial_z^2G_{i,j,\qpar}^{\text{mac}}(z,z^{\prime})=-\frac{1}{\epsilon_{i,\qpar}}\delta(z-z^{\prime}),
\end{equation}
where $z\in \mathcal V_i$ is the ``point of observation'' within the thin semiconductor, cf.\ Fig.~\ref{fig:five_dielectric_volumes}, and $z^{\prime}\in \mathcal V_j$ is the ``source point''.\\
The solution for the potential in the $i$-th volume induced by sources in the $j$-th volumes then reads:
\begin{align}
	\phi_{i,\qpar}^{\text{mac}}(z)=\sum_{j=1}^{n}\int_{\mathcal V_j}\mathrm dz^{\prime}G_{i,j,\qpar}^{\text{mac}}(z,z^{\prime})\rho_{j,\qpar}(z^{\prime}).
\end{align}
The boundary conditions (which follow from $\bf n\times\bf E_i=\bf n\times\bf E_{i+1}$ and $\bf n\cdot\bf D_i=\bf n\cdot\bf D_{i+1}$ at the boundary between adjacent volumes $\mathcal V_i$ and $\mathcal V_{i+1}$ as well as from $\bf E(\pm\infty)=0$ \cite{jackson1999classical}) read:
\begin{align}
\begin{split}
G_{1,j,\qpar}^{\text{mac}}(-\infty,z^{\prime})= &\, G_{n,j,\qpar}^{\text{mac}}(\infty,z^{\prime})=0,\\
	G_{i,j,\qpar}^{\text{mac}}(z_i,z^{\prime})= &\, G_{i+1,j,\qpar}^{\text{mac}}(z_i,z^{\prime}),\\
\epsilon_{i}\partial_zG_{i,j,\qpar}^{\text{mac}}(z,z^{\prime})\Big|_{z=z_i}={}&\epsilon_{i+1}\partial_zG_{i+1,j,\qpar}^{\text{mac}}(z,z^{\prime})\Big|_{z=z_i}.
\end{split}
\label{eq:Poisson_boundary_all}
\end{align}
By solving the equations for the Green's function (\eqref{eq:greens_function_equation_volumes}) we distinguish two cases. 
For $i=j$ (source point and point of observation in the same volume) we have to solve the inhomogeneous equations:
\begin{align}
	\label{eq:greens_function_equation_volumes_inhom}
    -|\qpar|^2G_{i,i,\qpar}^{\text{mac}}(z,z^{\prime}) + \partial_z^2G_{i,i,\qpar}^{\text{mac}}(z,z^{\prime})=-\frac{1}{\epsilon_{i}}\delta(z-z^{\prime}).
\end{align}
For $i\neq j$ (source point and point of observation in different volumes) we have to solve the homogeneous equations:
\begin{align}
	\label{eq:greens_function_equation_volumes_hom}
    -|\qpar|^2G_{i,j,\qpar}^{\text{mac}}(z,z^{\prime}) + \partial_z^2G_{i,j,\qpar}^{\text{mac}}(z,z^{\prime})=0,\quad i\neq j.
\end{align}
The particular solution of \eqref{eq:greens_function_equation_volumes_inhom} obtained via complex integration reads:
\begin{align}
    G_{i,i,\qpar}^{\text{mac,p}}(z,z^{\prime})=\frac{1}{2\epsilon_{i} |\qpar|}\efun{-|\qpar||z-z^{\prime}|}.
\end{align}
The homogeneous solution of \eqref{eq:greens_function_equation_volumes_hom} reads:
\begin{align}
    G_{i,j,\qpar}^{\text{mac,h}}(z,z^{\prime})=A_{i,j}\efun{|\qpar|z}+B_{i,j}\efun{-|\qpar|z}.
\end{align}
The complete solution reads:
\begin{align}
	G_{i,j,\qpar}^{\text{mac}}(z,z^{\prime})=\delta_{i,j}G_{i,i,\qpar}^{\text{mac,p}}(z,z^{\prime})+G_{i,j,\qpar}^{\text{mac,h}}(z,z^{\prime}).
\end{align}
Since we want to study electrons within the semiconductor monolayer, we have to calculate the Green's function whose source point and point of observation are both located within the volume of the monolayer semiconductor $\text{M}$, i.e., \mbox{$i=j=\text{M}$} (cf.\ Fig.~\ref{fig:five_dielectric_volumes}), which reads:
\begin{multline}
    G_{\text{M},\text{M},\qpar}^{\text{mac}}(z,z^{\prime})\\
    =\frac{1}{2\epsilon_0\epsilon_{\text{M}}|\qpar|}\mathrm e^{-|\qpar||z-z^{\prime}|}
    + A_{\text{M},\text{M}}\mathrm e^{|\qpar|z} + B_{\text{M},\text{M}}\mathrm e^{-|\qpar|z}.
\end{multline}
The coefficients $A_{\text{M},\text{M}}$ and $B_{\text{M},\text{M}}$ are obtained by solving the linear system of equations provided by the boundary conditions in \eqref{eq:Poisson_boundary_all}. 
We obtain for the Green's function in the semiconductor layer \mbox{$G_{\text{M},\text{M},\qpar}^{\text{mac}}(z,z^{\prime})\equiv G_{\qpar}^{\text{mac}}(z,z^{\prime})$} \cite{Mathematica}:
\begin{widetext}
\begin{multline}
    G_{\qpar}^{\text{mac}}(z,z^{\prime}) =  \frac{1}{2\epsilon_0\epsilon_{\text{M}}|\qpar|}\mathrm e^{-|\qpar||z-z^{\prime}|}\\
     +\frac{1}{2\epsilon_0\epsilon_{\text{M}}|\qpar|g_{\qpar}}\left(\left(\epsilon_{\text{M},-}\epsilon_{2,+}\mathrm e^{h|\qpar|}-\epsilon_{\text{M},+}\epsilon_{2,-}\mathrm e^{-h|\qpar|}\right)\right.
     \left.\left(\epsilon_{1,+}\epsilon_{\text{M},-}\mathrm e^{h|\qpar|}-\epsilon_{1,-}\epsilon_{\text{M},+}\mathrm e^{-h|\qpar|}\right)\mathrm e^{-d|\qpar|}\right.\left(\mathrm e^{|\qpar|(z-z^{\prime})}+\mathrm e^{-|\qpar|(z-z^{\prime})}\right)\\
    \left.
    +\left(\epsilon_{\text{M},-}\epsilon_{2,+}\mathrm e^{h|\qpar|}-\epsilon_{\text{M},+}\epsilon_{2,-}\mathrm e^{-h|\qpar|}\right)\left(\epsilon_{1,+}\epsilon_{\text{M},+}\mathrm e^{h|\qpar|}-\epsilon_{1,-}\epsilon_{\text{M},-}\mathrm e^{-h|\qpar|}\right)\mathrm e^{|\qpar|(z+z^{\prime})}\right.\\
    \left.
    +\left(\epsilon_{\text{M},+}\epsilon_{2,+}\mathrm e^{h|\qpar|}-\epsilon_{\text{M},-}\epsilon_{2,-}\mathrm e^{-h|\qpar|}\right)\left(\epsilon_{1,+}\epsilon_{\text{M},-}\mathrm e^{h|\qpar|}-\epsilon_{1,-}\epsilon_{\text{M},+}\mathrm e^{-h|\qpar|}\right)\mathrm e^{-|\qpar|(z+z^{\prime})}
    \right),
    \label{eq:GreensFunctionFiveVolumes}
\end{multline}
\end{widetext}
with the helper function $g_{\qpar}$:
\begin{multline}
    g_{\qpar}  =  
    -\epsilon_{\text{M},-}\epsilon_{\text{M},-}\\
    \times\left(\epsilon_{1,+}\epsilon_{2,+}\mathrm e^{-d|\qpar|}\mathrm e^{2h|\qpar|} 
    - \epsilon_{1,-}\epsilon_{2,-}\mathrm e^{d|\qpar|}\mathrm e^{-2h|\qpar|}\right)\\
    -2\epsilon_{\text{M},+}\epsilon_{\text{M},-}\left(\epsilon_1\epsilon_2-1\right)\left(\mathrm e^{d|\qpar|}-\mathrm e^{-d|\qpar|}\right)\\
    + 
    \epsilon_{\text{M},+}\epsilon_{\text{M},+}\\
    \times\left(\epsilon_{1,+}\epsilon_{2,+}\mathrm e^{d|\qpar|}\mathrm e^{2h|\qpar|} - \epsilon_{1,-}\epsilon_{2,-}\mathrm e^{-d|\qpar|}\mathrm e^{-2h|\qpar|}\right),
\end{multline}
where we defined:
\begin{align}
    \epsilon_{i,\pm}=\epsilon_i\pm 1.
\end{align}
Note that the limit of vanishing interlayer distance is discussed in \eqref{eq:Greens_Function_monolayer_zerodistance}. 
To calculate the quantum-confined Coulomb potential, \eqref{eq:CoulombPotentialGeneral}, we make the following \textit{ansatz} for the confinement wave function in $z$ direction:
\begin{align}
    \zeta(z) = \sqrt{\frac{2}{d}}\cos\mleft(\frac{\pi}{d}z\mright),
    \label{eq:confinement_wave_function}
\end{align}
which is the ground state of an infinitely deep potential well. 
Note that, in principle, we cannot assume that \eqref{eq:confinement_wave_function} for the ground state of an infinitely deep potential well \mbox{$n=1$} holds for the charge distributions in $z$ direction in every thin semiconductor: While \eqref{eq:confinement_wave_function} is a good approximation for conventional quantum wells \cite{malic2007coulomb,liu1994local} and TMDCs \cite{latini2015excitons} in the lowest conduction and highest valence bands, for h-BN, the first excited state with $n=2$ would be a better approximation to actual \textit{ab initio} calculations of the charge distribution in $z$ direction \cite{latini2015excitons}. 

The $z$ integrals occurring in \eqref{eq:CoulombPotentialGeneral} by employing \eqref{eq:GreensFunctionFiveVolumes} and \eqref{eq:confinement_wave_function} then read:
\begin{multline}
    \frac{4}{d^2}\int_{-\frac{d}{2}}^{\frac{d}{2}}\mathrm dz\,\mathrm dz^{\prime}\,\cos^2\mleft(\frac{\pi}{d}z\mright)\mathrm e^{-|\qpar||z-z^{\prime}|}\cos^2\mleft(\frac{\pi}{d}z^{\prime}\mright) \\
    = \frac{1}{|\qpar|d\left(4\pi^2+|\qpar|^2d^2\right)}\\
    \times\left(8\pi^2+3|\qpar|^2d^2 - \frac{32\pi^4\left(1-\mathrm e^{-|\qpar|d}\right)}{|\qpar|d\left(4\pi^2+|\qpar|^2d^2\right)}\right),
\end{multline}
and:
\begin{multline}
    \frac{4}{d^2}\int_{-\frac{d}{2}}^{\frac{d}{2}}\mathrm dz\,\mathrm dz^{\prime}\,\cos^2\mleft(\frac{\pi}{d}z\mright)\mathrm e^{\pm|\qpar|(\pm z\pm z^{\prime})}\cos^2\mleft(\frac{\pi}{d}z^{\prime}\mright)  \\
    = \left(\frac{8\pi^2\sinh\mleft(\frac{|\qpar|d}{2}\mright)}{|\qpar|d\left(4\pi^2+|\qpar|^2d^2\right)}\right)^2.
\end{multline}
With these ingredients, an explicit expression for the macroscopically screened Coulomb potential from \eqref{eq:CoulombPotentialGeneral} can be derived:
\begin{widetext}
\begin{shaded}
\begin{multline}
    V_{\qpar} =  \frac{e^2}{2\mathcal A\epsilon_0\epsilon_{\text{M}}|\qpar|}\frac{1}{|\mathbf q|d\left(4\pi^2+|\qpar|^2d^2\right)}\left(8\pi^2+3|\qpar|^2d^2 - \frac{32\pi^4\left(1-\mathrm e^{-|\qpar|d}\right)}{|\qpar|d\left(4\pi^2+|\qpar|^2d^2\right)}\right)\\ 
    +\frac{e^2}{2\mathcal A\epsilon_0\epsilon_{\text{M}}|\qpar|}\frac{1}{g_{\qpar}}
    \left(\frac{8\pi^2\sinh\mleft(\frac{|\qpar|d}{2}\mright)}{|\qpar|d\left(4\pi^2+|\qpar|^2d^2\right)}\right)^2\bigg(\left(\epsilon_{\text{M},-}\epsilon_{2,+}\mathrm e^{h|\qpar|}-\epsilon_{\text{M},+}\epsilon_{2,-}\mathrm e^{-h|\qpar|}\right)\left(\epsilon_{1,+}\epsilon_{\text{M},-}\mathrm e^{h|\qpar|}-\epsilon_{1,-}\epsilon_{\text{M},+}\mathrm e^{-h|\qpar|}\right)\\
    \times 2\mathrm e^{-d|\qpar|}
    +\left(\epsilon_{\text{M},-}\epsilon_{2,+}\mathrm e^{h|\qpar|}-\epsilon_{\text{M},+}\epsilon_{2,-}\mathrm e^{-h|\qpar|}\right)\left(\epsilon_{1,+}\epsilon_{\text{M},+}\mathrm e^{h|\qpar|}-\epsilon_{1,-}\epsilon_{\text{M},-}\mathrm e^{-h|\qpar|}\right)\\
    +\left(\epsilon_{\text{M},+}\epsilon_{2,+}\mathrm e^{h|\qpar|}-\epsilon_{\text{M},-}\epsilon_{2,-}\mathrm e^{-h|\qpar|}\right)
    \left(\epsilon_{1,+}\epsilon_{\text{M},-}\mathrm e^{h|\qpar|}-\epsilon_{1,-}\epsilon_{\text{M},+}\mathrm e^{-h|\qpar|}\right)
    \bigg).
    \label{eq:CoulombPotentialFiveVolumes}
\end{multline}
\end{shaded}
For a vanishing vacuum gap (\mbox{$h=0\,$nm}), the macroscopic Green's function can also be expressed as:
\begin{multline}
    G_{\mathbf q}^{\text{mac}}(z,z^{\prime})=\frac{1}{2\epsilon_0\epsilon_{\text{M}}|\qpar|}\mathrm e^{-|\qpar||z-z^{\prime}|}\\
     + \frac{1}{2\epsilon_0\epsilon_{\text{M}}|\qpar|}\frac{1}{\delta_-\mathrm e^{-|\qpar|d}-\gamma_+\mathrm e^{|\qpar|d}}
    \left(\gamma_-\mathrm e^{|\qpar|(z+z^{\prime})}
    +\delta_+\mathrm e^{-|\qpar|(z+z^{\prime})} - \delta_-\mathrm e^{-|\qpar|(z-z^{\prime})}\mathrm e^{-|\qpar|d}
    - \delta_-\mathrm e^{|\qpar|(z-z^{\prime})}\mathrm e^{-|\qpar|d}\right),
\label{eq:Greens_Function_monolayer_zerodistance}
\end{multline}
\end{widetext}
with:
\begin{align}
\begin{split}
    \gamma_{\pm} = &\, \left(\epsilon_{1}+\epsilon_{\text{M}}\right)\left(\epsilon_{2}\pm\epsilon_{\text{M}}\right),\\
    \delta_{\pm} = &\, \left(\epsilon_{1}-\epsilon_{\text{M}}\right)\left(\epsilon_{2}\pm\epsilon_{\text{M}}\right).
    \end{split}
\end{align}
The first term in \eqref{eq:Greens_Function_monolayer_zerodistance} 
is the inhomogeneous contribution and the second term in \eqref{eq:Greens_Function_monolayer_zerodistance} 
is the homogeneous contribution due to the dielectric environment, i.e., due to image charges. If we assume $\epsilon_1=\epsilon_2=\epsilon_{\text{M}}$, the second term in \eqref{eq:Greens_Function_monolayer_zerodistance} 
vanishes and the Green's function is only given by the inhomogeneous contribution in the first term in \eqref{eq:Greens_Function_monolayer_zerodistance}. 
For the simplest case of a Dirac delta confinement of the charges, i.e., $|\zeta(z)|^2 = \delta(z)$, the quantum-confined macroscopic Coulomb potential $V_{\qpar} = \frac{e^2}{\mathcal A}\int\mathrm dz\,\mathrm dz^{\prime}\,\left|\zeta(z)\right|^2 G_{\qpar}^{\text{mac}}(z,z^{\prime})\left|\zeta(z^{\prime})\right|^2$ reads:
\begin{align}
\begin{split}
    V_{\qpar}    =
    \frac{e^2}{2\mathcal A\epsilon_0\epsilon_{\text{M}}|\qpar|}\frac{\gamma_- + \delta_+ -\gamma_+\mathrm e^{|\qpar|d} - \delta_-\mathrm e^{-|\qpar|d}}{\delta_-\mathrm e^{-|\qpar|d}-\gamma_+\mathrm e^{|\qpar|d}}.
    \end{split}
    \label{eq:Coulomb_potential_no_vacuum_gap_delta_confinement}
\end{align}
Comparing our results with the literature, we observe that \eqref{eq:Greens_Function_monolayer_zerodistance} 
is not equal to but similar to Rytova's formula derived in Refs.~\cite{rytova1965coulomb,rytova1967coulomb}. Interestingly, the expression in Ref.~\cite{rytova1965coulomb} differs from the one in Ref.~\cite{rytova1967coulomb}, although both are derived with the same assumptions. If we assume $z\geq z^{\prime}$, 
\eqref{eq:Greens_Function_monolayer_zerodistance} corresponds exactly to Keldysh's formula in Ref.~\cite{keldysh1979coulomb}. Regarding more recent calculations, our result matches exactly with Ref.~\cite{neuhaus2020microscopic} or, if $z=z^{\prime}=0$, 
with Ref.~\cite{scharf2019dynamical} (excluding the $3\chi$ model). This validates our expression of the Green's function in \eqref{eq:Greens_Function_monolayer_zerodistance} 
and it strongly indicates that we have to slightly modify Rytova's expressions from Refs.~\cite{rytova1965coulomb,rytova1967coulomb}.

\FloatBarrier
\subsubsection{Analytical Model of Screening}
\label{sec:screening_analytical_model}
\textit{Ab initio} calculations suggest that the constant approximation, i.e., $\mathbf q=\mathbf 0$ approximation, of the quasi-3D material dielectric function $ \epsilon_{\text{M},\mathbf q=\mathbf 0}^{\text{3D}} = \epsilon_{\text{M}}^{}$, cf.\ Fig.~\ref{fig:five_dielectric_volumes}, significantly overestimates the screening at large momenta \cite{qiu2016screening,latini2015excitons,huser2013dielectric,qiu2013optical}. While this is tolerable in the evaluation of interaction processes, which do not depend too strongly on the Coulomb interaction strength such as exciton--phonon scattering, it underestimates biexciton and trion binding energies significantly and fails to describe the strength of the Coulomb interaction with large momentum transfers, e.g., Dexter coupling \cite{berghauser2018inverted,bernal2018exciton,timmer2024ultrafast,dogadov2026diss}, Auger coupling of momentum-dark excitons \cite{erkensten2021dark} or short-range exchange interaction \cite{qiu2015nonanalyticity,yu2014valley}, correctly.

To capture this property in an analytical theory, we use the following analytic model of the 3D material dielectric function $\epsilon_{\text{M},\mathbf q}^{\text{3D}}$ from Refs.~\cite{trolle2017model,cappellini1993model} in the static (or adiabatic) limit with $\omega=0$:
\begin{align}
\label{eq:dielectric_function_material_q_dependent}
    \epsilon_{\text{M},\mathbf q}^{\text{3D}} = 1 + \frac{1}{\left(
    \epsilon_{\text{M}}-1\right)^{-1} + \alpha_{\text{TF}}\frac{|\mathbf q|^2}{q_{\text{TF}}^2} + \frac{\hbar^2|\mathbf q|^4}{4m_0^2\omega_{\text{pl}}^2}}.
\end{align}
\eqref{eq:dielectric_function_material_q_dependent} is designed to model the macroscopic three-dimensional interband dielectric function in RPA \cite{walter1970wave} from actual \textit{ab initio} calculations at a temperature of $T=0\,$K, i.e, \eqref{eq:dielectric_function_microscopic} for $\lambda\neq\lambda^{\prime}$ in the macroscopic limit. 
We note that there are other, more rigorous approaches \cite{levine1982new,hybertsen1988model}, which fulfill the Kramers--Kronig relations and can even be used to construct the full microscopic dielectric matrix with all local fields \cite{hybertsen1988model}. 
In \eqref{eq:dielectric_function_material_q_dependent}, 
$\epsilon_{\text{M}}$ is the static bulk dielectric constant, cf.\ Fig.~\ref{fig:five_dielectric_volumes}, the contribution ``$\alpha_{\text{TF}}\frac{|\mathbf q|^2}{q_{\text{TF}}^2}$'' resembles the semiconductor Thomas--Fermi limit with fitting parameter $\alpha_{\text{TF}}$ and Thomas--Fermi wave vector 
$q_{\text{TF}}^3 = \sqrt{\frac{3\omega_{\text{pl}}^{2}m_0^{4}}{\hbar^{6}\pi^{4}\epsilon_0^{2}}}$, 
and the contribution ``$\frac{\hbar^2|\mathbf q|^4}{4m_0^2\omega_{\text{pl}}^2}$'' describes a single plasmon pole contribution, where $\omega_{\text{pl}}$ is the bulk plasmon frequency. 
Note that the Thomas--Fermi contribution in \eqref{eq:dielectric_function_material_q_dependent} should not be confused with the Thomas--Fermi limit of a quasi-free electron gas in metals or doped semiconductors derived from the \textit{intraband} Lindhard function ($\lambda=\lambda^{\prime}$ in \eqref{eq:dielectric_function_microscopic}), as it describes the limit for small $\mathbf q$ of the \textit{interband} Lindhard function ($\lambda\neq \lambda^{\prime}$ in \eqref{eq:dielectric_function_microscopic}), i.e., the screening induced by an interband polarization cloud and not by real carriers \cite{penn1962wave}. 
Consequently, $\omega_{\text{pl}}$ is related to the \textit{interband} plasmon, which describes the collective mode of all possible transitions from fully occuppied bands below the Fermi level including core bands to the empty conduction bands \cite{bechstedt1979electronic}. 

\eqref{eq:dielectric_function_material_q_dependent} serves as an analytical description of the macroscopic limit ($\mathbf G=\mathbf G^{\prime}=\mathbf 0$) of the \textit{ab initio} theory of screening and replaces the constant approximation of the dielectric function of the material $\epsilon_{\text{M}}\rightarrow \epsilon_{\text{M},\mathbf q}^{\text{3D}}$ in \eqref{eq:CoulombPotentialFiveVolumes} and \eqref{eq:Coulomb_potential_no_vacuum_gap_delta_confinement}. 
Note that substrate and superstrate material can be treated in an analog way.

To construct a two-dimensional material dielectric function $\epsilon_{\text{M},\qpar}^{\text{2D}}$ from the three-dimensional dielectric function $\epsilon_{\text{M},\mathbf q}^{\text{3D}}$ of the bulk material, which we include as the material ``background'' in our treatment of the layered dielectric structure, cf.~Fig.~\ref{fig:five_dielectric_volumes}, we employ \eqref{eq:macroscopic_dielectric_function_2D_definition} and \eqref{eq:dielectric_function_2d_louie_def_realspace} and write:
\begin{align}
\label{eq:effective_confined_screening}
    \left(\epsilon_{\text{M},\qpar}^{\text{2D}}\right)^{-1} = \frac{V_{\qpar}}
    {V_{0,\qpar}}
\end{align}
where $V_{\qpar}$ is the macroscopically screened Coulomb potential from \eqref{eq:CoulombPotentialGeneral} and $V_{0,\qpar} = V_{\qpar}|_{\epsilon_{\text{M},\qpar}^{\text{3D}}=1}$ is the unscreened Coulomb potential obtained from \eqref{eq:CoulombPotentialGeneral} by setting the background material dielectric function $\epsilon_{\text{M},\qpar}^{\text{3D}}$, cf.~\eqref{eq:dielectric_function_material_q_dependent}, to unity. 
\begin{figure}[h!]
\centering
\begin{tabular}{c}
     \hspace{-0.205cm}
     \vspace{-0.1cm}
		\includegraphics[width=0.989\linewidth]{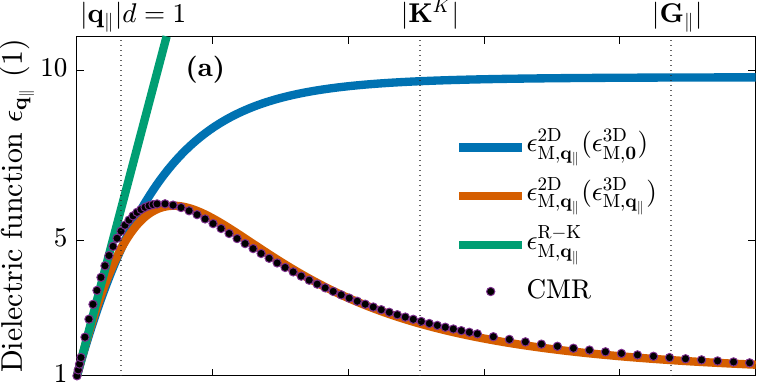}\\
		\includegraphics[width=1\linewidth]{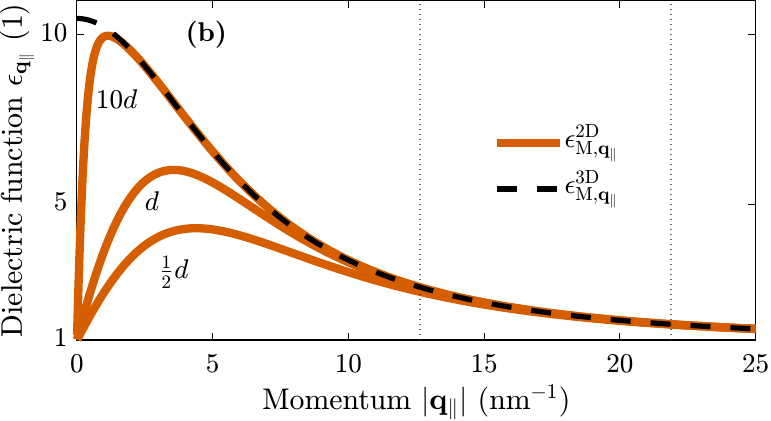}
\end{tabular}
	
	\caption[]{(a) The two-dimensional dielectric function $\epsilon_{\text{M},\qpar}^{\text{2D}}$ from \eqref{eq:effective_confined_screening} of an example MoS$_2$ monolayer ($\epsilon_{\text{M},\parallel}=15.5$, $\epsilon_{\text{M},{\bot}}=6.2$, $\epsilon_{\text{M},\text{bulk}}=10.5$ \cite{laturia2018dielectric}, $d=0.618\,$nm \cite{kylanpaa2015binding}, $\hbar\omega_{\text{pl}}=22.5\,$eV \cite{kumar2012tunable}) suspended in vacuum ($\epsilon_1=\epsilon_2=1$) is shown. ``$\epsilon_{\text{M},\qpar}^{\text{2D}}(\epsilon_{\text{M},\mathbf 0}^{\text{3D}})$'' uses a constant approximation of the three-dimensional material background dielectric function $\epsilon_{\text{M},\mathbf 0}^{\text{3D}}$ and ``$\epsilon_{\text{M},\qpar}^{\text{2D}}(\epsilon_{\text{M},\qpar}^{\text{3D}})$'' uses the $\qpar$ dependent model from \eqref{eq:dielectric_function_material_q_dependent}. ``$\epsilon_{\text{M},\qpar}^{\text{R--K}}$'' denotes the Rytova--Keldysh thin-film limit in vacuum from \eqref{eq:epsilon_rytova_keldysh} and ``CMR'' denotes \textit{ab initio} calculations extracted from the Computational Materials Repository (CMR) \cite{andersen2015dielectric}. $\mathbf K^K$ is a $K$ valley momentum, cf.~Sec.~\ref{sec:expansion_band_extrema}, 
 and $\Gpar$ is a reciprocal lattice vector. (b) The two-dimensional dielectric function $\epsilon_{\text{M},\qpar}^{\text{2D}}$ from \eqref{eq:effective_confined_screening} for three different layer thicknesses and the three-dimensional dielectric function $\epsilon_{\text{M},\qpar}^{\text{3D}}$ from \eqref{eq:dielectric_function_material_q_dependent} are shown.
 }
	\label{fig:effective_screening}
\end{figure}
In Fig.~\ref{fig:effective_screening}(a), the effective two-dimensional dielectric function $\epsilon_{\text{M},\qpar}^{\text{2D}}$ from \eqref{eq:effective_confined_screening} for a constant ``$\epsilon_{\text{M},\qpar}^{\text{2D}}(\epsilon_{\text{M},\mathbf 0}^{\text{3D}})$'' (blue solid line) and $\mathbf q$ dependent ``$\epsilon_{\text{M},\qpar}^{\text{2D}}(\epsilon_{\text{M},\qpar}^{\text{3D}})$'' (red solid line) three-dimensional material dielectric background function from \eqref{eq:dielectric_function_material_q_dependent} is shown. Additionally, ``$\epsilon_{\text{M},\qpar}^{\text{R--K}}$'' (green solid line) depicts the Rytova--Keldysh thin-film limit in vacuum \cite{keldysh1979coulomb}, which can be obtained in first order Taylor expansion in $|\qpar|d$:
\begin{align}
    \epsilon_{\text{M},\qpar}^{\text{R--K}} = 1 + \frac{1}{2}\epsilon_{\text{M},\mathbf 0}^{\text{3D}}|\qpar|d,
    \label{eq:epsilon_rytova_keldysh}
\end{align}
and the black dots represent the dielectric function calculated by \textit{ab initio} methods extracted from the Computational Materials Repository (CMR) \cite{andersen2015dielectric}. 
Fig.~\ref{fig:effective_screening}(a) shows that the constant approximation $\epsilon_{\text{M},\qpar}^{\text{2D}}(\epsilon_{\text{M},\mathbf 0}^{\text{3D}})$ significantly overestimates the screening at large momenta and we are able to obtain a good agreement of the analytical model $\epsilon_{\text{M},\qpar}^{\text{2D}}(\epsilon_{\text{M},\qpar}^{\text{3D}})$ from \eqref{eq:dielectric_function_material_q_dependent} with \textit{ab initio} calculations by adjusting the Thomas--Fermi parameter $\alpha_{\text{TF}}$ accordingly. The linear Rytova--Keldysh limit (green solid line) is only a good approximation at crystal momenta $|\qpar|\leq \frac{1}{d}$ and becomes even worse than the constant approximation at larger $\qpar$ (blue solid line). In Fig.~\ref{fig:effective_screening}(b), we depict the two-dimensional dielectric function from \eqref{eq:effective_confined_screening} for three different layer thicknesses $d$ and the three-dimensional dielectric function from \eqref{eq:dielectric_function_material_q_dependent}. We observe that the two-dimensional dielectric function approaches the three-dimensional dielectric function, if $d$ increases, which resembles actual \textit{ab initio} results \cite{huser2013dielectric,latini2015excitons}, validating our approach.

In general, the attenuation of the dielectric function at large momenta stems predominantly from the local-field effects, i.e., from the contributions with $\mathbf G\neq\mathbf 0$ and $\mathbf G^{\prime}\neq\mathbf 0$. 
Since screening affects the Coulomb interaction via the inverse $\left(\epsilon^{-1}\right)_{\mathbf q,\mathbf G,\mathbf G^{\prime}}$ all local fields with $\mathbf G$ and $\mathbf G^{\prime}$ affect the $\mathbf G=\mathbf G^{\prime}=\mathbf 0$ component of the inverse dielectric matrix $\left(\epsilon^{-1}\right)_{\mathbf q,\mathbf 0,\mathbf 0}$, which is not just the inverse of the $\mathbf G=\mathbf G^{\prime}=\mathbf 0$ component of the dielectric matrix $\left(\epsilon_{\mathbf q,\mathbf 0,\mathbf 0}\right)^{-1}$.

Hence, with a properly parametrized analytical model of the bulk material dielectric function $\epsilon_{\text{M},\mathbf q}^{\text{3D}}$ from \eqref{eq:dielectric_function_material_q_dependent} combined with the analytical dielectric slab model, cf.~Fig.~\ref{fig:five_dielectric_volumes}, we obtain a good agreement with \textit{ab initio} approaches including environmental screening and the evaluation of the microscopic dielectric function from \eqref{eq:dielectric_function_microscopic} via \textit{ab initio} methods can be circumvented, as long as we do not explicitly need local-field effects with $\mathbf G\neq\mathbf G^{\prime}$.

We want to highlight that the model used here and in Ref.~\cite{trolle2017model} is only \textit{one} possibility to describe screening beyond the Rytova--Keldysh limit. For example, the $3\chi$ model used in, e.g., Refs.~\cite{van2018coulomb,scharf2019dynamical}, the simplified numerical model in Ref.~\cite{ninhos2026microscopic}, or the Wannier-function-continuum-electrostatics model in Refs.~\cite{florian2018dielectric,steinhoff2017exciton,steinhoff2014influence,rosner2015wannier} are other valid approaches.

\FloatBarrier

\section{Excitons and Optical Spectra}
\label{sec:excitons}

In this section, we discuss the occurrence of bound electron--hole pairs, the excitons. 
First, we derive an effective Bethe--Salpeter equation, which can be used in \textit{ab initio} calculations, then we discuss a simplified version for few-band effective-mass approaches, the Wannier equation, and excitonic signatures in optical spectra.

From \eqref{eq:Coulomb_Hamiltonian_SecondQuantized_ValleyExpanded_adiabatic_approx}, we extract the following contributions, 
the hole--hole interaction:
\begin{multline}
    \hat H_{\text{Coul-hh}}^{}= \frac{1}{2}\sum_{\substack{\mathbf k,\mathbf k^{\prime},\mathbf q,\mathbf G,\mathbf G^{\prime},\\v_1,v_2,v_3,v_4}} V^{}_{\mathbf q,\mathbf G,\mathbf G^{\prime}}\\
    \times 
    \overline \Upsilon_{\mathbf k+\mathbf q+\mathbf G,\mathbf k}^{v_1,v_4}\overline \Upsilon_{\mathbf k^{\prime}-\mathbf q-\mathbf G^{\prime},\mathbf k^{\prime}}^{v_2,v_3}
    \vdag{1,\mathbf k+\mathbf q}\vdag{2,\mathbf k^{\prime}-\mathbf q}\vndag{3,\mathbf k^{\prime}}\vndag{4,\mathbf k},
    \label{eq:CoulombHamiltonianHoleHole}
\end{multline}
the electron--electron interaction:
\begin{multline}
    \hat H_{\text{Coul-ee}}^{}= \frac{1}{2}\sum_{\substack{\mathbf k,\mathbf k^{\prime},\mathbf q,\mathbf G,\mathbf G^{\prime},\\c_1,c_2,c_3,c_4}} V^{}_{\mathbf q,\mathbf G,\mathbf G^{\prime}}\\
    \times 
    \overline \Upsilon_{\mathbf k+\mathbf q+\mathbf G,\mathbf k}^{c_1,c_4}\overline \Upsilon_{\mathbf k^{\prime}-\mathbf q-\mathbf G^{\prime},\mathbf k^{\prime}}^{c_2,c_3}
    \cdag{1,\mathbf k+\mathbf q}\cdag{2,\mathbf k^{\prime}-\mathbf q}\cndag{3,\mathbf k^{\prime}}\cndag{4,\mathbf k},
    \label{eq:CoulombHamiltonianElectronElectron}
\end{multline}
the direct electron--hole interaction:
\begin{multline}
    \hat H_{\text{Coul-eh}}^{\text{dir}}= \sum_{\substack{\mathbf k,\mathbf k^{\prime},\mathbf q,\mathbf G,\mathbf G^{\prime},\\v_1,v_2,c_1,c_2}} \frac{1}{2}\left(V^{}_{\mathbf q,\mathbf G,\mathbf G^{\prime}} + V^{}_{-\mathbf q,-\mathbf G^{\prime},-\mathbf G} \right)\\
    \times 
    \overline \Upsilon_{\mathbf k+\mathbf q+\mathbf G,\mathbf k}^{c_1,c_2}\overline \Upsilon_{\mathbf k^{\prime}-\mathbf q-\mathbf G^{\prime},\mathbf k^{\prime}}^{v_1,v_2}
    \cdag{1,\mathbf k+\mathbf q}\vdag{1,\mathbf k^{\prime}-\mathbf q}\vndag{2,\mathbf k^{\prime}}\cndag{2,\mathbf k},
    \label{eq:CoulombHamiltonianElectronHole}
\end{multline}
the exchange electron--hole interaction:
\begin{multline}
    \hat H_{\text{Coul-eh}}^{\text{exch}}= \sum_{\substack{\mathbf k,\mathbf k^{\prime},\mathbf q,\mathbf G,\mathbf G^{\prime},\\v_1,v_2,c_1,c_2}} \frac{1}{2}\left(V^{}_{\mathbf q,\mathbf G,\mathbf G^{\prime}} + V^{}_{-\mathbf q,-\mathbf G^{\prime},-\mathbf G} \right)\\
    \times 
    \overline \Upsilon_{\mathbf k+\mathbf q+\mathbf G,\mathbf k}^{c_1,v_2}\overline \Upsilon_{\mathbf k^{\prime}-\mathbf q-\mathbf G^{\prime},\mathbf k^{\prime}}^{v_1,c_2}
    \cdag{1,\mathbf k+\mathbf q}\vdag{1,\mathbf k^{\prime}-\mathbf q}\cndag{2,\mathbf k^{\prime}}\vndag{2,\mathbf k},
    \label{eq:CoulombHamiltonianElectronHole}
\end{multline}
and the single-particle contributions arising from normal ordering:
\begin{align}
\begin{split}
    &\hat H_{\text{Coul},\text{SP}} \\
    & = \frac{1}{2}\sum_{\substack{\mathbf k,\mathbf q,\mathbf G,\mathbf G^{\prime},\\v,\lambda}}V^{}_{\mathbf q,\mathbf G,\mathbf G^{\prime}} \overline \Upsilon_{\mathbf k,\mathbf k-\mathbf q-\mathbf G}^{v,\lambda}\overline \Upsilon_{\mathbf k-\mathbf q-\mathbf G^{\prime},\mathbf k}^{\lambda,v}\vdag{\mathbf k}{}\vndag{\mathbf k}{}\\
    &\quad + \frac{1}{2}\sum_{\substack{\mathbf k,\mathbf q,\mathbf G,\mathbf G^{\prime},\\c,\lambda}}V^{}_{\mathbf q,\mathbf G,\mathbf G^{\prime}} \overline \Upsilon_{\mathbf k,\mathbf k-\mathbf q-\mathbf G}^{c,\lambda}\overline \Upsilon_{\mathbf k-\mathbf q-\mathbf G^{\prime},\mathbf k}^{\lambda,c}\cdag{\mathbf k}{}\cndag{\mathbf k}{}.
    \end{split}
    \label{eq:CoulombHamiltonianSelfEnergy}
\end{align}
Note that we use a compound index for the bands and spins $\lambda \equiv \lambda,s$ here and restricted ourselves the the ground state in the out-of-plane direction: $V^{}_{\mathbf q,\mathbf G,\mathbf G^{\prime}}  \equiv V^{1,1,1,1}_{\mathbf q,\mathbf G,\mathbf G^{\prime}}(\omega=0) $ in the static limit, cf.~\eqref{eq:Coulomb_Potential_Energy_SecondQuantized}. Moreover, we assume three-dimensional quasi-momenta for notational simplicity. 
Then, we calculate the Heisenberg equations of motion for the optically active interband transition $\langle \vdag{\mathbf q}\cndag{\mathbf q} \rangle$ at zero center-of-mass momentum:
\begin{align}
    \mathrm i\hbar\partial_t \langle \vdag{\mathbf q}\cndag{\mathbf q} \rangle = \langle [\vdag{\mathbf q}\cndag{\mathbf q},\hat H_{\text{tot}}^{}] \rangle,
    \label{eq:heisenberg_eom_interband_transition}
\end{align}
with the total (full) solid-state Hamiltonian in Born--Oppenheimer approximation:
\begin{multline}
    \hat H_{\text{tot}}^{}
    = \hat H_0^{} \\
    + \hat H_{\text{Coul-hh}}^{} + \hat H_{\text{Coul-ee}}^{}
    + \hat H_{\text{Coul-eh}}^{\text{dir}} + \hat H_{\text{Coul-eh}}^{\text{exch}}
    +\hat H_{\text{Coul},\text{SP}},
\end{multline}
where the unrenormalized single-particle Hamiltonian reads:
\begin{align}
    \hat H_0^{} = \sum_{\mathbf k,c}E_{c,\mathbf k}\cdag{\mathbf k}\cndag{\mathbf k} + \sum_{\mathbf k,v}E_{v,\mathbf k}\vdag{\mathbf k}\vndag{\mathbf k},
\end{align}
with Bloch-electron dispersion $E_{\lambda,\mathbf k}$ for valence-band states ($\lambda=v$) and conduction-band states ($\lambda=c$), 
and perform a Hartree--Fock approximation (not to be confused with the Hartree--Fock theory for calculating ground states) assuming completely filled valence bands and empty conduction bands, which corresponds to the limit of linear optics in a charge-neutral semiconductor. Note that under these circumstances, the electron--electron interaction Hamiltonian $\hat H_{\text{Coul-ee}}^{}$ does not contribute at all. 
Also note that, since we start from the effective Coulomb Hamiltonian from \eqref{eq:Coulomb_Hamiltonian_SecondQuantized_ValleyExpanded_adiabatic_approx}, screening (via background states) already occurs on a Hartree--Fock level. 
At last, we expand the electron--hole transitions as follows:
\begin{align}
    \langle \vdag{\mathbf q}\cndag{\mathbf q} \rangle = \sum_{\mu}\Phi_{\mu,v,c,\mathbf q} \Pol{\mu},
    \label{eq:exciton_expansion_BSE}
\end{align}
For demonstration purposes, here, we assume only optically bright transitions with zero center-of-mass momentum. In \eqref{eq:exciton_expansion}, where we discuss the effective-mass approach, we depict the general expansion, which also considers excitonic transitions with nonzero center-of-mass momenta. 
The expansion coefficients $\Phi_{\mu,v,c,\mathbf q}$ in \eqref{eq:exciton_expansion_BSE} satisfy the following eigenvalue equation:
\begin{shaded}
\begin{multline}
    \left(E_{c,\mathbf q}-E_{v,\mathbf q} + \Sigma_{c,v,\mathbf q}^{\text{H}} + \Sigma_{c,v,\mathbf q}^{\text{SEX}} + \Sigma_{c,v,\mathbf q}^{\text{COH}}
     \right)\Phi_{\mu,v,c,\mathbf q}\\
    + \sum_{\mathbf q^{\prime},v^{\prime},c^{\prime}}\left(K_{\text{eh}}^{\text{dir}}\,\vphantom{K}_{v,c,\mathbf q}^{v^{\prime},c^{\prime},\mathbf q^{\prime}} + K_{\text{eh}}^{\text{exch}}\,\vphantom{K}_{v,c,\mathbf q}^{v^{\prime},c^{\prime},\mathbf q^{\prime}}\right)\Phi_{\mu,v^{\prime},c^{\prime},\mathbf q^{\prime}}\\
    = E_{\mu}\Phi_{\mu,v,c,\mathbf q}.
    \label{eq:BSE}
\end{multline}
\end{shaded}
\eqref{eq:BSE} is an effective \textit{Bethe--Salpeter equation} for optically active electron--hole pairs in a static and Tamm--Dancoff approximation. The latter occurs naturally within our equations-of-motion approach, since we already neglected those exchange electron--hole Hamiltonians containing operator products such as $\vdag{}\vdag{}\cndag{}\cndag{}$ or $\cdag{}\cdag{}\vndag{}\vndag{}$, which are responsible for a coupling between $\langle \vdag{\mathbf q}\cndag{\mathbf q}\rangle$ and $\langle \cdag{\mathbf q}\vndag{\mathbf q}\rangle$, i.e., the anti-resonant contributions. Resonant--anti-resonant coupling due to direct electron--hole interaction does not occur in Hartree--Fock approximation. Note that we also neglected any off-diagonal self-energy contributions in the first line in \eqref{eq:BSE}, as they are expected to introduce only minor corrections \cite{rohlfing2000electron,hybertsen1986electron}.

In \eqref{eq:BSE}, $E_{c/v,\mathbf q}$ are the unrenormalized electron/hole single-particle dispersions dependent on the quasi-momenta $\mathbf q$ covering the entire Brillouin zone. They are renormalized via the Hartree self-energy $\Sigma_{c,v,\mathbf q}^{\text{H}}$:
\begin{align}
    \Sigma_{c,v,\mathbf q}^{\text{H}} 
    = \sum_{\mathbf q^{\prime},\mathbf G,\mathbf G^{\prime},v^{\prime}}\overline V_{\mathbf 0,\mathbf G,\mathbf G^{\prime}} \overline \Upsilon_{\mathbf q^{\prime}+\mathbf G,\mathbf q^{\prime}}^{v^{\prime},v^{\prime}}\left( \overline \Upsilon_{\mathbf q-\mathbf G^{\prime},\mathbf q}^{c,c} - \overline \Upsilon_{\mathbf q -\mathbf G^{\prime},\mathbf q}^{v,v} \right),
\end{align}
which vanishes for $\mathbf G=\mathbf G^{\prime}=\mathbf 0$, 
via $\Sigma_{c,v,\mathbf q}^{\text{SEX}}$, which we identify as the screened exchange self-energy \cite{hybertsen1986electron}:
\begin{multline}
    \Sigma_{c,v,\mathbf q}^{\text{SEX}} = - \sum_{\mathbf q^{\prime},\mathbf G,\mathbf G^{\prime},v^{\prime}}\overline V_{\mathbf q^{\prime},\mathbf G,\mathbf G^{\prime}}
    \left(\overline \Upsilon_{\mathbf q,\mathbf q-\mathbf q^{\prime}-\mathbf G}^{c,v^{\prime}}\overline \Upsilon_{\mathbf q-\mathbf q^{\prime}-\mathbf G^{\prime},\mathbf q}^{v^{\prime},c}\right.\\
    \left.- 
    \overline \Upsilon_{\mathbf q,\mathbf q-\mathbf q^{\prime}-\mathbf G}^{v,v^{\prime}}\overline \Upsilon_{\mathbf q-\mathbf q^{\prime}-\mathbf G^{\prime},\mathbf q}^{v^{\prime},v}
    \right),
    \label{eq:COH_exchange_self_energy}
\end{multline}
and $\Sigma_{c,v,\mathbf q}^{\text{COH}}$, which we identify as the Coulomb-hole self-energy \cite{hybertsen1986electron}:
\begin{multline}
    \Sigma_{c,v,\mathbf q}^{\text{COH}} = \frac{1}{2}\sum_{\mathbf q^{\prime},\mathbf G,\mathbf G^{\prime},\lambda}V_{\mathbf q^{\prime},\mathbf G,\mathbf G^{\prime}}^{}\left(\overline \Upsilon_{\mathbf q,\mathbf q-\mathbf q^{\prime}-\mathbf G}^{c,\lambda} \overline \Upsilon_{\mathbf q-\mathbf q^{\prime}-\mathbf G^{\prime},\mathbf q}^{\lambda,c}\right.\\
    \left. - \overline \Upsilon_{\mathbf q,\mathbf q-\mathbf q^{\prime}-\mathbf G}^{v,\lambda} \overline \Upsilon_{\mathbf q-\mathbf q^{\prime}-\mathbf G^{\prime},\mathbf q}^{\lambda,v}\right)
    .
    \label{eq:COH_direct_self_energy}
\end{multline}
In contrast to the screened exchange self-energy in \eqref{eq:COH_exchange_self_energy}, which only includes a sum over occupied states $v^{\prime}$, the Coulomb-hole self-energy in \eqref{eq:COH_direct_self_energy}, which originates from the single-particle Coulomb Hamiltonian in \eqref{eq:CoulombHamiltonianSelfEnergy}, includes a sum over all occupied as well as unoccupied states $\lambda$. 
The electron--hole interaction is governed by the direct electron--hole interaction kernel $K_{\text{eh}}^{\text{dir}}\,\vphantom{K}_{v,c,\mathbf q}^{v^{\prime},c^{\prime},\mathbf q^{\prime}}$:
\begin{align}
    K_{\text{eh}}^{\text{dir}}\,\vphantom{K}_{v,c,\mathbf q}^{v^{\prime},c^{\prime},\mathbf q^{\prime}} = - \sum_{\mathbf G,\mathbf G^{\prime}}\overline V_{\mathbf q - \mathbf q^{\prime},\mathbf G,\mathbf G^{\prime}} \overline \Upsilon_{\mathbf q,\mathbf q^{\prime}-\mathbf G}^{c,c^{\prime}}\overline \Upsilon_{\mathbf q^{\prime}-\mathbf G^{\prime},\mathbf q}^{v^{\prime},v},
    \label{eq:direct_eh_kernel}
\end{align}
which acts attractive, 
and the exchange electron--hole interaction kernel $K_{\text{eh}}^{\text{exch}}\,\vphantom{K}_{v,c,\mathbf q}^{v^{\prime},c^{\prime},\mathbf q^{\prime}}$:
\begin{align}
    K_{\text{eh}}^{\text{exch}}\,\vphantom{K}_{v,c,\mathbf q}^{v^{\prime},c^{\prime},\mathbf q^{\prime}} = \sum_{\mathbf G,\mathbf G^{\prime}}\overline V_{\mathbf 0,\mathbf G,\mathbf G^{\prime}} \overline \Upsilon_{\mathbf q,\mathbf q-\mathbf G}^{c,v}\overline \Upsilon_{\mathbf q^{\prime}-\mathbf G^{\prime},\mathbf q^{\prime}}^{v^{\prime},c^{\prime}},
    \label{eq:exchange_eh_kernel}
\end{align}
which usually acts repulsive. 
Here, we use the following abbreviation:
\begin{align}
    \overline V_{\mathbf q,\mathbf G,\mathbf G^{\prime}} = \frac{1}{2}\left( V_{\mathbf q,\mathbf G,\mathbf G^{\prime}} + V_{-\mathbf q,-\mathbf G^{\prime},-\mathbf G} \right).
    \label{eq:coulomb_potential_symmetric}
\end{align}
Note that all interaction mechanisms, direct and exchange, are screened due to the background of inactive states, cf.~Sec.~\ref{sec:microscopic_screening}, since we only include a few active states, which are optically addressed, explicitly in the Bethe--Salpeter equation in \eqref{eq:BSE}. Also note the appearance of the Hartree self-energy $\Sigma_{v,c,\mathbf q}^{\text{H}}$ in \eqref{eq:BSE}, since we expand the field operator in \eqref{eq:field_operator} using Bloch functions $\psi_{\lambda,\mathbf k}^s(\mathbf r)$ solving the electron--ion problem only. If we expand the field operator in Kohn--Sham orbitals $\psi_{\lambda,\mathbf k}^s(\mathbf r)\rightarrow \psi_{\lambda,\mathbf k}^{\text{KS},s}(\mathbf r)$ solving the Kohn--Sham equation, the Hartree self-energy $\Sigma_{v,c,\mathbf q}^{\text{H}}$ disappears, since it is already included in the Kohn--Sham dispersions $E_{\lambda,\mathbf k}^{s}\rightarrow E_{\lambda,\mathbf k}^{\text{KS},s}$. Instead, an exchange--correlation self-energy $\Sigma_{v,c,\mathbf q}^{\text{xc}}$ occurs as a compensation.

If we reduce \eqref{eq:BSE} to a two-band model, expand around the extrema of the Brillouin zone, where an effective-mass approximation of the band structure is applicable, perform a low-wavenumber approximation in the Bloch form factors, which removes the coupling between different valence bands ($v\leftrightarrow v^{\prime}$) and conduction bands ($c\leftrightarrow c^{\prime}$), disregard scattering terms involving a large valley-momentum transfer (Dexter interaction) and exchange contributions coupling distinct excitonic configurations (Förster interaction), 
the Bethe--Salpeter equation can be reduced to a Schr\"odinger-like equation \cite{sham1966many}, the \textit{Wannier equation} \cite{wannier1937structure} for $\Phi_{\mu,v,c,\mathbf q}\rightarrow \ExWFtwo{\mu,\mathbf q}{\xi,\xi^{\prime},s,s^{\prime}}$:
\begin{shaded}
\begin{multline}
    \left(\tilde E_{\text{gap}}^{\xi,\xi^{\prime},s,s^{\prime}} + \frac{\hbar^2\mathbf q^2}{2m_{\text{r},s,s^{\prime}}^{\xi,\xi^{\prime}}}\right)\ExWFtwo{\mu,\mathbf q}{\xi,\xi^{\prime},s,s^{\prime}}
    - \sum_{\mathbf q^{\prime}} K_{\text{eh}}^{\text{dir}}\,\vphantom{K}_{\mathbf q^{\prime}}^{\xi,\xi^{\prime},s,s^{\prime}}  \ExWFtwo{\mu,\mathbf q-\mathbf q^{\prime}}{\xi,\xi^{\prime},s,s^{\prime}} \\
    = E_{\mu}^{\xi,\xi^{\prime},s,s^{\prime}}\ExWFtwo{\mu,\mathbf q}{\xi,\xi^{\prime},s,s^{\prime}}.
    \label{eq:WannierEquation}
\end{multline}
\end{shaded}
Here, $\tilde E_{\text{gap}}^{\xi,\xi^{\prime},s,s^{\prime}}$ is the renormalized band gap:
\begin{multline}
    \tilde E_{\text{gap}}^{\xi,\xi^{\prime},s,s^{\prime}}\\
    = 
    E_{c,\mathbf 0}^{\xi^{\prime},s^{\prime}} - E_{v,\mathbf 0}^{\xi,s} + \Sigma_{c,v,\mathbf 0}^{\text{H},\xi,\xi^{\prime},s,s^{\prime}} + \Sigma_{c,v,\mathbf 0}^{\text{SEX},\xi,\xi^{\prime},s,s^{\prime}} + \Sigma_{c,v,\mathbf 0}^{\text{COH},\xi,\xi^{\prime},s,s^{\prime}},
    \label{eq:renormalized_band_gap}
\end{multline}
where $E_{c/v,\mathbf 0}^{\xi,s}$ is the unrenormalized conduction/valence band edge with respect to valley $\xi$ and spin $s$, $\Sigma_{c,v,\mathbf 0}^{\text{H},\xi,\xi^{\prime},s,s^{\prime}}$ is the (reduced) Hartree self-energy:
\begin{multline}
    \Sigma_{c,v,\mathbf 0}^{\text{H},\xi,\xi^{\prime},s,s^{\prime}} \\
    = \sum_{\substack{\mathbf G,\mathbf G^{\prime},\\\xi^{\prime\prime},s^{\prime\prime}}}\overline V_{\mathbf 0,\mathbf G,\mathbf G^{\prime}} \overline \Upsilon_{\mathbf G,\mathbf 0}^{v,v,\xi^{\prime\prime},\xi^{\prime\prime},s^{\prime\prime}}\left( \overline \Upsilon_{\mathbf 0,\mathbf G^{\prime}}^{c,c,\xi^{\prime},\xi^{\prime},s^{\prime}} - \overline \Upsilon_{ \mathbf 0, \mathbf G^{\prime}}^{v,v,\xi,\xi,s} \right),
    \label{eq:hartree_self_energy_reduced}
\end{multline}
$\Sigma_{c,v,\mathbf 0}^{\text{SEX},\xi,\xi^{\prime},s,s^{\prime}}$ is the (reduced) screened exchange self-energy:
\begin{multline}
    \Sigma_{c,v,\mathbf 0}^{\text{SEX},\xi,\xi^{\prime},s,s^{\prime}}\\
    = - \sum_{\mathbf q,\mathbf G,\mathbf G^{\prime},\xi^{\prime\prime}}\overline V_{\mathbf q+\mathbf K^{\xi^{\prime\prime}},\mathbf G,\mathbf G^{\prime}}^{} \left(\overline \Upsilon_{\mathbf 0,-\mathbf q-\mathbf G}^{c,v,\xi^{\prime},\xi^{\prime}-\xi^{\prime\prime},s^{\prime}} \overline \Upsilon_{-\mathbf q-\mathbf G^{\prime},\mathbf 0}^{v,c,\xi^{\prime}-\xi^{\prime\prime},\xi^{\prime},s^{\prime}}\right.\\
    \left. - \overline \Upsilon_{\mathbf 0,-\mathbf q-\mathbf G}^{v,v,\xi,\xi-\xi^{\prime\prime},s} \overline \Upsilon_{-\mathbf q-\mathbf G^{\prime},\mathbf 0}^{v,v,\xi-\xi^{\prime\prime},\xi,s} \right),
    \label{eq:screened_exchange_self_energy_reduced}
\end{multline}
and $\Sigma_{c,v,\mathbf 0}^{\text{COH},\xi,\xi^{\prime},s,s^{\prime}}$ is the (reduced) Coulomb-hole self-energy:
\begin{multline}
    \Sigma_{c,v,\mathbf 0}^{\text{COH},\xi,\xi^{\prime},s,s^{\prime}} \\
    = \frac{1}{2}\sum_{\substack{\mathbf q,\mathbf G,\mathbf G^{\prime},\xi^{\prime\prime},\\\lambda=c,v}}V_{\mathbf q+\mathbf K^{\xi^{\prime\prime}},\mathbf G,\mathbf G^{\prime}}^{} \left(\overline \Upsilon_{\mathbf 0,-\mathbf q-\mathbf G}^{c,\lambda,\xi^{\prime},\xi^{\prime}-\xi^{\prime\prime},s^{\prime}} \overline \Upsilon_{-\mathbf q-\mathbf G^{\prime},\mathbf 0}^{\lambda,c,\xi^{\prime}-\xi^{\prime\prime},\xi^{\prime},s^{\prime}}\right.\\
    \left. - \overline \Upsilon_{\mathbf 0,-\mathbf q-\mathbf G}^{v,\lambda,\xi,\xi-\xi^{\prime\prime},s} \overline \Upsilon_{-\mathbf q-\mathbf G^{\prime},\mathbf 0}^{\lambda,v,\xi-\xi^{\prime\prime},\xi,s} \right)
    .
    \label{eq:coulomb_hole_self_energy_reduced}
\end{multline}
In \eqref{eq:WannierEquation}, $\mathbf q$ is the relative momentum, $m_{\text{r},s,s^{\prime}}^{\xi,\xi^{\prime}} = \left( \frac{1}{m_h^{\xi,s}} + \frac{1}{m_{e}^{\xi^{\prime},s^{\prime}}} \right)^{-1}$ is the reduced mass with effective masses of the hole/electron $m_{h/e}^{\xi,s}$ and $\ExWFtwo{\mu,\mathbf q}{\xi,\xi^{\prime},s,s^{\prime}}$ is the excitonic wave function with excitonic quantum number $\mu$, where the first (second) valley and spin index always belongs to the hole (electron), and $E_{\mu}^{\xi,\xi^{\prime},s,s^{\prime}}$ is the corresponding excitonic energy. The excitonic wave functions are normalized according to:
\begin{align}
    \sum_{\mathbf q}\ExWFstartwo{\mu,\mathbf q}{\xi,\xi^{\prime},s,s^{\prime}}\ExWFtwo{\nu,\mathbf q}{\xi,\xi^{\prime},s,s^{\prime}} = \delta_{\mu,\nu},
    \label{eq:normalization_condition}
\end{align}
and obey the completeness relation:
\begin{align}
    \sum_{\mu}\ExWFstartwo{\mu,\mathbf q}{\xi,\xi^{\prime},s,s^{\prime}}\ExWFtwo{\mu,\mathbf q^{\prime}}{\xi,\xi^{\prime},s,s^{\prime}} = \delta_{\mathbf q,\mathbf q^{\prime}}.
    \label{eq:completeness_relation}
\end{align}

In \eqref{eq:WannierEquation}, the (reduced) direct electron--hole interaction kernel reads:
\begin{align}
    K_{\text{eh}}^{\text{dir}}\,\vphantom{K}_{\mathbf q^{\prime}}^{\xi,\xi^{\prime},s,s^{\prime}} = \sum_{\mathbf G,\mathbf G^{\prime}}\overline V_{\mathbf q^{\prime},\mathbf G,\mathbf G^{\prime}}
    \overline \Upsilon_{\mathbf 0,-\mathbf q^{\prime}-\mathbf G}^{c,c,\xi^{\prime},\xi^{\prime},s^{\prime}}
    \overline \Upsilon_{-\mathbf q^{\prime}-\mathbf G^{\prime},\mathbf 0}^{v,v,\xi,\xi,s}.
    \label{eq:electron_hole_kernel_direct_reduced}
\end{align}
By neglecting all local fields and applying a Taylor expansion in the form factors with respect to the momentum transfer $\mathbf q^{\prime}$ in zeroth order, cf.~\eqref{eq:formfactor_zeroth}, the electron--hole kernel from \eqref{eq:electron_hole_kernel_direct_reduced} can be written as:
\begin{align}
    K_{\text{eh}}^{\text{dir}}\,\vphantom{K}_{\mathbf q^{\prime}}^{\xi,\xi^{\prime},s,s^{\prime}} \approx  V_{\mathbf q^{\prime}},
    \label{eq:electron_hole_kernel_direct_reduced_taylor_expanded}
\end{align}
where $V_{\mathbf q} \equiv \overline V_{\mathbf q,\mathbf 0,\mathbf 0}$, cf.~\eqref{eq:coulomb_potential_symmetric}. 

Note that, while we provided the Bethe--Salpeter equation only for optically bright excitons in \eqref{eq:BSE}---which can be straightforwardly extended to account for all optically dark excitons as well---the Wannier equation in \eqref{eq:WannierEquation} explicitly takes into account valley-momentum-bright intravalley excitons ($\xi=\xi^{\prime}$) and valley-momentum-dark intervalley excitons ($\xi\neq\xi^{\prime}$), as well as spin-bright ($s=s^{\prime}$) and spin-dark ($s\neq s^{\prime}$) excitons.

\begin{figure}[h!]
    \centering
    \subfigure[]{\includegraphics[width=0.277\linewidth]{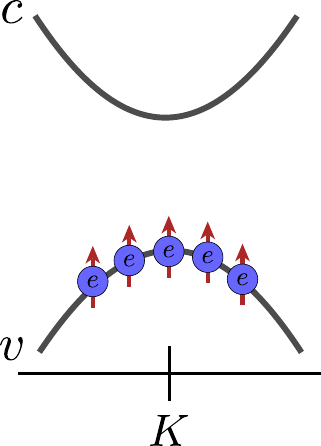}}
    \subfigure[]{\includegraphics[width=0.346\linewidth]{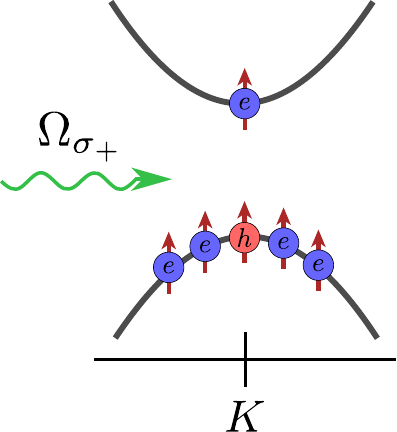}}
    \subfigure[]{\includegraphics[width=0.334\linewidth]{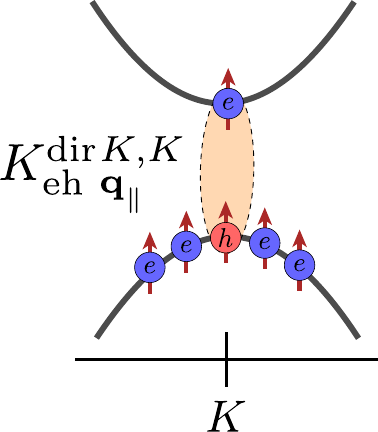}}
    \caption{Fully occupied valence band (a), optical excitation with Rabi frequency $\Omega_{\sigma_+}$ in $\sigma_+$ polarization creates an electron--hole pair at the $K$ valley (b) \cite{article:THEORY_spin_valley_selective_excitation_Yao2012}, which binds via electron--hole Coulomb attraction $K_{\text{eh}}^{\text{dir}}\vphantom{K}_{\qpar}^{K,K}$, cf.~\eqref{eq:electron_hole_kernel_direct_reduced}, forming an exciton at zero center-of-mass momentum $\Pndagtwo{\mu,\mathbf 0}{K,K,\uparrow,\uparrow}$ (c).}
    \label{fig:exciton}
\end{figure}

In the few-band effective-mass excitonic picture, electron--hole (interband) transitions are transformed into excitonic transitions as follows \cite{katsch2018theory}:
\begin{shaded}
\begin{align}
    \Pndagtwo{\mu,\mathbf Q}{\xi,\xi^{\prime},s,s^{\prime}} = \sum_{\mathbf q}\ExWFstartwo{\mu,\mathbf q}{\xi,\xi^{\prime},s,s^{\prime}} \vdagtwo{\mathbf q+\beta_{\xi,\xi^{\prime}}^{s,s^{\prime}}\mathbf Q}{\xi,s}  \cndagtwo{\mathbf q-\alpha_{\xi,\xi^{\prime}}^{s,s^{\prime}}\mathbf Q}{\xi^{\prime},s^{\prime}},
    \label{eq:exciton_expansion}
\end{align}
\end{shaded}
\noindent
where $\Pndagtwo{\mu,\mathbf Q}{\xi,\xi^{\prime},s,s^{\prime}}$ ($\Pdagtwo{\mu,\mathbf Q}{\xi,\xi^{\prime},s,s^{\prime}}$) annihilates (creates) an exciton at excitonic state $\mu$ and center-of-mass momentum $\mathbf Q$ with spin-valley configuration $\xi,\xi^{\prime},s,s^{\prime}$, cf.~Fig.~\ref{fig:exciton}. The first (second) index belongs to the hole (electron). The effective-mass ratios read:
\begin{align}
    \alpha_{\xi,\xi^{\prime}}^{s,s^{\prime}} = \frac{m_{e}^{\xi^{\prime},s^{\prime}}}{ m_h^{\xi,s} + m_{e}^{\xi^{\prime},s^{\prime}} },\quad 
    \beta_{\xi,\xi^{\prime}}^{s,s^{\prime}} = \frac{m_h^{\xi,s}}{ m_h^{\xi,s} + m_{e}^{\xi^{\prime},s^{\prime}} }.
    \label{eq:effective_mass_ratios}
\end{align}
In \eqref{eq:exciton_expansion}, we explicitly take into account momentum-dark electron--hole pairs with $\mathbf Q\neq \mathbf 0$. 

By coupling the excitons to a classical transverse Maxwell field \cite{knorr1996theory,book:graphene_carbon_nanotubes_malic2013,jahnke1997linear} in a thin-film and rotating-wave approximation, we can derive an expression for the linear absorption $\alpha(\omega)$:
\begin{shaded}
    \begin{align}
        \alpha(\omega) = \frac{\omega}{c_0}\Im \mleft( \frac{\mathcal A}{\epsilon_0}\sum_{\mu}\frac{\big|\mathbf d_{\mathbf 0}^{c,v,K,K,\uparrow}\ExWFstartwo{\mu}{K,K,\uparrow,\uparrow}(\rpar=\mathbf 0)\big|^2}{E_{\mu}^{K,K,\uparrow,\uparrow} - \hbar\omega - \mathrm i\hbar\gamma_{\mu}^{K,K,\uparrow,\uparrow}}\mright),
        \label{eq:AbsorptionLinearOptics}
    \end{align}
\end{shaded}
\noindent
for $\sigma_+$-polarized light without loss of generality, which corresponds to the Elliott formula \cite{elliott1957intensity}. In \eqref{eq:AbsorptionLinearOptics}, the (half) homogeneous linewidth reads: $\hbar\gamma_{\mu}^{\xi,\xi^{\prime},s,s^{\prime}} = \hbar\gamma_{\text{rad},\mu}^{\xi,\xi^{\prime},s,s^{\prime}} + \hbar\gamma_{\text{nrad},\mu}^{\xi,\xi^{\prime},s,s^{\prime}}$, where $\hbar\gamma_{\text{rad},\mu}^{\xi,\xi^{\prime},s,s^{\prime}}$ is the radiative (half) linewidth:
\begin{multline}
    \hbar \gamma_{\text{rad},\mu}^{\xi,\xi^{\prime},s,s^{\prime}}\\
    = \frac{E_{\mu}^{\xi,\xi,s,s}\mu_0c_0\mathcal A}{\hbar n_{\text{ref}}}
    \big|\mathbf d_{\mathbf 0}^{cv,\xi,\xi,s}\ExWFstartwo{\mu}{\xi,\xi,s,s}(\rpar=\mathbf 0)\big|^2\delta_{\xi,\xi^{\prime}}^{s,s^{\prime}},
    \label{eq:radiative_damping}
\end{multline}
where $n_{\text{ref}}$ is the mean refractive index of the dielectric environment, and $\hbar\gamma_{\text{nrad},\mu}^{\xi,\xi^{\prime},s,s^{\prime}}$ is the nonradiative (half) linewidth due to, e.g., exciton--phonon interaction \cite{selig2016excitonic,lengers2020theory} or uniformly distributed impurities \cite{ma2014charge,jena2007enhancement}. In principle, inhomogeneous contributions to the linewidth can also be included \cite{glutsch1994theory,thranhardt2003interplay,raja2019dielectric}, but entail ensemble averages, which we do not consider here. 
The oscillator strength of $\alpha(\omega)$ is determined by the transition dipole moment $\mathbf d_{\mathbf 0}^{c,v,K,K,\uparrow}$ in low-wavenumber approximation and by the excitonic wave function $\ExWFstartwo{\mu}{K,K,\uparrow,\uparrow}(\rpar=\mathbf 0)$ denoting the probability of finding an electron and a hole at the same position in real space. The transition dipole moment can be analytically calculated via the eigenfunctions of the few-band \textit{k}$\cdot$\textit{p} Hamiltonian from Ref.~\cite{kormanyos2015k}:
\begin{align}
    \mathbf d_{\mathbf 0}^{c v,\xi,\xi,s} = (\delta_{\xi,K}-\delta_{\xi,K^{\prime}})\frac{\mathrm ie\gamma_{\text{\textit{k}$\cdot$\textit{p}}}\sqrt{2}}{E_{c,\mathbf 0}^{\xi,s}-E_{v,\mathbf 0}^{\xi,s}}\mathbf e_{\delta_{\xi,K^{\prime}}-\delta_{\xi,K}},
    \label{eq:dipole_moment_evaluated}
\end{align}
with momentum matrix element $\gamma_{\text{\textit{k}$\cdot$\textit{p}}}$ and Jones vectors $\mathbf e_{\pm}=\frac{1}{\sqrt{2}}\begin{pmatrix}
    1 & \pm\mathrm i
\end{pmatrix}^{\top}$.

\begin{figure}[h!]
    \centering
    \includegraphics[width=1\linewidth]{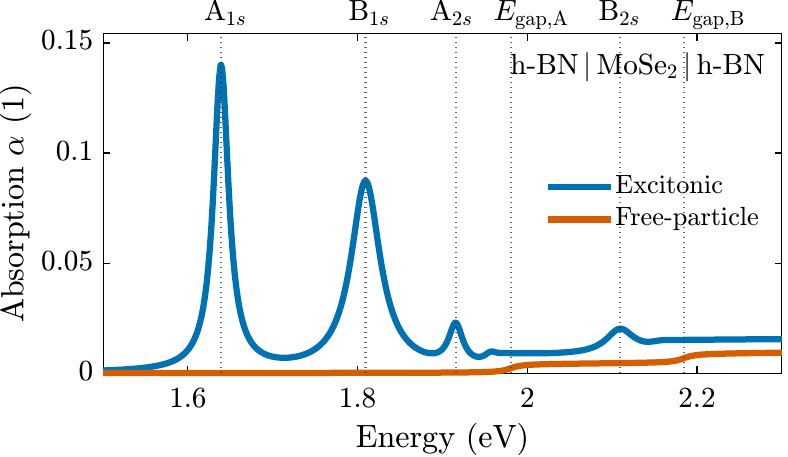}
    \caption{Linear absorption $\alpha(\omega)$ from \eqref{eq:AbsorptionLinearOptics} of a h-BN-encapsulated monolayer MoSe$_2$ by taking the A and B series into account. Parameters are shown in Tab.~\ref{tab:linear_absorption_parameters}.}
    \label{fig:linear_absorption}
\end{figure}

In Fig.~\ref{fig:linear_absorption}, we depict the linear absorption spectrum $\alpha(\omega)$, cf.~\eqref{eq:AbsorptionLinearOptics}, of a h-BN-encapsulated MoSe$_2$ monolayer by taking the A and B series into account. 
The blue solid line in Fig.~\ref{fig:linear_absorption} depicts the full calculations using \eqref{eq:WannierEquation}, the ``excitonic'' limit, and the red solid line depicts calculations without electron--hole Coulomb attraction, i.e., \eqref{eq:WannierEquation} with $K_{\text{eh}}^{\text{dir}}\,\vphantom{K}_{\qparp}^{\xi,\xi^{\prime},s,s^{\prime}}\approx V_{\qparp} = 0$, the ``free-particle'' limit.

There are several features to discuss.

First, by comparing the free-particle limit to the excitonic case, additional resonances below the free-particle band gap $E_{\text{gap}}$ appear, the \textit{excitonic resonances}. They are labeled as $1s,2s\dots$, in analogy to the hydrogen atom, since they denote bound eigenstates of the problem of electrons and holes interacting via an attractive Coulomb potential.

Second, at and above the free-particle band gap, the excitonic absorption is larger than the free-particle absorption. This is called \textit{Sommerfeld enhancement} \cite{sommerfeld1931beugung,zimmermann1995excitonic,winkler1995excitons} and stems from the fact that the unbound eigenstates of the interacting electron--hole problem above the band gap do not fully behave as free plane waves but are still perturbed by the Coulomb interaction and form so-called \textit{Coulomb waves} \cite{elliott1957intensity,article:THEORY_BoundUnboundAbsorption_Normfactor_Shinada1966,gozem2015photoelectron}. It can be quantified as:
\begin{align}
    S(\omega) = \frac{\alpha_{\text{excitonic}}(\omega)}{\alpha_{\text{free-particle}}(\omega)}.
    \label{eq:Sommerfeld_enhancement_factor}
\end{align}
For the conditions in Fig.~\ref{fig:linear_absorption}, we obtain a Sommerfeld enhancement factor of \mbox{$S\mleft(\omega>\frac{E_{\text{gap,B}}}{\hbar}\mright)\approx 1.6$}, similar to conventional quantum wells \cite{winkler1995excitons}.

Third, the spectrum depicts two distinct excitonic series, the \textit{A series} and the $\textit{B series}$, which are energetically different. The A series corresponds to electron--hole pairs at the $K$ valley, where both electrons and holes are in the spin-up state, whereas the B series corresponds to electron--hole pairs at the $K$ valley with spin-down states. The energy difference between, e.g., the A$_{1s}$ and the B$_{1s}$ exciton is caused by the conduction and valence band splitting due to spin--orbit interaction via intrinsic in-plane electric fields \cite{book:Spin_orbit_coupling_Winkler2003,kormanyos2015k,article:THEORY_Bychkov_Rashba_Coupling_Kormanyos2014,kosmider2013large,article:THEORY_spin_valley_selective_excitation_Yao2012,marsili2021spinorial} as well as differences in their respective binding energies due to different band curvatures, i.e., different effective masses. In case of $\sigma_-$ excitation, the spin and valley labeling swaps.

Note that we assume a slightly idealized spin structure with pure spin-up and spin-down states. \textit{Ab initio} calculations show \cite{deilmann2020ab,junior2022first} that even directly at the band extrema, the spins are slightly mixed due to spin--orbit interaction via intrinsic out-of-plane ionic electric fields, and electron--electron fields can further enrich the picture \cite{mittenzwey2026many,mittenzwey2025ultrafast,gindikin2018spin,breit1929effect}.

Moreover, one has to be careful to compare the exact energetic ordering of the excitonic peaks in Fig.~\ref{fig:linear_absorption}, e.g., the relative position of the A$_{2s}$ and B$_{1s}$ excitons, to real-world measurements \cite{liu2021exciton}, since unintentional doping, impurities and substrate effects beyond a h-BN encapsulation may significantly alter the dielectric function, which the binding energies of higher excitonic states sensitively depend upon, in a nontrivial way.

Also note that our calculated spectrum in Fig.~\ref{fig:linear_absorption} does not include other excitonic contributions due to, e.g., higher-lying C excitons, which are delocalized along the $\Gamma$-$K$ direction in the Brillouin zone \cite{qiu2016screening,erben2022optical} and significantly affect the absorption spectrum at higher energies. These cannot be properly described in an effective-mass approximation and have to be described via the full Bethe--Salpeter equation in \eqref{eq:BSE}.

\section{Coulomb Scattering Processes}
Since the Wannier equation in \eqref{eq:WannierEquation} takes---next to the Coulomb-induced band gap renormalization---into account only one specific Coulomb interaction process, i.e., electron--hole Coulomb attraction involving momentum transfers in the vicinity of a band extrema (valley), which are small compared to the distance between different valleys, 
\eqref{eq:WannierEquation} lacks the description of other exciton scattering processes relevant in time-dependent optical experiments. Hence, its eigenstates rather provide a suitable basis in materials with strong Coulomb interaction as a starting point. In this section, we describe exciton scattering processes at all center-of-mass momenta $\Qpar$ potentially reaching well outside the light cone formulated in a basis constructed from the eigenstates of the Wannier equation, cf.~\eqref{eq:WannierEquation}.

In the following, we disregard the subscript ``$\parallel$'' for notational simplicity. Even though we discuss two-dimensional materials, the concepts are equally valid in three dimensions.

Recall the many-body Coulomb Hamiltonian from \eqref{eq:Coulomb_Hamiltonian_SecondQuantized_ValleyExpanded} in a valley-expanded description:
\begin{multline}
\label{eq:Coulomb_Hamiltonian_SecondQuantized_ValleyExpanded_2}
    \hat H_{\text{Coul}}\\ = \frac{1}{2}\sum_{\substack{\lambda_1,\dots\lambda_4,\\\mathbf k,\mathbf k^{\prime},\mathbf q,\\\mathbf G,\mathbf G^{\prime},\\\xi,\xi^{\prime},\xi^{\prime\prime},s,s^{\prime}}} V^{}_{\mathbf q+ \mathbf K^{\xi^{\prime\prime}},\mathbf G,\mathbf G^{\prime}}
    \overline \Upsilon_{\mathbf k+\mathbf q+\mathbf G,\mathbf k}^{\lambda_1,\lambda_4,\xi+\xi^{\prime\prime},\xi,s}\overline \Upsilon_{\mathbf k^{\prime}-\mathbf q-\mathbf G^{\prime},\mathbf k^{\prime}}^{\lambda_2,\lambda_3,\xi^{\prime}-\xi^{\prime\prime},\xi^{\prime},s^{\prime}}\\
    \times\adagtwo{\lambda_1,\mathbf k+\mathbf q}{\xi+\xi^{\prime\prime},s}\adagtwo{\lambda_2,\mathbf k^{\prime}-\mathbf q}{\xi^{\prime}-\xi^{\prime\prime},s^{\prime}}\andagtwo{\lambda_3,\mathbf k^{\prime}}{\xi^{\prime},s^{\prime}}\andagtwo{\lambda_4,\mathbf k}{\xi,s}.
\end{multline}
Here, we already restricted the out-of-plane confinement to the respective ground state with $V^{1,1,1,1}_{\mathbf q,\mathbf G,\mathbf G^{\prime}} \equiv V^{}_{\mathbf q,\mathbf G,\mathbf G^{\prime}}$ and assumed a statically screened Coulomb potential. With \eqref{eq:Coulomb_Hamiltonian_SecondQuantized_ValleyExpanded_2}, we can calculate the dynamics of an arbitrary excitonic operator $\hat A$ via the Heisenberg equations of motion:
\begin{align}
    \mathrm i\hbar\partial_t \hat A = [\hat A,\hat H_{\text{Coul}}].
    \label{eq:heisenberg_eom_general}
\end{align}
Here, $\hat A$ can be, e.g., an excitonic transition $\Pndagtwo{\mu,\mathbf Q}{\xi,\xi^{\prime},s,s^{\prime}}$ or an excitonic occupation/intraexcitonic coherence $\hat N\vphantom{N}_{\mu,\nu,\mathbf Q}^{\xi,\xi^{\prime},\xi^{\prime\prime},\xi^{\prime\prime\prime},s,s^{\prime},s^{\prime\prime},s^{\prime\prime\prime}} = \Pdagtwo{\mu,\mathbf Q}{\xi,\xi^{\prime},s,s^{\prime}}\Pndagtwo{\nu,\mathbf Q}{\xi^{\prime\prime},\xi^{\prime\prime\prime},s^{\prime\prime},s^{\prime\prime\prime}}$. 
The operator products appearing on the right-hand side in \eqref{eq:heisenberg_eom_general} can be evaluated by assuming bosonic commutation relations:
\begin{align}
    [\Pndagtwo{\mu,\mathbf Q}{\xi,\xi^{\prime},s,s^{\prime}},\Pdagtwo{\nu,\mathbf Q^{\prime}}{\xi^{\prime\prime},\xi^{\prime\prime\prime},s^{\prime\prime},s^{\prime\prime\prime}}] = \delta_{\mu,\nu}^{\mathbf Q,\mathbf Q^{\prime}}\delta_{\xi,\xi^{\prime\prime}}^{\xi^{\prime},\xi^{\prime\prime\prime}}\delta_{s,s^{\prime\prime}}^{s^{\prime},s^{\prime\prime\prime}},
    \label{eq:excitonic_commutation_relation_bosonic}
\end{align}
which is exact as long as we stay linear in the exciton density $N$ \cite{katsch2018theory}. The corresponding expectation values and their dynamics can then be obtained via applying the cluster-expansion technique \cite{fricke1996transport,kira2006many} to \eqref{eq:heisenberg_eom_general}. However, to apply \eqref{eq:excitonic_commutation_relation_bosonic} for a given excitonic correlation of interest, we first need to obtain an excitonic description of the Hamiltonian in \eqref{eq:Coulomb_Hamiltonian_SecondQuantized_ValleyExpanded_2}. The dissection of individual exciton scattering processes by discussing the corresponding Hamiltonians is the topic of this section. 

In the following, we illustrate only direct electron--hole and exchange electron--hole processes in the \textit{density-independent} limit, i.e., we disregard all density-dependent processes such as nonlinear energy renormalizations, linewidth broadening or Auger scattering \cite{mittenzwey2026excitonic,vosco2025exciton,deckert2025coherent,steinhoff2024exciton,erkensten2022microscopic,steinhoff2021microscopic,erkensten2021dark,katsch2020exciton,katsch2019theory}. In this limit, electron--electron processes do not occur and hole--hole processes only renormalize the band gap via the second term of the screened exchange self-energy in \eqref{eq:screened_exchange_self_energy_reduced}. Hence, the processes outlined in the following are only sufficient up to a strict $\chi^{(3)}$ limit, e.g., weakly nonlinear pump--probe experiments with energy-averaged timetraces around well separated excitonic resonances, where all spectral signatures besides Pauli-blocking-induced bleaching have been averaged out.

\subsection{Direct Electron--Hole Interaction}
\label{sec:direct_interaction}
\begin{figure}[h!]
    \centering
    \subfigure[]{\includegraphics[width=0.6\linewidth]{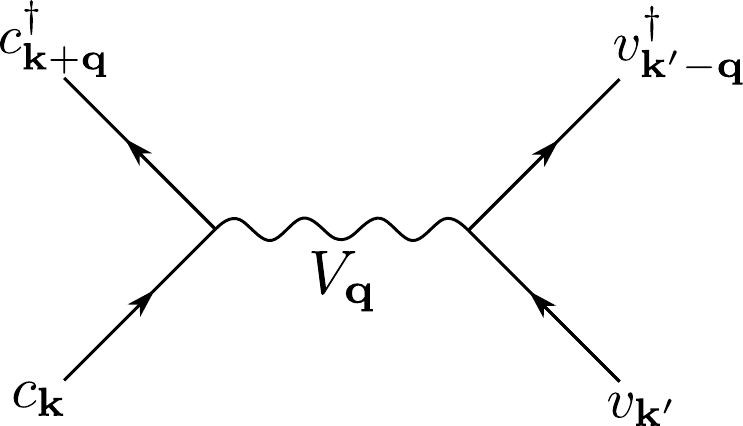}}
    \caption{Direct electron--hole Coulomb interaction.}
    \label{fig:direct_interaction}
\end{figure}

From \eqref{eq:Coulomb_Hamiltonian_SecondQuantized_ValleyExpanded_2}, we consider electron--hole interaction processes, where the respective electrons and holes remain in their bands ($\lambda_1=\lambda_4$ and $\lambda_2=\lambda_3$). These processes are usually referred to as direct Coulomb interactions. In a two-band model, they are described by:
\begin{multline}
    \hat H_{\text{Coul-eh}}^{\text{dir}} \\
    = \sum_{\substack{\mathbf k,\mathbf k^{\prime},\mathbf q,\mathbf G,\mathbf G^{\prime},\\\xi,\xi^{\prime},\xi^{\prime\prime},s,s^{\prime}}}
    \overline V^{}_{\mathbf q+ \mathbf K^{\xi^{\prime\prime}},\mathbf G,\mathbf G^{\prime}}
    \overline \Upsilon_{\mathbf k+\mathbf q+\mathbf G,\mathbf k}^{c,c,\xi+\xi^{\prime\prime},\xi,s}\overline \Upsilon_{\mathbf k^{\prime}-\mathbf q-\mathbf G^{\prime},\mathbf k^{\prime}}^{v,v,\xi^{\prime}-\xi^{\prime\prime},\xi^{\prime},s^{\prime}}\\
    \times\cdagtwo{\mathbf k+\mathbf q}{\xi+\xi^{\prime\prime},s}\vdagtwo{\mathbf k^{\prime}-\mathbf q}{\xi^{\prime}-\xi^{\prime\prime},s^{\prime}}\vndagtwo{\mathbf k^{\prime}}{\xi^{\prime},s^{\prime}}\cndagtwo{\mathbf k}{\xi,s}.
    \label{eq:CoulombHamiltonianDirectElectronHole}
\end{multline}
In Fig.~\ref{fig:direct_interaction}, we display the direct electron--hole interaction in a two-band model for arbitrary momentum transfers $\mathbf q$.

\subsubsection*{Small Momentum Transfer}
Direct electron--hole Coulomb interaction processes with small momentum transfer ($\mathbf K^{\xi^{\prime\prime}}=\mathbf 0$) 
give rise to the binding of electron--hole pairs and are diagonalized together with the free-particle Hamiltonian within the transformation into the excitonic picture, cf.~\eqref{eq:exciton_expansion}, by means of solving the Wannier equation in \eqref{eq:WannierEquation}. 
This procedure results in the free excitonic Hamiltonian:
\begin{shaded}
\begin{align}
    \hat H_{\text{X-0}} = \sum_{\mu,\mathbf Q,\xi,\xi^{\prime},s,s^{\prime}}E_{\mu,\mathbf Q}^{\xi,\xi^{\prime},s,s^{\prime}}\Poldagtwo{\mu,\mathbf Q}{\xi,\xi^{\prime},s,s^{\prime}}\Poltwo{\mu,\mathbf Q}{\xi,\xi^{\prime},s,s^{\prime}},
\end{align}
\end{shaded}
\noindent
where:
\begin{align}
    E_{\mu,\mathbf Q}^{\xi,\xi^{\prime},s,s^{\prime}} = E_{\mu}^{\xi,\xi^{\prime},s,s^{\prime}} + \frac{\hbar^2\mathbf Q^2}{2(m_{h}^{\xi,s} + m_{e}^{\xi^{\prime},s^{\prime}} )},
\end{align}
is the exciton dispersion with exciton energy $E_{\mu}^{\xi,\xi^{\prime},s,s^{\prime}}$ obtained by solving the Wannier equation in \eqref{eq:WannierEquation} and center-of-mass momentum $\mathbf Q$.

\FloatBarrier
\subsubsection*{Large Momentum Transfer}
The direct electron--hole interaction for large momentum transfer (\mbox{$\mathbf K^{\xi^{\prime\prime}}\neq\mathbf 0$}) represents the valley analog in momentum space of Coulomb-induced charge transfer in real space, i.e., Dexter interaction \cite{dexter1953theory}. It occurs between distinct valleys in the Brillouin zone and is always spin-conserving with regard to the individual electron and hole scattering.
The corresponding Hamiltonian is given by \eqref{eq:CoulombHamiltonianDirectElectronHole} with the condition \mbox{$\mathbf K^{\xi^{\prime\prime}}\neq \mathbf 0$}:
\begin{multline}
    \hat H_{\text{Dex}}^{\text{dir}}\\
    = \sum_{\substack{\mathbf k,\mathbf k^{\prime},\mathbf q,\mathbf G,\mathbf G^{\prime},\\\xi,\xi^{\prime},\xi^{\prime\prime},\\ (\xi^{\prime\prime}\neq \Gamma),s,s^{\prime}}} \overline V^{}_{\mathbf q+ \mathbf K^{\xi^{\prime\prime}},\mathbf G,\mathbf G^{\prime}}
    \overline \Upsilon_{\mathbf k+\mathbf q+\mathbf G,\mathbf k}^{c,c,\xi+\xi^{\prime\prime},\xi,s}\overline \Upsilon_{\mathbf k^{\prime}-\mathbf q-\mathbf G^{\prime},\mathbf k^{\prime}}^{v,v,\xi^{\prime}-\xi^{\prime\prime},\xi^{\prime},s^{\prime}}\\
    \times\cdagtwo{\mathbf k+\mathbf q}{\xi+\xi^{\prime\prime},s}\vdagtwo{\mathbf k^{\prime}-\mathbf q}{\xi^{\prime}-\xi^{\prime\prime},s^{\prime}}\vndagtwo{\mathbf k^{\prime}}{\xi^{\prime},s^{\prime}}\cndagtwo{\mathbf k}{\xi,s}.
    \label{eq:CoulombHamiltonianDirectElectronHoleLargeMomentumTransfer}
\end{multline}
In the excitonic picture with \eqref{eq:exciton_expansion}, the Dexter Hamiltonian reads:
\begin{shaded}
\begin{multline}
    \hat H_{\text{X-Dex}}\\
    =  \sum_{\substack{\mu,\nu,\mathbf Q,\\\xi,\xi^{\prime},\xi^{\prime\prime},\\(\xi^{\prime\prime}\neq \Gamma),s,s^{\prime}}}
    D_{\mu,\nu,\mathbf Q,\mathbf K^{\xi^{\prime\prime}}}^{
    \xi,\xi^{\prime},\xi^{\prime\prime},s,s^{\prime}}
    \Poldagtwo{\mu,\mathbf Q}{\xi,\xi^{\prime}+\xi^{\prime\prime},s,s^{\prime}}
    \Poloptwo{\nu,\mathbf Q}{\xi-\xi^{\prime\prime},\xi^{\prime},s,s^{\prime}},
    \label{eq:dexter_hamiltonian_excitonic}
\end{multline}
\end{shaded}
\noindent
with matrix element:
\begin{multline}
    D_{\mu,\nu,\mathbf Q,\mathbf K^{\xi^{\prime\prime}}}^{
    \xi,\xi^{\prime},\xi^{\prime\prime},s,s^{\prime}} \\
    = - \mathcal C_{\text{Dex},\mu,\nu,\mathbf Q}^{\text{dir},\xi,\xi^{\prime},\xi^{\prime\prime},s,s^{\prime}} V^{}_{\mathbf K^{\xi^{\prime\prime}}}\sum_{\mathbf q}\ExWFstartwo{\mu,\mathbf q}{\xi,\xi^{\prime}+\xi^{\prime\prime},s,s^{\prime}}\sum_{\mathbf q^{\prime}}\ExWFtwo{\nu,\mathbf q^{\prime}}{\xi-\xi^{\prime\prime},\xi^{\prime},s,s^{\prime}},
    \label{eq:DexterMatrixElement}
\end{multline}
where:
\begin{multline}
\mathcal C_{\text{Dex},\mu,\nu,\mathbf Q}^{\text{dir},\xi,\xi^{\prime},\xi^{\prime\prime},s,s^{\prime}}\\
    = \left(V^{}_{\mathbf K^{\xi^{\prime\prime}}}\sum_{\mathbf q}\ExWFstartwo{\mu,\mathbf q}{\xi,\xi^{\prime}+\xi^{\prime\prime},s,s^{\prime}}\sum_{\mathbf q^{\prime}}\ExWFtwo{\nu,\mathbf q^{\prime}}{\xi-\xi^{\prime\prime},\xi^{\prime},s,s^{\prime}}\right)^{-1}\\
    \times \sum_{\mathbf q,\mathbf q^{\prime},\mathbf G,\mathbf G^{\prime}} \overline V_{\mathbf q-\mathbf q^{\prime}+\mathbf Q+\mathbf K^{\xi^{\prime\prime}},\mathbf G,\mathbf G^{\prime}}
    \overline \Upsilon_{\mathbf q+\mathbf Q+\mathbf G,\mathbf q^{\prime}}^{c,c,\xi^{\prime}+\xi^{\prime\prime},\xi^{\prime},s^{\prime}}\overline \Upsilon_{\mathbf q^{\prime}-\mathbf Q-\mathbf G^{\prime},\mathbf q}^{v,v,\xi-\xi^{\prime\prime},\xi,s}\\
    \times \ExWFstartwo{\mu,\mathbf q+\beta_{\xi,\xi^{\prime}+\xi^{\prime\prime}}^{s,s^{\prime}}\mathbf Q}{\xi,\xi^{\prime}+\xi^{\prime\prime},s,s^{\prime}}\ExWFtwo{\nu,\mathbf q^{\prime}-\alpha_{\xi-\xi^{\prime\prime},\xi^{\prime}}^{s,s^{\prime}}\mathbf Q}{\xi-\xi^{\prime\prime},\xi^{\prime},s,s^{\prime}},
    \label{eq:dexter_correction_factor}
\end{multline}
carries all corrections including possible local-field effects compared to the rough estimate $V^{}_{\mathbf K^{\xi^{\prime\prime}}}$. 
We note that Dexter interaction already affects linear spectra by introducing minor energy shifts and broadening due to intervalley A--B coupling \cite{bernal2018exciton}. In particular, 
even though the Dexter matrix element acting on electrons and holes directly at the band edges for $\mathbf q=\mathbf q^{\prime}=\mathbf 0$ vanishes in $C_3$-symmetric TMDCs for $K$ valley momentum transfers $\xi^{\prime\prime}=K$ \cite{glazov_private}:
\begin{align}
    \sum_{\substack{\mathbf G}} V_{\mathbf K^{\xi^{\prime\prime}}+\mathbf G}\overline \Upsilon_{\mathbf G,\mathbf 0}^{c,c,\xi^{\prime}+\xi^{\prime\prime},s^{\prime}}\overline \Upsilon_{-\mathbf G,\mathbf 0}^{v,v,\xi-\xi^{\prime\prime},\xi,s} = 0,
\end{align}
we cannot conclude that the Dexter matrix element for \textit{excitons} in \eqref{eq:DexterMatrixElement} at zero COM momenta $\mathbf Q=\mathbf 0$ vanishes as well, since it encodes a momentum convolution in the relative momenta $\mathbf q$ and $\mathbf q^{\prime}$, which are in general nonzero even if $\mathbf Q=\mathbf 0$.

\begin{figure}[h!]
    \centering
    \subfigure[]{\includegraphics[width=0.49\linewidth]{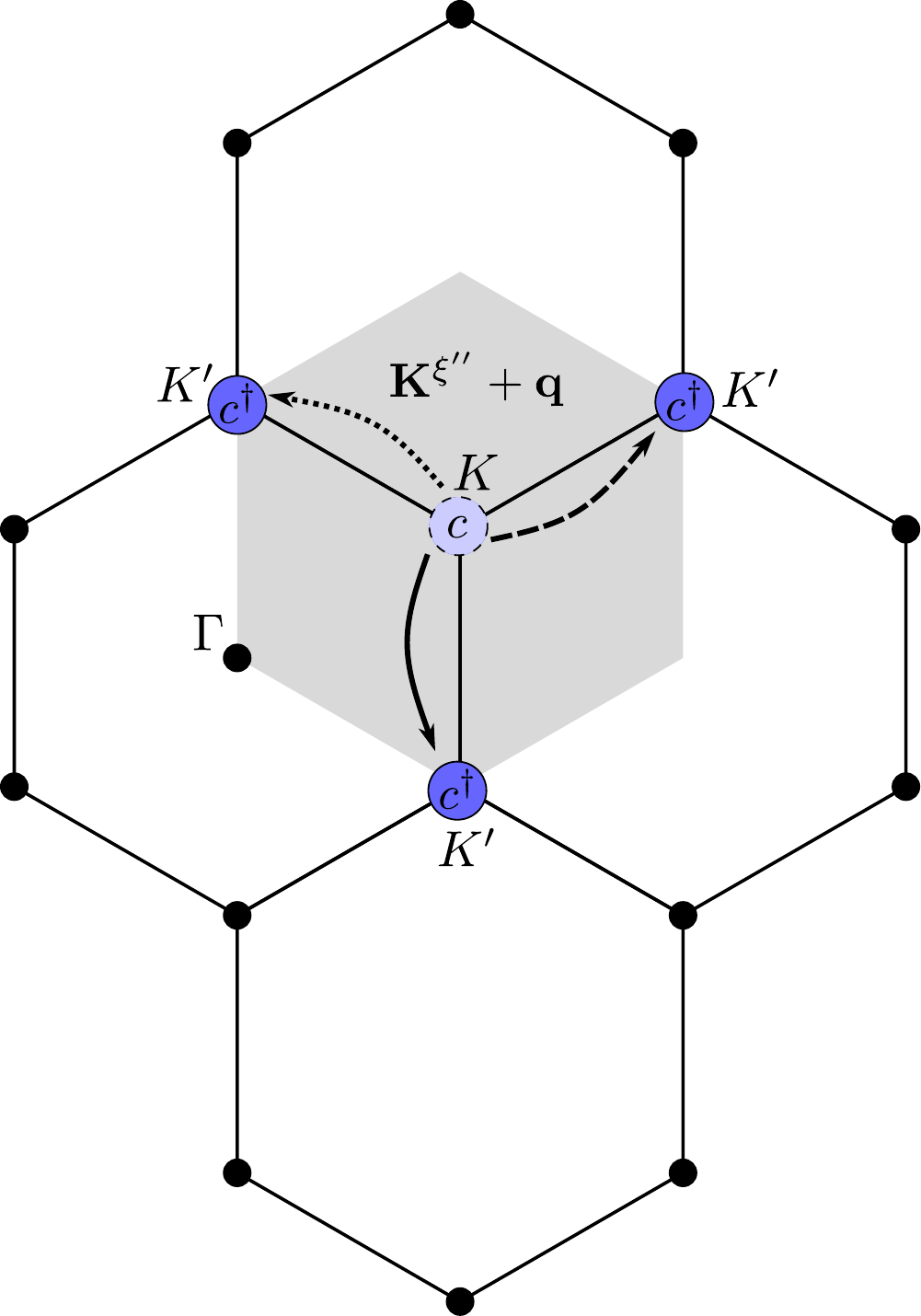}}
    \subfigure[]{\includegraphics[width=0.49\linewidth]{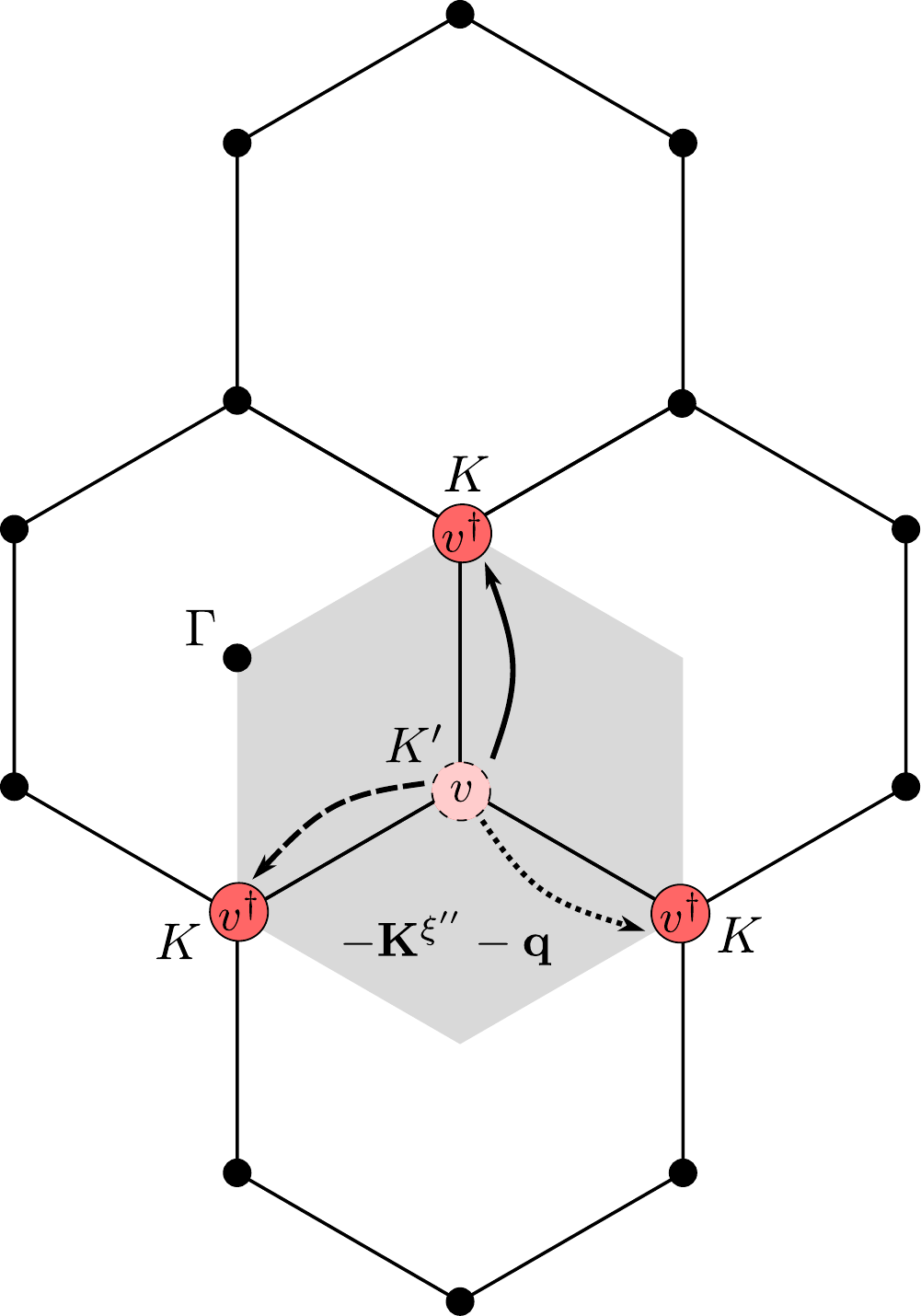}}
    \subfigure[]{\includegraphics[width=0.55\linewidth]{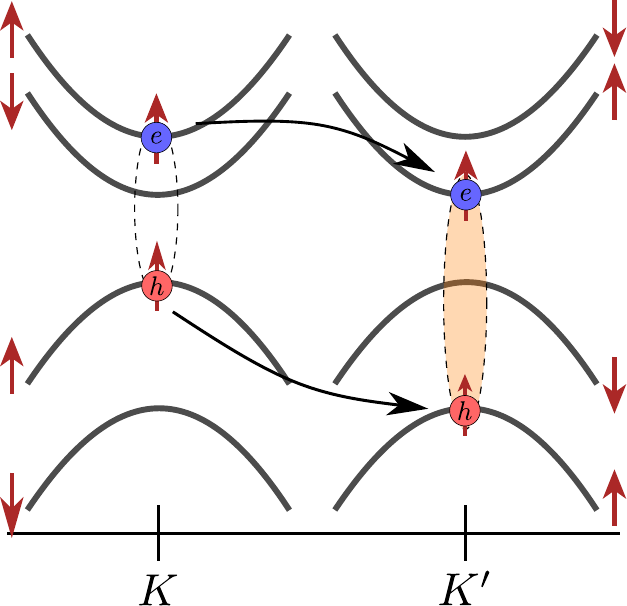}}
    \caption{Example process of intervalley interaction ($\xi^{\prime\prime}\neq \Gamma$) of intravalley excitons ($\xi=\xi^{\prime}$) involving simultaneous $K^{\prime} \rightarrow K$ and $K \rightarrow K^{\prime}$ intraband scattering (Dexter): Three equivalent processes denoted by solid, dashed and dotted arrows involving electron (a) and hole scattering (b). Grey-shaded areas denote the total set of possible momentum transfers $\mathbf K^{\xi^{\prime\prime}}+\mathbf q$ (conduction-band electron scattering) or $-\mathbf K^{\xi^{\prime\prime}}-\mathbf q$ (valence-band electron scattering). Combined process visualized in electron--hole pairs in a reduced band structure (c), cf.~\eqref{eq:dexter_hamiltonian_excitonic}. Note that the spin polarization is idealized, as spin--orbit interaction slightly mixes spins even directly at the band extrema \cite{deilmann2020ab,junior2022first}. }
    \label{fig:interint_intraexc}
\end{figure}

First, we discuss intervalley scattering of \textit{intravalley} excitons. If we restrict ourselves to the energetically lowest excitons, 
there exists one process:
\begin{align}
    K\text{-}K\leftrightarrow K^{\prime}\text{-}K^{\prime},
\end{align}
of which we depict the direction $K$-$K$$\rightarrow$$K^{\prime}$-$K^{\prime}$ in Fig.~\ref{fig:interint_intraexc} for the spin-up--spin-up configuration.
For the example Dexter process depicted in Fig.~\ref{fig:interint_intraexc}, the reduced form factor $\overline \Upsilon$, \eqref{eq:reduced_form_factor}, has to be calculated between different valleys: 
\begin{align}
\overline \Upsilon_{\mathbf k+\mathbf q+\mathbf G,\mathbf k}^{c,c,K^{\prime},K,s}\quad\text{and}\quad\overline \Upsilon_{\mathbf k^{\prime}-\mathbf q-\mathbf G^{\prime},\mathbf k^{\prime}}^{v,v,K,K^{\prime},s}.
\label{eq:formfactors_dexter_normalprocess}
\end{align}
For an exact evaluation of \eqref{eq:dexter_correction_factor}, first-principle methods are needed. 
However, in many cases, it is sufficient to disregard the local fields and approximate the form factors as follows: $\overline \Upsilon_{\mathbf k+\mathbf q,\mathbf k}^{c,c,K^{\prime},K,s} = \overline \Upsilon_{\mathbf k^{\prime}-\mathbf q,\mathbf k^{\prime}}^{v,v,K,K^{\prime},s} \approx 1$, which entails:
\begin{align}
    \mathcal C_{\text{Dex}}^{\text{dir}} \approx 1,
    \label{eq:dexter_correction_factor_estimate}
\end{align}
serving as an upper estimate. 
In Fig.~\ref{fig:interint_intraexc}(a,b) we depict the scattering of the conduction-band electron (a) and the valence-band electron (b) of the corresponding Dexter process, where the grey-shaded areas denote the total set of possible momentum transfers $\mathbf K^{\xi^{\prime\prime}}+\mathbf q$. Here, even though there are three equivalent processes, the total scattering amplitude is not multiplied by three, since each $K$ or $K^{\prime}$ valley as a final state is only covered by one third. Note that we refrain from drawing specific reciprocal lattice vectors $\mathbf G^{\prime\prime}$ and $\mathbf G^{\prime\prime\prime}$ denoting possible Umklapp processes \textit{in the narrower sense} (not to be confused with the local-field contributions $\mathbf G$ and $\mathbf G^{\prime}$), as the form factors $\overline \Upsilon$ do not depend on $\mathbf G^{\prime\prime}$ and $\mathbf G^{\prime\prime\prime}$, cf.~Sec.~\ref{sec:formfactors} for a discussion. The combined process is depicted in Fig.~\ref{fig:interint_intraexc}(c) for electron--hole pairs in a reduced band structure. Note that the hole in Fig.~\ref{fig:interint_intraexc}(c) scatters in the opposite direction compared to the valence-band electron in Fig.~\ref{fig:interint_intraexc}(b). 
The estimate in \eqref{eq:dexter_correction_factor_estimate} already yields a reasonable Dexter interaction strength, which has back-up from measurements \cite{dogadov2026diss,berghauser2018inverted,bernal2018exciton,timmer2024ultrafast}. One reason is that the interaction strength of Dexter processes is mainly governed by a Coulomb potential carrying a large argument, i.e.~$V_{\mathbf q + \mathbf K^{\xi^{\prime\prime}}}\approx V_{\mathbf K^{\xi^{\prime\prime}}}$, where $\mathbf K^{\xi^{\prime\prime}}$ is a valley momentum. The resulting interaction strength is of the order of tens of meV, which is one order of magnitude smaller than the excitonic binding energy governed by a Coulomb potential $V_{\mathbf q}$ carrying a small momentum argument $|\mathbf q|\ll |\mathbf K^{\xi^{\prime\prime}}|$.

\begin{figure}[h!]
    \centering
    \subfigure[]{\includegraphics[width=0.49\linewidth]{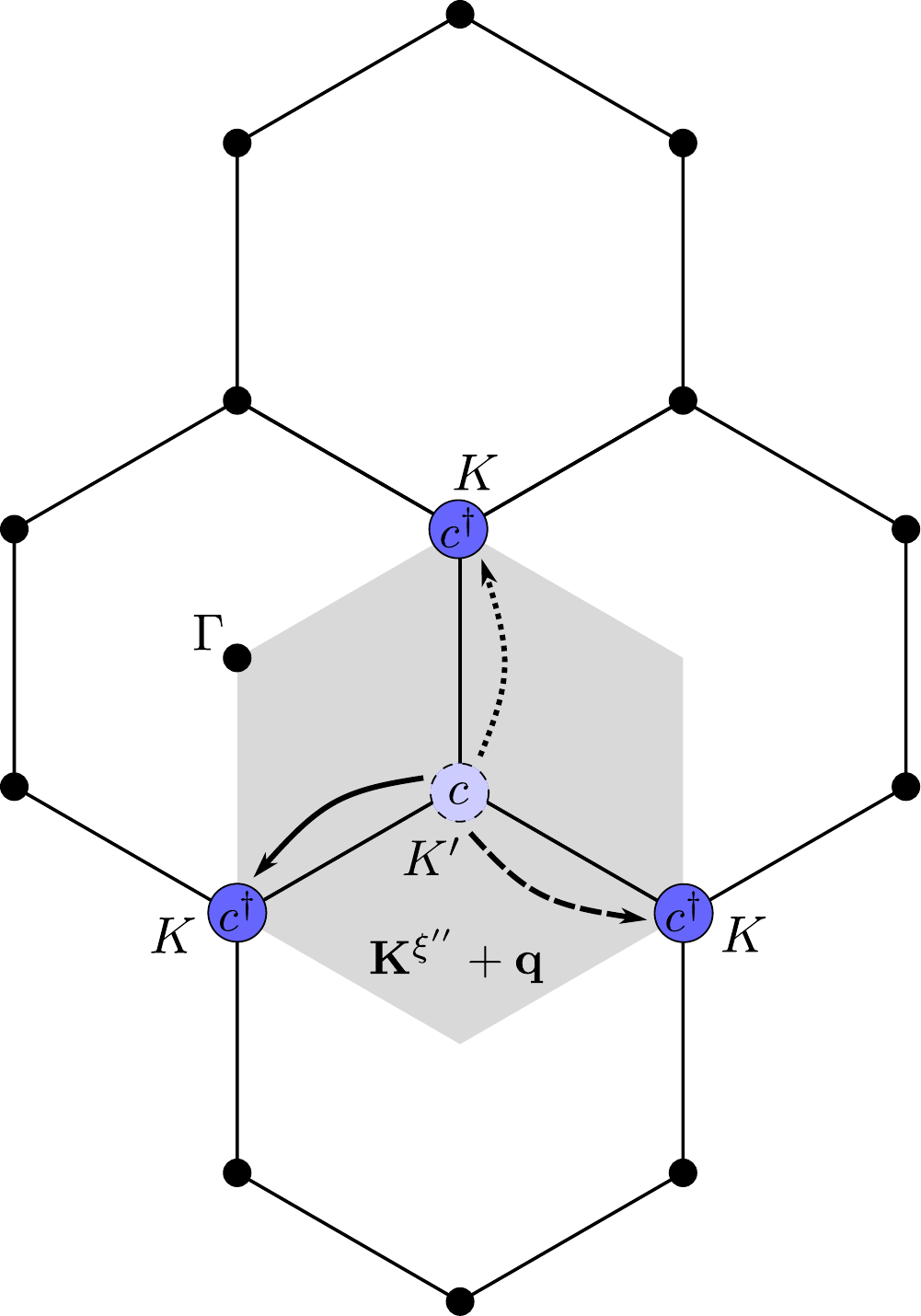}}
    \subfigure[]{\includegraphics[width=0.49\linewidth]{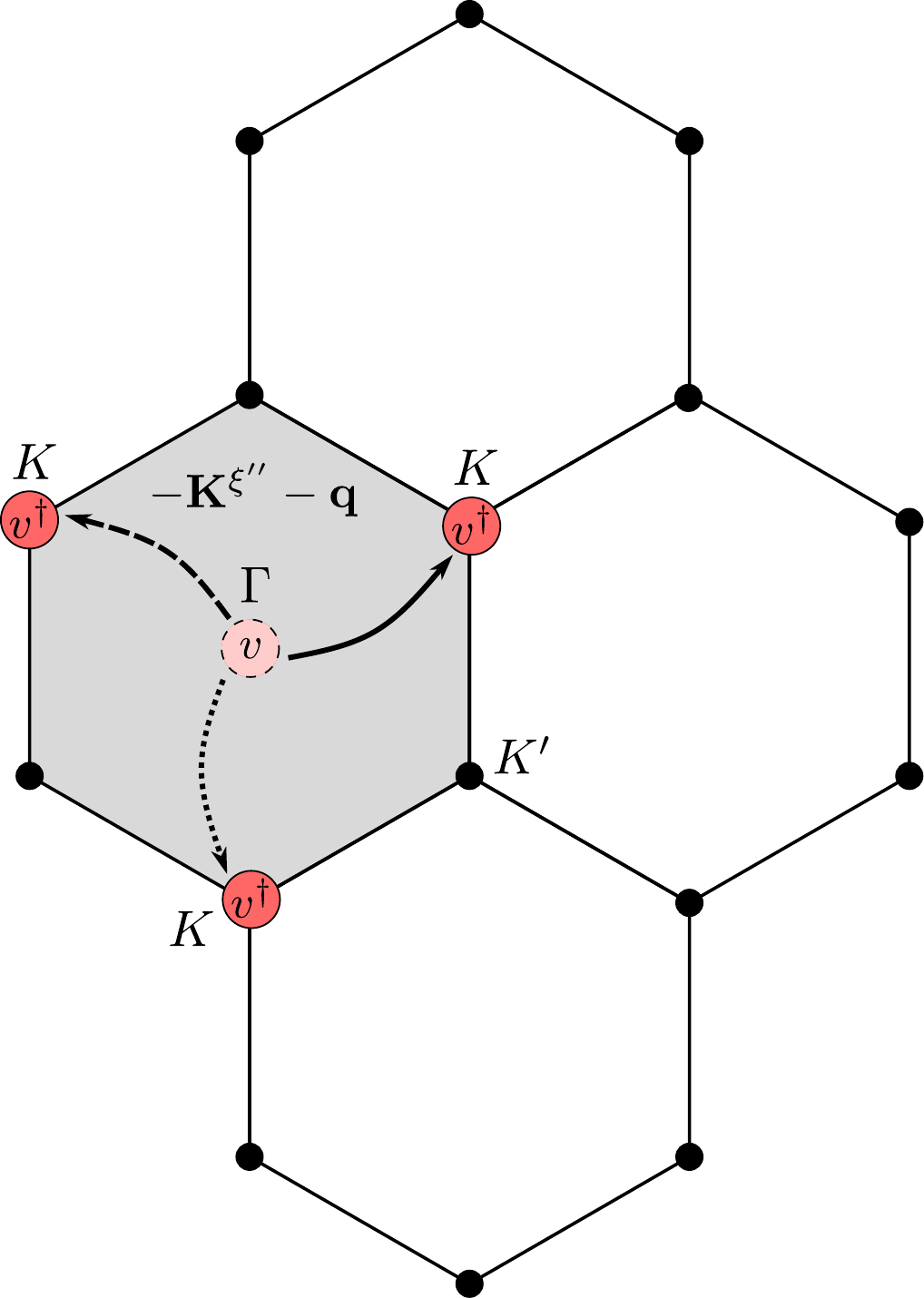}}
    \subfigure[]{\includegraphics[width=0.8\linewidth]{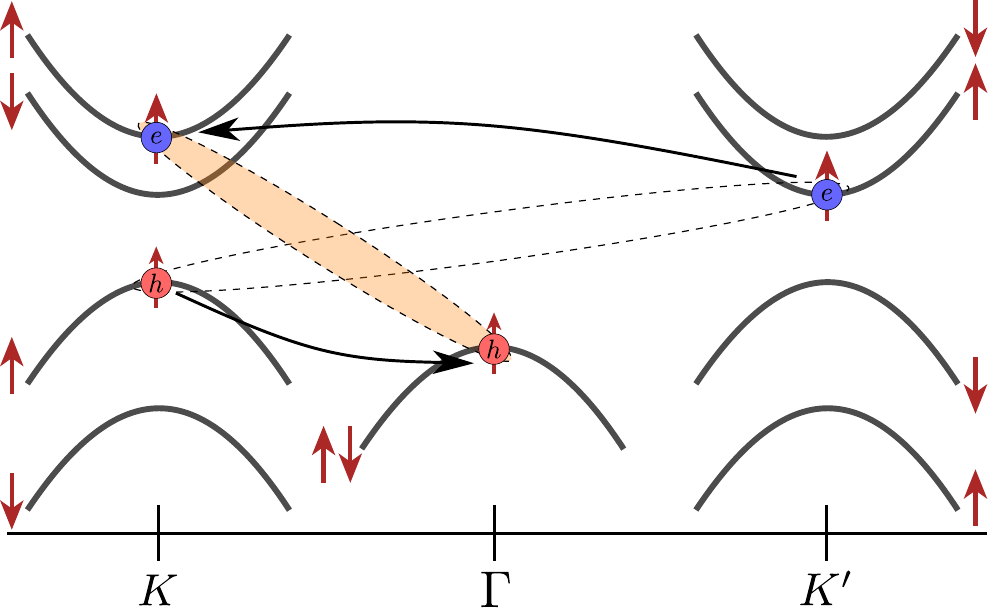}}
    \caption{Example process of intervalley interaction ($\xi^{\prime\prime}\neq \Gamma$) of intervalley excitons ($\xi\neq\xi^{\prime}$) involving simultaneous $\Gamma \rightarrow K$ and $K^{\prime}\rightarrow K$ intraband scattering (Dexter): Three equivalent processes denoted by solid, dashed and dotted arrows involving electron (a) and hole scattering (b). Grey-shaded areas denote the total set of possible momentum transfers $\mathbf K^{\xi^{\prime\prime}}+\mathbf q$ (conduction-band electron scattering) or $-\mathbf K^{\xi^{\prime\prime}}-\mathbf q$ (valence-band electron scattering). Combined process visualized in electron--hole pairs in a reduced band structure (c), cf.~\eqref{eq:dexter_hamiltonian_excitonic}.}
    \label{fig:interint_interexc}
\end{figure}

Next to intervalley scattering of \textit{intravalley} excitons as depicted in Fig.~\ref{fig:interint_intraexc}, \eqref{eq:dexter_hamiltonian_excitonic} also encodes intervalley scattering of \textit{intervalley} excitons. 
There are four possible processes for the energetically lowest electron--hole pairs involving holes at the $K$ and $\Gamma$ valleys and electrons at the $K$, $K^{\prime}$, $\Lambda$ and $\Lambda^{\prime}$ valleys:
\begin{align}
\begin{split}
    K\text{-}K^{\prime}&\leftrightarrow\Gamma\text{-}K~\text{($K$~momentum~transfer)},\\
    K^{\prime}\text{-}K&\leftrightarrow\Gamma\text{-}K^{\prime}~\text{($K$~momentum~transfer)},\\
    K\text{-}\Lambda&\leftrightarrow \Gamma\text{-}\Lambda^{\prime}~\text{($\Lambda$~momentum~transfer)},\\
    K^{\prime}\text{-}\Lambda^{\prime}&\leftrightarrow \Gamma\text{-}\Lambda~\text{($\Lambda$~momentum~transfer)}.
    \end{split}
\end{align}
In Fig.~\ref{fig:interint_interexc}, we depict the process $K$-$K^{\prime}$\,$\rightarrow$\,$\Gamma$-$K$ for the spin-up--spin-up configuration exemplarily.
As before, there exist three equivalent processes, which are weighted each by one third. 
The reduced form factors for the process depicted in Fig.~\ref{fig:interint_interexc} read:
\begin{align}
\overline \Upsilon_{\mathbf k+\mathbf q+\mathbf G,\mathbf k}^{c,c,K,K^{\prime},s}\quad\text{and}\quad\overline \Upsilon_{\mathbf k^{\prime}-\mathbf q-\mathbf G^{\prime},\mathbf k^{\prime}}^{v,v,K,\Gamma,s},
\end{align}
which have to be calculated via \textit{ab initio} methods. 
It has been shown in Ref.~\cite{dogadov2026diss} that the Dexter process involving intervalley excitons depicted in Fig.~\ref{fig:interint_interexc} plays a crucial role in the valley depolarization after ultrafast optical excitation of momentum-dark materials such as monolayer WSe$_2$ by breaking the quasi-stable \mbox{$K$-$K^{\prime}$} configuration. Here, a good agreement with the experiment has already been obtained by neglecting all local fields and using a rough approximation of $\overline \Upsilon_{\mathbf k+\mathbf q+\mathbf G,\mathbf k}^{c,c,K,K^{\prime},s} \approx \overline \Upsilon_{\mathbf k^{\prime}-\mathbf q-\mathbf G^{\prime},\mathbf k^{\prime}}^{v,v,K,\Gamma,s}\approx 1$, which entails $\mathcal C_{\text{Dex}}^{\text{dir}}\approx 1$ from \eqref{eq:dexter_correction_factor} in \eqref{eq:DexterMatrixElement}.

\FloatBarrier
\subsection{Electron--Hole Exchange Interaction}
\label{sec:exchange_interaction}
\begin{figure}[h!]
    \centering
    \subfigure[]{\includegraphics[width=0.6\linewidth]{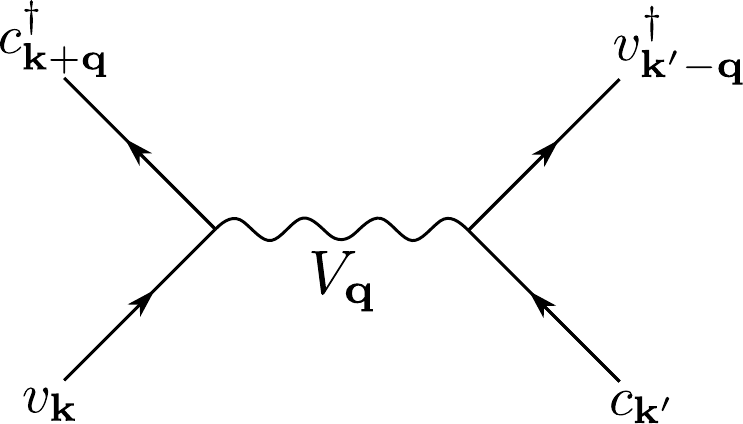}}
    \caption{Electron--hole Coulomb exchange interaction.}
    \label{fig:exchange_interaction}
\end{figure}

From \eqref{eq:Coulomb_Hamiltonian_SecondQuantized_ValleyExpanded_2}, we consider electron--hole interaction processes, where the respective electrons and holes flip their bands, cf.~Fig.~\ref{fig:exchange_interaction}. These processes are usually referred to as Coulomb exchange interactions:
\begin{multline}
    \hat H_{\text{Coul-eh}}^{\text{exch}}\\
    = \sum_{\substack{\mathbf k,\mathbf k^{\prime},\mathbf q,\mathbf G,\mathbf G^{\prime},\\\xi,\xi^{\prime},\xi^{\prime\prime},s,s^{\prime}}}
    \overline V^{}_{\mathbf q+ \mathbf K^{\xi^{\prime\prime}},\mathbf G,\mathbf G^{\prime}}
    \overline \Upsilon_{\mathbf k+\mathbf q+\mathbf G,\mathbf k}^{c,v,\xi+\xi^{\prime\prime},\xi,s}\overline \Upsilon_{\mathbf k^{\prime}-\mathbf q-\mathbf G^{\prime},\mathbf k^{\prime}}^{v,c,\xi^{\prime}-\xi^{\prime\prime},\xi^{\prime},s^{\prime}}\\
    \times\cdagtwo{\mathbf k+\mathbf q}{\xi+\xi^{\prime\prime},s}\vdagtwo{\mathbf k^{\prime}-\mathbf q}{\xi^{\prime}-\xi^{\prime\prime},s^{\prime}}\cndagtwo{\mathbf k^{\prime}}{\xi^{\prime},s^{\prime}}\vndagtwo{\mathbf k}{\xi,s}.
    \label{eq:CoulombHamiltonianExchangeElectronHole}
\end{multline}

As before, we distinguish between processes involving small ($\mathbf K^{\xi^{\prime\prime}}=\mathbf 0$) and large ($\mathbf K^{\xi^{\prime\prime}} \neq \mathbf 0$) momentum transfers.

\subsubsection*{Small Momentum Transfer}
The exchange Hamiltonian for small momentum transfer (\mbox{$\mathbf K^{\xi^{\prime\prime}}=\mathbf 0$}) reads:
\begin{multline}
    \hat H_{\text{Coul-eh}}^{\text{exch}} 
    = \sum_{\substack{\mathbf k,\mathbf k^{\prime},\mathbf q,\mathbf G,\mathbf G^{\prime},\\\xi,\xi^{\prime},s,s^{\prime}}} \overline V^{}_{\mathbf q,\mathbf G,\mathbf G^{\prime}}
    \overline \Upsilon_{\mathbf k+\mathbf q+\mathbf G,\mathbf k}^{c,v,\xi,\xi,s}\overline \Upsilon_{\mathbf k^{\prime}-\mathbf q-\mathbf G^{\prime},\mathbf k^{\prime}}^{v,c,\xi^{\prime},\xi^{\prime},s^{\prime}}\\
    \times 
    \cdagtwo{\mathbf k+\mathbf q}{\xi,s}\vdagtwo{\mathbf k^{\prime}-\mathbf q}{\xi^{\prime},s^{\prime}}\cndagtwo{\mathbf k^{\prime}}{\xi^{\prime},s^{\prime}}\vndagtwo{\mathbf k}{\xi,s}.
    \label{eq:hamiltonian_exchange_shortrange}
\end{multline}
Due to the nature of the exchange interaction being an interband scattering process and the orthogonality of the Bloch factors at equal momenta, $\overline \Upsilon_{\mathbf k,\mathbf k}^{\lambda,\lambda^{\prime},\xi,\xi,s} = \delta_{\lambda,\lambda^{\prime}}$, it is beneficial to further distinguish between short-range ($\mathbf G,\mathbf G^{\prime}\neq\mathbf 0$) and long-range contributions ($\mathbf G=\mathbf G^{\prime}=\mathbf 0$). The former originates from spatially varying Coulomb fields on the scale of the unit cell, while the latter originates from spatially varying Coulomb fields on the scale of the total lattice.\\

\textit{Short-range interactions:}

\begin{figure}[h!]
    \centering
    \subfigure[]{\includegraphics[width=0.32\linewidth]{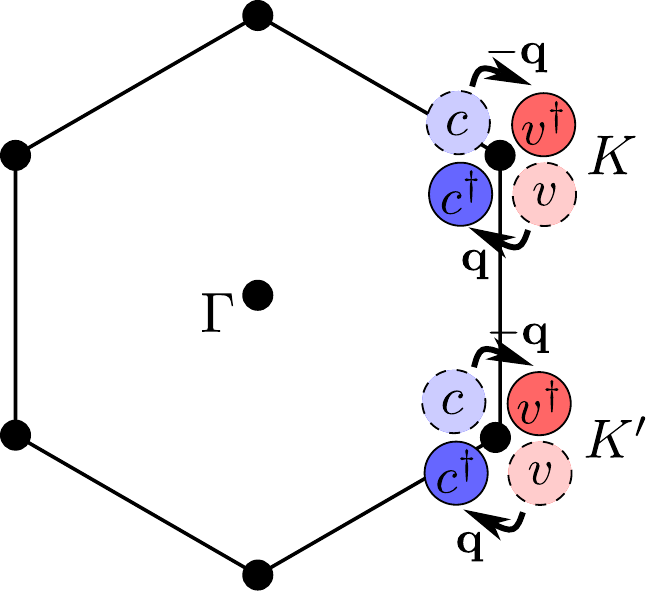}}\\
    \subfigure[]{\includegraphics[width=0.49\linewidth]{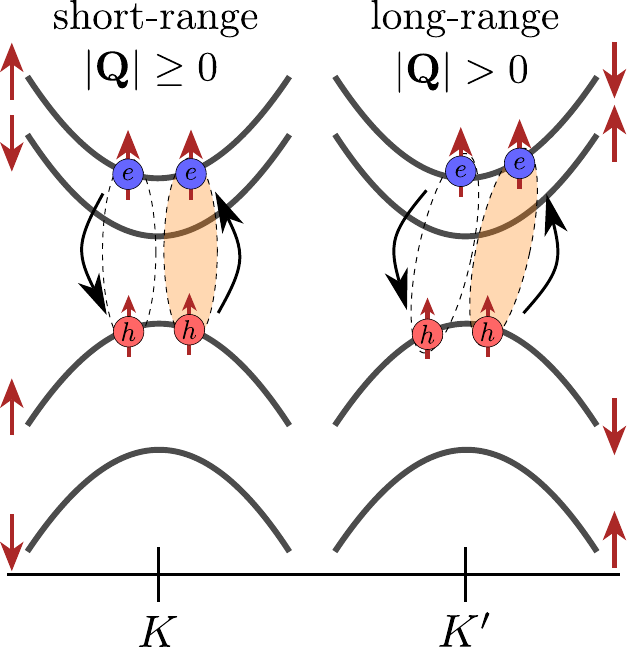}}
    \subfigure[]{\includegraphics[width=0.49\linewidth]{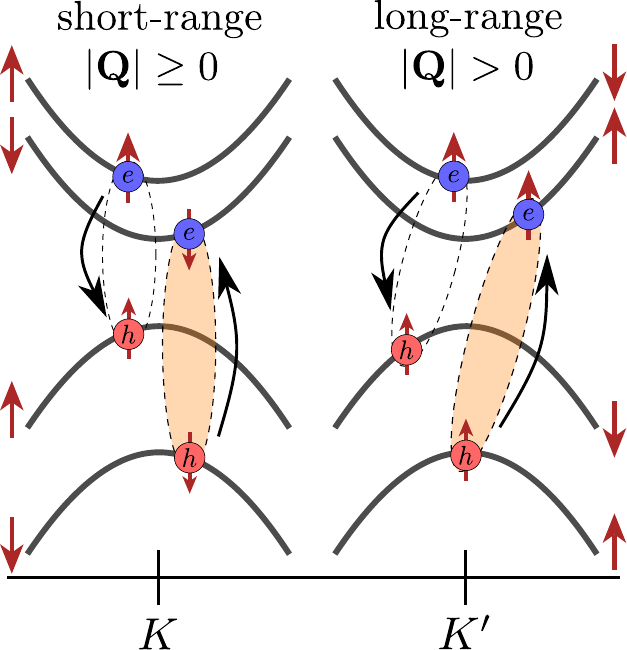}}
    \caption{Intravalley exchange interaction of intravalley excitons ($\xi=\xi^{\prime}$) involving simultaneous $K\rightarrow K$ interband scattering (a): Spin-conserving ($s=s^{\prime}$) (b) and spin-flip ($s\neq s^{\prime}$) (c) coupling within the same valley ($\xi=\xi^{\prime}$) due to short-range ($\mathbf G\neq \mathbf 0$), cf.~\eqref{eq:HamiltonianExchange_ExcitonPicture_shortrange}, and long-range ($\mathbf G=\mathbf 0$), cf.~\eqref{eq:HamiltonianExchange_ExcitonPicture_longrange}, interactions.}
    \label{fig:intraint_intraexc_exchange}
\end{figure}

Contributions with $\mathbf G,\mathbf G^{\prime}\neq\mathbf 0$ in \eqref{eq:hamiltonian_exchange_shortrange} are short-range interactions due to spatially varying local fields within a unit cell and they do not vanish at $\mathbf q=\mathbf 0$, since $\overline \Upsilon_{\mathbf k+\mathbf G,\mathbf k}^{\lambda,\lambda^{\prime},\xi,\xi,s} \neq 0$ for $\lambda\neq\lambda^{\prime}$ due to the momentum mismatch with $\mathbf G\neq\mathbf 0$. At $\mathbf q=\mathbf 0$,  these interactions involve reduced form factors as follows:
\begin{align}
\overline \Upsilon_{\mathbf k+\mathbf G,\mathbf k}^{c,v,\xi,\xi,s},
\end{align}
which have to be evaluated via first-principle methods,
and a Coulomb potential with a large argument: $V_{\mathbf q,\mathbf G,\mathbf G^{\prime}}\approx V_{\mathbf G}$ if $\mathbf G=\mathbf G^{\prime}$. 
With \eqref{eq:exciton_expansion}, the corresponding excitonic Hamiltonian reads:
\begin{shaded}
\begin{align}
    \hat H_{\text{X-SR}}^{\text{exch}} =  \sum_{\substack{\mu,\nu,\mathbf Q,\\\xi,\xi^{\prime},s,s^{\prime}}}X_{\text{SR},\mu,\nu,\mathbf Q}^{\xi,\xi^{\prime},s,s^{\prime}}\Poldagtwo{\mu,\mathbf Q}{\xi,\xi,s,s}\Poloptwo{\nu,\mathbf Q}{\xi^{\prime},\xi^{\prime},s^{\prime},s^{\prime}},
    \label{eq:HamiltonianExchange_ExcitonPicture_shortrange}
\end{align}
\end{shaded}
\noindent
with matrix element:
\begin{align}
    X_{\text{SR},\mu,\nu,\mathbf Q}^{\xi,\xi^{\prime},s,s^{\prime}} = \mathcal C_{\text{SR},\mu,\nu,\mathbf Q}^{\text{exch},\xi,\xi^{\prime},s,s^{\prime}} V_{\mathbf G} 
    \sum_{\mathbf q}
    {\varphi^*}_{\mu,\mathbf q}^{\xi,\xi,s,s}\sum_{\mathbf q^{\prime}}\varphi_{\nu,\mathbf q^{\prime}}^{\xi^{\prime},\xi^{\prime},s^{\prime}, s^{\prime}},
    \label{eq:ExchangeMatrixElement_shortrange}
\end{align}
where:
\begin{multline}
    \mathcal C_{\text{SR},\mu,\nu,\mathbf Q}^{\text{exch},\xi,\xi^{\prime},s,s^{\prime}} = 
    \left(V_{\mathbf G} 
    \sum_{\mathbf q}
    {\varphi^*}_{\mu,\mathbf q}^{\xi,\xi,s,s}\sum_{\mathbf q^{\prime}}\varphi_{\nu,\mathbf q^{\prime}}^{\xi^{\prime},\xi^{\prime},s^{\prime}, s^{\prime}}\right)^{-1}\\
    \times\sum_{\substack{\mathbf q,\mathbf q^{\prime},\mathbf G,\mathbf G^{\prime},\\(\mathbf G,\mathbf G^{\prime}\neq\mathbf 0)}} \overline V_{\mathbf Q,\mathbf G,\mathbf G^{\prime}}^{} \overline \Upsilon_{\mathbf q+\mathbf Q+\mathbf G,\mathbf q}^{c,v,\xi,\xi,s}\overline \Upsilon_{\mathbf q^{\prime}-\mathbf Q-\mathbf G^{\prime},\mathbf q^{\prime}}^{v,c,\xi^{\prime},\xi^{\prime},s^{\prime}}\\
    \times
    \ExWFstartwo{\mu,\mathbf q+\beta_{\xi,\xi}^{s,s}\mathbf Q}{\xi,\xi,s,s}\ExWFtwo{\nu,\mathbf q^{\prime}-\alpha_{\xi^{\prime},\xi^{\prime}}^{s^{\prime},s^{\prime}}\mathbf Q}{\xi^{\prime},\xi^{\prime},s^{\prime},s^{\prime}}
    ,
    \label{eq:exchange_shortrange_correction_factor}
\end{multline}
carries all corrections evaluated by \textit{ab initio} methods \cite{qiu2015nonanalyticity} compared to a simple estimate of $V_{\mathbf G}$, where $\mathbf G$ is a first reciprocal lattice vector. 
Short-range exchange interactions in \eqref{eq:HamiltonianExchange_ExcitonPicture_shortrange} only affect spin-bright intravalley excitons causing singlet--triplet splitting \cite{qiu2021solving,qiu2015nonanalyticity}, cf.~Fig.~\ref{fig:intraint_intraexc_exchange}(b)---in a slightly idealized picture, as spin--orbit interaction weakly mixes spins even directly at the band extrema \cite{deilmann2020ab,junior2022first})---or double-spin-flip processes within equal valleys, cf.~Fig.~\ref{fig:intraint_intraexc_exchange}(c). Hence, it is possible to obtain a rough estimate of $\mathcal C_{\text{SR}}^{\text{exch}}$ in a few-band effective-mass model without \textit{ab initio} methods, if the energy splitting of the corresponding spin-bright and spin-dark excitons is known from experiments.

An intriguing empirical fact is the existence of an \textit{intravalley} short-range contribution ($\xi=\xi^{\prime}$ in \eqref{eq:ExchangeMatrixElement_shortrange}), but the vanishing of an \textit{intervalley} short-range contribution ($\xi\neq\xi^{\prime}$ in \eqref{eq:ExchangeMatrixElement_shortrange}), since, otherwise, a valley hybridization between the $\Poltwo{\mu,\mathbf Q}{K,K,\uparrow,\uparrow}$ and $\Poltwo{\mu,\mathbf Q}{K^{\prime},K^{\prime},\downarrow,\downarrow}$ excitons would occur already at zero center-of-mass momentum $\mathbf Q=\mathbf 0$, which should produce observable features such as valley-split exciton lines in linear absorption measurements or near-instantaneous valley depolarization in nonlinear pump--probe experiments. However, such features have not been observed experimentally, since valley-split exciton lines do not appear and valley depolarization usually takes some time (ps scale) \cite{dogadov2026diss,raiber2022ultrafast,dal2015ultrafast}. Hence, the matrix element for intervalley short-range exchange interaction must vanish or, at least, very small compared to its intravalley counterpart, which we illustrate in the following. Since the matrix element in \eqref{eq:ExchangeMatrixElement_shortrange} contains a summation over all $\mathbf G\neq\mathbf 0$, cf.~\eqref{eq:exchange_shortrange_correction_factor}, the total matrix element is built from a multitude of terms. Even if we restrict the summation to the most relevant first reciprocal lattice vectors and assume $\mathbf G=\mathbf G^{\prime}$, this would entail in total six terms. 
The individual intravalley exchange equal-spin overlaps in low-wavenumber approximation and $|\mathbf q|\ll |\mathbf G|$ read:
\begin{align}
    \sum_{\mathbf G\neq\mathbf 0}V_{\mathbf G}\overline \Upsilon_{\mathbf G,\mathbf 0}^{c,v,\xi,\xi,s}\overline \Upsilon_{-\mathbf G,\mathbf 0}^{v,c,\xi,\xi,s}\quad\text{(intravalley)},
    \label{eq:intravalley_sr}
\end{align}
while the intervalley exchange opposite-spin overlaps read:
\begin{align}
    \sum_{\mathbf G\neq\mathbf 0}V_{\mathbf G}\overline \Upsilon_{\mathbf G,\mathbf 0}^{c,v,\xi,\xi,s}\overline \Upsilon_{-\mathbf G,\mathbf 0}^{v,c,\bar \xi,\bar \xi,\bar s}\quad\text{(intervalley)},
\end{align}
Since the Bloch factors are connected via a time-reversal transformation $\mathbf k\rightarrow -\mathbf k$, $\xi\rightarrow \bar \xi$ (for $\xi\in \{K,K^{\prime}\}$), $s\rightarrow \bar s$, we can write:
\begin{multline}
    \sum_{\mathbf G\neq\mathbf 0}V_{\mathbf G}\overline \Upsilon_{\mathbf G,\mathbf 0}^{c,v,\xi,\xi,s}\overline \Upsilon_{-\mathbf G,\mathbf 0}^{v,c,\bar \xi,\bar \xi,\bar s}\\
    = \sum_{\mathbf G\neq\mathbf 0}V_{\mathbf G}\overline \Upsilon_{\mathbf G,\mathbf 0}^{c,v,\xi,\xi,s}\overline \Upsilon_{\mathbf G,\mathbf 0}^{v,c, \xi, \xi,s}\quad\text{(intervalley)}.
    \label{eq:intervalley_sr_pt}
\end{multline}
Hence, the only difference between the intravalley exchange element in \eqref{eq:intravalley_sr} and the intervalley exchange element in \eqref{eq:intervalley_sr_pt} is the sign of the phase factor in the second reduced Bloch factor ($\mathrm e^{\mathrm i\mathbf G\cdot\mathbf r}$ vs.\ $\mathrm e^{-\mathrm i\mathbf G\cdot\mathbf r}$). Thus, by summing over all relevant $\mathbf G\neq\mathbf 0$ vectors, the intervalley contributions should cancel each other due to their phase mismatch, while the intravalley contributions do not. Here, explicit \textit{ab initio} calculations can shed more light on this issue.\\

\textit{Long-range interactions:}

\begin{figure}[h!]
    \centering
    \subfigure[]{\includegraphics[width=0.32\linewidth]{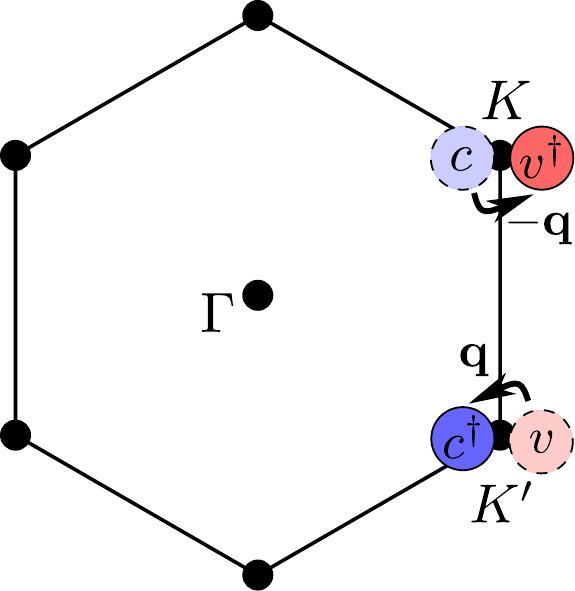}}\\
    \subfigure[]{\includegraphics[width=0.49\linewidth]{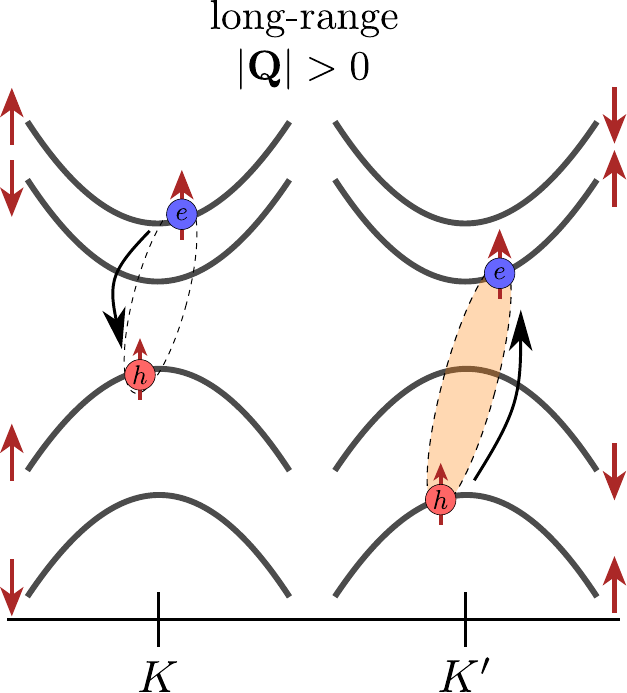}}
    \subfigure[]{\includegraphics[width=0.49\linewidth]{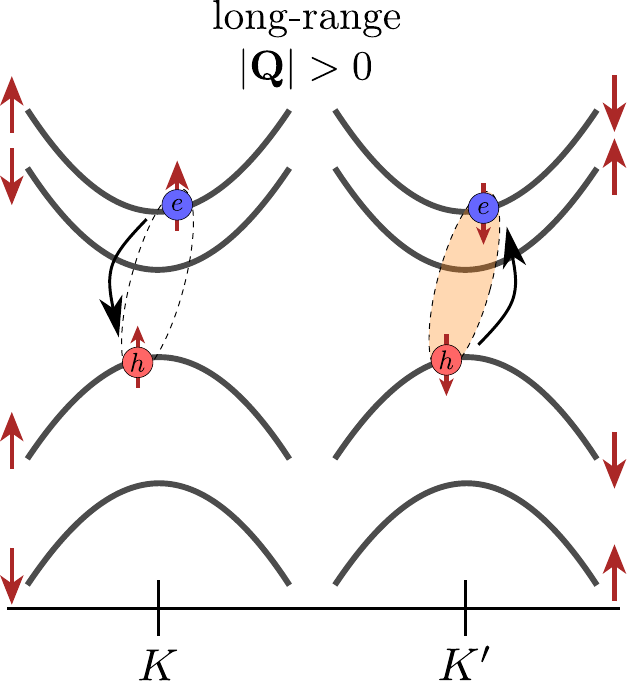}}
    \caption{Intervalley ($\xi\neq\xi^{\prime}$) exchange interaction of intravalley excitons involving simultaneous $K^{\prime}\rightarrow K^{\prime}$ and $K\rightarrow K$ interband scattering (a): Spin-conserving ($s=s^{\prime}$) valley transfer (b) and spin-flip ($s\neq s^{\prime}$) valley transfer (c) via nonlocal fields ($\mathbf G=\mathbf 0$) (Förster transfer), cf.~\eqref{eq:HamiltonianExchange_ExcitonPicture_longrange}.}
    \label{fig:interint_intraexc_exchange}
\end{figure}

Contributions with $\mathbf G=\mathbf G^{\prime}=\mathbf 0$ in \eqref{eq:hamiltonian_exchange_shortrange} are long-range interactions, as the interacting Coulombic fields can vary over multiple unit cells. These processes occur within equal valleys ($\xi=\xi^{\prime}$), cf.~Fig.~\ref{fig:intraint_intraexc_exchange}, and distinct valleys ($\xi\neq\xi^{\prime}$), cf.~Fig.~\ref{fig:interint_intraexc_exchange}. 
The corresponding excitonic Hamiltonian with \eqref{eq:exciton_expansion} reads:
\begin{shaded}
\begin{align}
    \hat H_{\text{X-LR}}^{\text{exch}} =  \sum_{\substack{\mu,\nu,\mathbf Q,\\\xi,\xi^{\prime},s,s^{\prime}}}X_{\text{LR},\mu,\nu,\mathbf Q}^{\xi,\xi^{\prime},s,s^{\prime}}\Poldagtwo{\mu,\mathbf Q}{\xi,\xi,s,s}\Poloptwo{\nu,\mathbf Q}{\xi^{\prime},\xi^{\prime},s^{\prime},s^{\prime}},
    \label{eq:HamiltonianExchange_ExcitonPicture_longrange}
\end{align}
\end{shaded}
\noindent
with matrix element:
\begin{multline}
X_{\text{LR},\mu,\nu,\mathbf Q}^{\xi,\xi^{\prime},s,s^{\prime}} = \sum_{\mathbf q,\mathbf q^{\prime}} \overline V_{\mathbf Q,\mathbf 0,\mathbf 0}^{} \overline \Upsilon_{\mathbf q+\mathbf Q,\mathbf q}^{c,v,\xi,\xi,s}\overline \Upsilon_{\mathbf q^{\prime}-\mathbf Q,\mathbf q^{\prime}}^{v,c,\xi^{\prime},\xi^{\prime},s^{\prime}}\\
    \times
    \ExWFstartwo{\mu,\mathbf q+\beta_{\xi,\xi}^{s,s}\mathbf Q}{\xi,\xi,s,s}\ExWFtwo{\nu,\mathbf q^{\prime}-\alpha_{\xi^{\prime},\xi^{\prime}}^{s^{\prime},s^{\prime}}\mathbf Q}{\xi^{\prime},\xi^{\prime},s^{\prime},s^{\prime}}.
    \label{eq:ExchangeMatrixElement_longrange_general}
\end{multline}
In contrast to the short-range contributions, the long-range contributions vanish at a center-of-mass momentum \mbox{$\mathbf Q=\mathbf 0$}. Hence, it is suitable to Taylor expand the form factors in $\mathbf Q=\mathbf 0$ up to the first non-vanishing order, cf.~\eqref{eq:formfactor_first}:
\begin{align}
\overline \Upsilon_{\mathbf k+\mathbf Q,\mathbf k}^{\lambda,\bar \lambda,\xi,\xi,s} \approx -\frac{\mathrm i}{e}
    \mathbf Q\cdot\mathbf d_{\mathbf k}^{\lambda,\bar \lambda,\xi,\xi,s} \approx -\frac{\mathrm i}{e}
    \mathbf Q\cdot\mathbf d_{\mathbf 0}^{\lambda,\bar \lambda,\xi,\xi,s},
\end{align}
where $\mathbf d_{\mathbf 0}^{\lambda,\bar \lambda,\xi,\xi,s}$ is the dipole matrix element in low-wavenumber approximation, cf.~\eqref{eq:dipole_moment}. The long-range exchange matrix element from \eqref{eq:ExchangeMatrixElement_longrange_general} can then be approximated as:
\begin{multline}
    X_{\text{LR},\mu,\nu,\mathbf Q}^{\xi,\xi^{\prime},s,s^{\prime}}\\
    \approx  
    V_{\mathbf Q}^{}\left(\frac{1}{e^2}\left(\mathbf Q\cdot \mathbf d_{\mathbf 0}^{cv,\xi,s}\vphantom{d_{}^{cv,\xi,s}}\right)\left(\mathbf Q\cdot \mathbf d_{\mathbf 0}^{vc,\xi^{\prime},s^{\prime}}\right) + \mathcal O\mleft((\mathbf Q\cdot \mathbf d)^3\mright)\right)\\
    \times 
    \sum_{\mathbf q}
    {\varphi^*}_{\mu,\mathbf q}^{\xi,\xi,s,s}\sum_{\mathbf q^{\prime}}\varphi_{\nu,\mathbf q^{\prime}}^{\xi^{\prime},\xi^{\prime},s^{\prime}, s^{\prime}}.
    \label{eq:ExchangeMatrixElement_longrange}
\end{multline}

Long-range interactions between different excitonic configurations are the reciprocal-space analog to the Coulomb-induced energy transfer, i.e., Förster interaction \cite{forster1948zwischenmolekulare}, in real space: 
Nonresonant long-range interactions are responsible for the coupling of A and B excitons within equal valleys \cite{guo2019exchange} (double spin flip scattering, $\xi=\xi^{\prime}$, $s\neq s^{\prime}$ in \eqref{eq:HamiltonianExchange_ExcitonPicture_longrange}), cf.~Fig.~\ref{fig:intraint_intraexc_exchange}(c), and for the coupling of A and B excitons between the $K$/$K^{\prime}$ valleys (spin-conserving scattering, $\xi\neq\xi^{\prime}$, $s = s^{\prime}$ in \eqref{eq:HamiltonianExchange_ExcitonPicture_longrange}), cf.~Fig.~\ref{fig:interint_intraexc_exchange}(b). 
Moreover, resonant long-range exchange between the valleys gives rise to nonanalytic dispersions \cite{qiu2015nonanalyticity,kwong2021effect} and double spin-flip scattering between $K$ and $K^{\prime}$ valleys (Maialle--Silva--Sham- or Bir--Aronov--Pikus mechanism, $\xi\neq\xi^{\prime}$, $s\neq s^{\prime}$ in \eqref{eq:HamiltonianExchange_ExcitonPicture_longrange}) \cite{bir1975spin,maialle1993exciton,lechner2005spin,yu2014valley,selig2019ultrafast,selig2020suppression}, cf.~Fig.~\ref{fig:interint_intraexc_exchange}(c). 
Additionally, within the same valley, long-range exchange further impacts the exciton dispersion at nonzero center-of-mass momenta $\mathbf Q$ ($\xi=\xi^{\prime}$, $s = s^{\prime}$ in \eqref{eq:HamiltonianExchange_ExcitonPicture_longrange}), cf.~Fig.~\ref{fig:intraint_intraexc_exchange}(b). Hence, long-range exchange interaction causes a deviation from an otherwise purely quadratic effective-mass exciton dispersion possibly impairing biexciton binding \cite{kwong2021effect}.

\FloatBarrier

\subsubsection*{Large Momentum Transfer}
The exchange Hamiltonian for large momentum transfer ($\mathbf K^{\xi^{\prime\prime}}\neq\mathbf 0$) is given by \eqref{eq:CoulombHamiltonianExchangeElectronHole} with the condition $\xi^{\prime\prime}\neq \Gamma$. Since the momentum transfer is already on the scale of a large valley momentum $\mathbf K^{\xi^{\prime\prime}}$, a distinction in short-range and long-range contributions is meaningless. Hence, we distinguish two cases as follows: Intravalley scattering between excitonic configurations in equal valleys with $\xi^{\prime\prime} = \xi^{\prime}-\xi\neq \Gamma$ and intervalley scattering with $\xi^{\prime\prime} \neq \xi^{\prime}-\xi$ and $\xi^{\prime\prime} \neq \Gamma$ between excitonic configurations in unequal valleys.\\

\textit{Intravalley scattering with $\xi^{\prime\prime} = \xi^{\prime}-\xi\neq \Gamma$:}

\begin{figure}[h!]
    \centering
    \subfigure[]{\includegraphics[width=0.49\linewidth]{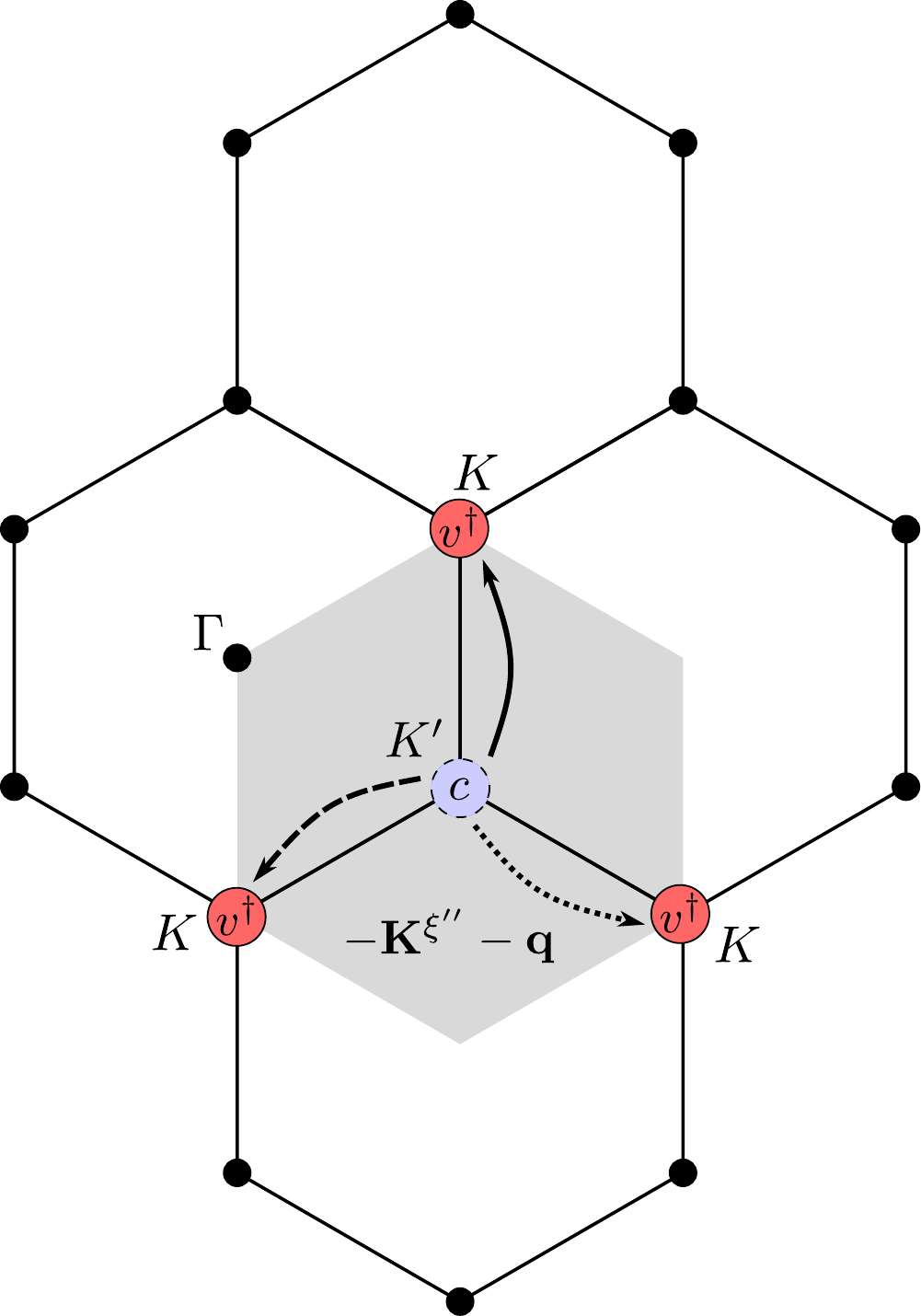}}
    \subfigure[]{\includegraphics[width=0.49\linewidth]{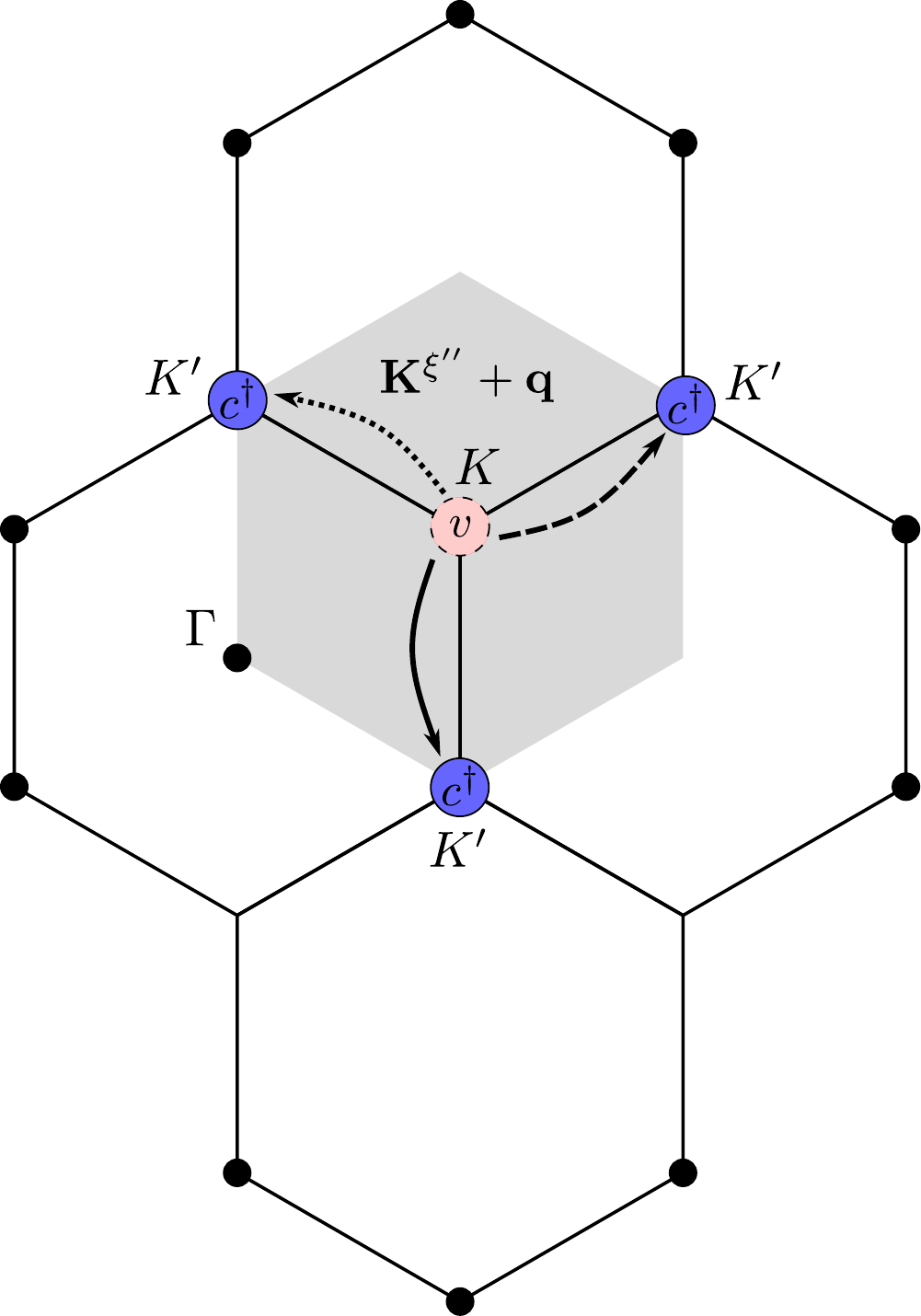}}
    \subfigure[]{\includegraphics[width=0.49\linewidth]{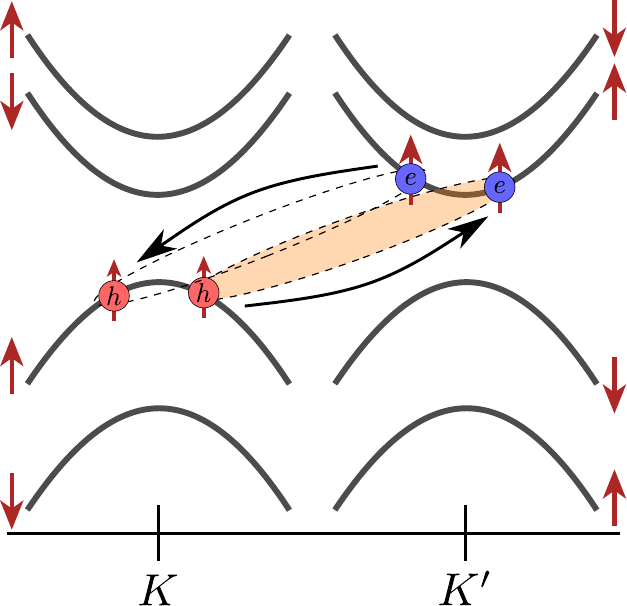}}
    \subfigure[]{\includegraphics[width=0.49\linewidth]{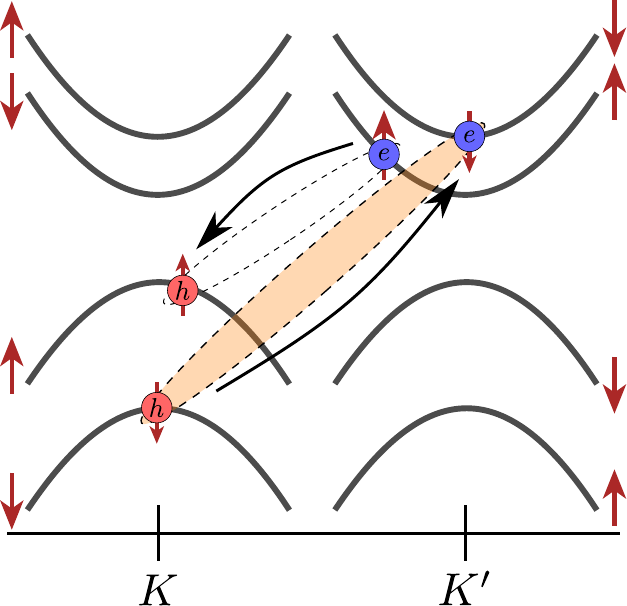}}
    \caption{Example scattering channel of intravalley exchange interaction $\xi^{\prime\prime} = \xi^{\prime}-\xi$ of intervalley excitons ($\xi\neq\xi^{\prime}$) involving simultaneous $K\leftrightarrow K^{\prime}$ interband scattering (a). (b): Three equivalent processes. Corresponding example spin-conserving ($s = s^{\prime}$) (c) and spin-flip ($s \neq s^{\prime}$) (d) process in the excitonic picture, cf.~\eqref{eq:HamiltonianExchange_ExcitonPicture_large_momentum_transfer}.}
    \label{fig:intraint_interexc_exchange1}
\end{figure}

The corresponding excitonic Hamiltonian reads:
\begin{shaded}
\begin{align}
    \hat H_{\text{X-K,intra}}^{\text{exch}} =  \sum_{\substack{\mu,\nu,\mathbf Q,\xi,\xi^{\prime},s,s^{\prime}\\ (\xi\neq\xi^{\prime})}}X_{\text{K,intra},\mathbf Q,\mu,\nu}^{\xi, \xi^{\prime},s,s^{\prime}}\Poldagtwo{\mu,\mathbf Q}{\xi, \xi^{\prime},s,s}\Poloptwo{\nu,\mathbf Q}{\xi,\xi^{\prime},s^{\prime},s^{\prime}},
    \label{eq:HamiltonianExchange_ExcitonPicture_large_momentum_transfer}
\end{align}
\end{shaded}
\noindent
with matrix element:
\begin{multline}
    X_{\text{K,intra},\mathbf Q,\mu,\nu}^{\xi, \xi^{\prime},s,s^{\prime}}\\
    = \mathcal C_{\text{K,intra},\mu,\nu,\mathbf Q}^{\text{exch},\xi,\xi^{\prime},s,s^{\prime}}V_{\mathbf K^{\xi^{\prime}}-\mathbf K^{\xi}}
    \sum_{\mathbf q} \ExWFstartwo{\mu,\mathbf q}{\xi,\xi^{\prime},s,s}
    \sum_{\mathbf q^{\prime}}
    \ExWFtwo{\nu,\mathbf q^{\prime}}{\xi,\xi^{\prime},s^{\prime},s^{\prime}}.
    \label{eq:ExchangeMatrixElement_large_momentum_transfer}
\end{multline}
where $\mathcal C_{\text{K,intra}}^{\text{exch}}$ is a correction to a simple estimate of $V_{\mathbf K^{\xi^{\prime}}-\mathbf K^{\xi}}$ obtained from \textit{ab initio} calculations:
\begin{multline}
    \mathcal C_{\text{K,intra},\mu,\nu,\mathbf Q}^{\text{exch},\xi,\xi^{\prime},s,s^{\prime}} = \left(V_{\mathbf K^{\xi^{\prime}}-\mathbf K^{\xi}}\sum_{\mathbf q} \ExWFstartwo{\mu,\mathbf q}{\xi,\xi^{\prime},s,s}
    \sum_{\mathbf q^{\prime}}
    \ExWFtwo{\nu,\mathbf q^{\prime}}{\xi,\xi^{\prime},s^{\prime},s^{\prime}}\right)^{-1} \\
    \times \sum_{\mathbf q,\mathbf q^{\prime},\mathbf G,\mathbf G^{\prime}} \overline V_{\mathbf Q+\mathbf K^{\xi^{\prime}}-\mathbf K^{\xi},\mathbf G,\mathbf G^{\prime}}^{} \overline \Upsilon_{\mathbf q+\mathbf Q+\mathbf G,\mathbf q}^{c,v,\xi^{\prime},\xi,s} \overline \Upsilon_{\mathbf q^{\prime}-\mathbf Q-\mathbf G^{\prime},\mathbf q^{\prime}}^{v,c,\xi,\xi^{\prime},s^{\prime}}\\
    \times \ExWFstartwo{\mu,\mathbf q+\beta_{\xi,\xi^{\prime}}^{s,s}\mathbf Q}{\xi,\xi^{\prime},s,s}\ExWFtwo{\nu,\mathbf q^{\prime}-\alpha_{\xi,\xi^{\prime}}^{s^{\prime},s^{\prime}}\mathbf Q}{\xi,\xi^{\prime},s^{\prime},s^{\prime}}.
\end{multline}
Large-momentum-transfer exchange interaction, where the corresponding electrons and holes do not change valleys ($\xi+\xi^{\prime\prime} = \xi^{\prime}$ and $\xi^{\prime}-\xi^{\prime\prime} = \xi$), occurs between all possible intervalley excitons with $\xi\neq\xi^{\prime}$, cf.~\eqref{eq:HamiltonianExchange_ExcitonPicture_large_momentum_transfer}. Spin-conserving large-momentum-transfer exchange causes a blue shift of spin-bright intervalley excitons, cf.~Fig.~\ref{fig:intraint_interexc_exchange1}(c), and spin-flip large-momentum-transfer exchange interaction couples intervalley A and B excitons, cf.~Fig.~\ref{fig:intraint_interexc_exchange1}(d), analog to the small-momentum-transfer counterpart acting on intravalley excitons, cf.\ Fig.~\ref{fig:intraint_intraexc_exchange}. In particular, the spin-conserving large-momentum-transfer exchange interaction is responsible for the blue shift of the spin-bright intervalley $K,K^{\prime}$-$\uparrow,\uparrow$ exciton over the spin-dark intravalley $K,K$-$\uparrow,\downarrow$ exciton \cite{li2022intervalley,yang2022relaxation,he2020valley}, which would be otherwise energy-degenerate, since both conduction bands considered here, $K^{\prime},\uparrow$ and $K,\downarrow$ have equal effective masses \cite{kormanyos2015k}.\\

\textit{Intervalley scattering with $\xi^{\prime\prime} \neq \xi^{\prime}-\xi$ and $\xi^{\prime\prime} \neq \Gamma$:}

The corresponding excitonic Hamiltonian reads:
\begin{shaded}
\begin{multline}
    \hat H_{\text{X-K,inter}}^{\text{exch}} \\
    =  \sum_{\substack{\mu,\nu,\mathbf Q,\xi,\xi^{\prime},\xi^{\prime\prime},s,s^{\prime}\\ (\xi^{\prime\prime}\neq \xi^{\prime}-\xi, \xi^{\prime\prime} \neq \Gamma)}}X_{\text{K,inter},\mathbf Q,\mu,\nu}^{\xi, \xi^{\prime}, \xi^{\prime\prime},s,s^{\prime}}\Poldagtwo{\mu,\mathbf Q}{\xi, \xi+\xi^{\prime\prime},s,s}\Poloptwo{\nu,\mathbf Q}{\xi^{\prime}-\xi^{\prime\prime},\xi^{\prime},s^{\prime},s^{\prime}},
    \label{eq:HamiltonianExchange_ExcitonPicture_large_momentum_transfer_inter}
\end{multline}
\end{shaded}
\noindent
with matrix element:
\begin{multline}
    X_{\text{K,inter},\mathbf Q,\mu,\nu}^{\xi, \xi^{\prime},\xi^{\prime\prime},s,s^{\prime}} = \mathcal C_{\text{K,inter},\mu,\nu,\mathbf Q}^{\text{exch}\xi,\xi^{\prime},\xi^{\prime\prime},s,s^{\prime}}V_{\mathbf K^{\xi^{\prime\prime}}}\\
    \times 
    \sum_{\mathbf q} \ExWFstartwo{\mu,\mathbf q}{\xi,\xi+\xi^{\prime\prime},s,s}
    \sum_{\mathbf q^{\prime}}
    \ExWFtwo{\nu,\mathbf q^{\prime}}{\xi^{\prime}-\xi^{\prime\prime},\xi^{\prime},s^{\prime},s^{\prime}},
    \label{eq:ExchangeMatrixElement_large_momentum_transfer_inter}
\end{multline}
where $\mathcal C_{\text{K,inter}}^{\text{exch}}$ is a correction to a simple estimate of $V_{\mathbf K^{\xi^{\prime\prime}}}$ obtained from \textit{ab initio} calculations:
\begin{multline}
    \mathcal C_{\text{K,inter},\mu,\nu,\mathbf Q}^{\text{exch},\xi,\xi^{\prime},\xi^{\prime\prime},s,s^{\prime}}\\
    = \left(V_{\mathbf K^{\xi^{\prime\prime}}}\sum_{\mathbf q} \ExWFstartwo{\mu,\mathbf q}{\xi,\xi+\xi^{\prime\prime},s,s}
    \sum_{\mathbf q^{\prime}}
    \ExWFtwo{\nu,\mathbf q^{\prime}}{\xi^{\prime}-\xi^{\prime\prime},\xi^{\prime},s^{\prime},s^{\prime}}\right)^{-1} \\
    \times \sum_{\mathbf q,\mathbf q^{\prime},\mathbf G,\mathbf G^{\prime}} \overline V_{\mathbf Q+\mathbf K^{\xi^{\prime\prime}},\mathbf G,\mathbf G^{\prime}}^{} \overline \Upsilon_{\mathbf q+\mathbf Q+\mathbf G,\mathbf q}^{c,v,\xi+\xi^{\prime\prime},\xi,s} \overline \Upsilon_{\mathbf q^{\prime}-\mathbf Q-\mathbf G^{\prime},\mathbf q^{\prime}}^{v,c,\xi^{\prime}-\xi^{\prime\prime},\xi^{\prime},s^{\prime}}\\
    \times \ExWFstartwo{\mu,\mathbf q+\beta_{\xi,\xi+\xi^{\prime\prime}}^{s,s}\mathbf Q}{\xi,\xi+\xi^{\prime\prime},s,s}
    \ExWFtwo{\nu,\mathbf q^{\prime}-\alpha_{\xi^{\prime}-\xi^{\prime\prime},\xi^{\prime}}^{s^{\prime},s^{\prime}}\mathbf Q}{\xi^{\prime}-\xi^{\prime\prime},\xi^{\prime},s^{\prime},s^{\prime}}.
\end{multline}

\begin{figure}[h!]
    \centering
    \subfigure[]{\includegraphics[width=0.49\linewidth]{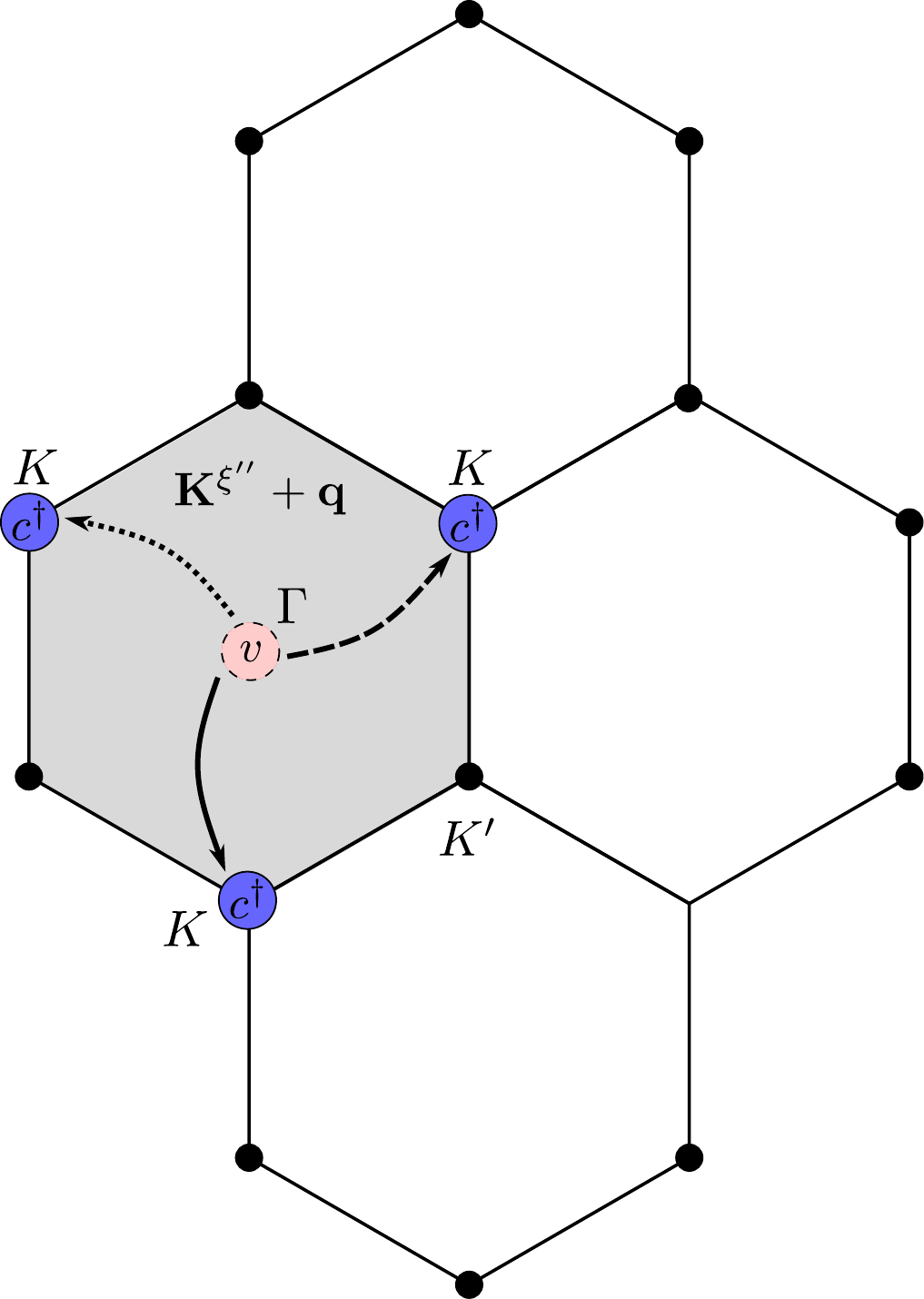}}
    \subfigure[]{\includegraphics[width=0.49\linewidth]{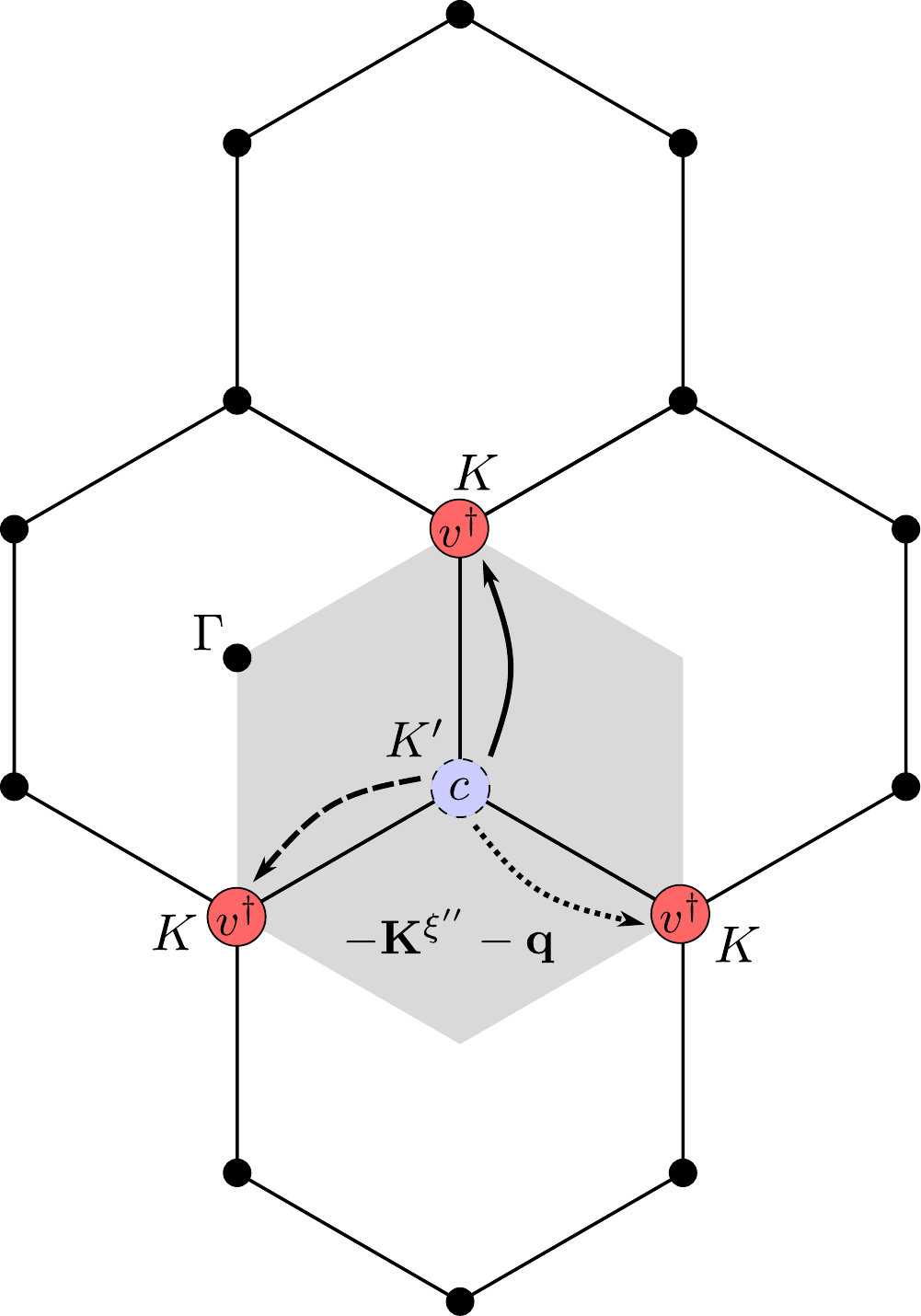}}
    \subfigure[]{\includegraphics[width=0.8\linewidth]{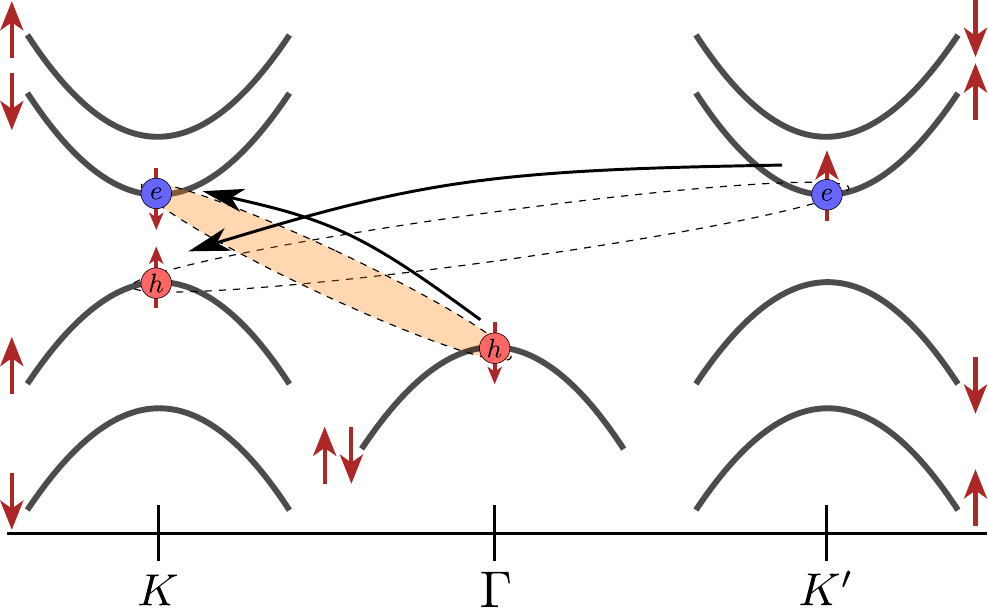}}
    \caption{Example scattering channel of intervalley exchange interaction ($\xi^{\prime\prime} \neq \Gamma$) of intervalley excitons ($\xi^{\prime\prime} \neq \xi^{\prime}-\xi$) involving simultaneous $\Gamma \rightarrow K$ and $K^{\prime}\rightarrow K$ and interband scattering (a) and (b): Three equivalent processes. Corresponding example spin-flip process ($s\neq s^{\prime}$) in the excitonic picture, cf.~\eqref{eq:HamiltonianExchange_ExcitonPicture_large_momentum_transfer_inter}.}
    \label{fig:interint_interexc_exchange1}
\end{figure}

On the other hand, large-momentum-transfer exchange interaction, where the corresponding electrons and holes change valleys ($\xi+\xi^{\prime\prime} \neq \xi^{\prime}$ and $\xi^{\prime}-\xi^{\prime\prime} \neq \xi$), occurs between the following spin-bright intervalley excitons, if we restrict ourselves to the energetically lowest holes at the $K$, $K^{\prime}$ and $\Gamma$ valleys and the electrons to the energetically lowest $K$, $K^{\prime}$ and $\Lambda$, $\Lambda^{\prime}$ valleys:
\begin{align}
\begin{split}
    K\text{-}K^{\prime}&\leftrightarrow\Gamma\text{-}K,\\
    K^{\prime}\text{-}K&\leftrightarrow\Gamma\text{-}K^{\prime},\\
    K\text{-}\Lambda&\leftrightarrow \Gamma\text{-}\Lambda^{\prime},\\
    K^{\prime}\text{-}\Lambda^{\prime}&\leftrightarrow \Gamma\text{-}\Lambda.
    \end{split}
\end{align}
These processes cause an energy transfer between distinct valleys, cf.~Fig.~\ref{fig:interint_interexc_exchange1}, where we depict the spin-flip process $K\text{-}K^{\prime}\rightarrow\Gamma\text{-}K$ exemplarily, which involves three equivalent processes analog to the charge-transfer process in Fig.~\ref{fig:interint_interexc}. This process is expected to be of minor importance, since the large valley-momentum transfer and the exchange nature strongly suppress the magnitude of this process compared to other scattering processes. However, it can contribute to a slight acceleration of the valley depolarization process in dark materials such as monolayer WS$_2$ or WSe$_2$ \cite{dogadov2026diss}, but no simulation has explicitly considered this process up to now.

\FloatBarrier
\section{Conclusion}

In this manuscript, we reviewed and discussed the second-quantized Coulomb interaction between Bloch electrons in 2D-confined semiconductors as a starting point for exciton quantum kinetics in a Heisenberg-equations-of-motion approach. We derived the many-body Coulomb interaction including all Umklapp processes, discussed the origin of microscopic screening and reviewed an effective macroscopic approach to screening including the dielectric environment for practical calculations. Further, we discussed the binding of electron--hole pairs, i.e., excitons, described by the Bethe--Salpeter equation and its few-band effective-mass manifestation, the Wannier equation in undoped semiconductors. At last, we provided a comprehensive discussion of relevant exciton scattering processes in the density-independent limit.

\FloatBarrier
\medskip
\acknowledgments
Funded by the Deutsche Forschungsgemeinschaft (DFG, German Research Foundation) -- Project No.\ 420760124 (H.M., A.K.); 556436549 (A.K.). H.M.\ acknowledges funding by project 21209528 (``proof of trust''). 
T.D.\ acknowledges financial support from the Deutsche Forschungsgemeinschaft (DFG, German Research Foundation) through Project No.\ 426726249 (DE 2749/2-1 and DE 2749/2-2).

H.M.\ thanks Alexander Steinhoff (Universität Oldenburg), Jonas Grumm (Technische Universität Berlin), Robert Lemke (Technische Universität Berlin), Michiel Snoeken (Technische Universität Berlin), and, especially, Mikhail M.\ Glazov (Ioffe Institute) for fruitful discussions.

\setcounter{table}{0}
\renewcommand{\thetable}{A\Roman{table}}
\appendix

\section{Supercell Hamiltonian}
\label{app:supercell_hamiltonian}
In this section, we derive the Coulomb Hamiltonian, which directly uses Bloch functions obtained from \textit{ab initio} caluculations using supercells. Since the Hamiltonian always describes the total system of interest in the first place, we need a description capturing the common \textit{ab initio} supercell-setup with many layers, which is then reduced to a single-layer description as a starting point for practical quantum-kinetical calculations. For reasons of simplicity, we assume a static screening. Implementing a time-dependent dynamical screening as in \eqref{eq:Coulomb_Hamiltonian_SecondQuantized_Formfactors_Evaluated} is straightforward.

The general Hamiltonian reads:
\begin{multline}
    H_{\text{Coul}}\\
    = \frac{1}{2}\int\mathrm d^2r_{\parallel}^{}\,\mathrm dz\,d^2r^{\prime}_{\parallel}\,\mathrm dz^{\prime}\,\rho(\rpar,z)G(\rpar,\rparp,z,z^{\prime})\rho(\rparp,z^{\prime}).
\end{multline}
The field operator reads:
\begin{align}
    \hat \Psi(\rpar,z,t)  = \sum_{\lambda,\kpar{},k_z,s}\psi_{\lambda,\kpar{},k_z}^{s}(\rpar,z)\chi_s\andagtwo{\lambda,\kpar{},k_z}{s}(t),
\end{align}
where:
\begin{align}
    \psi_{\lambda,\mathbf k_{\parallel},k_z}^{s}(\mathbf r_{\parallel},z) = &\, \frac{1}{\sqrt{\mathcal A L_z}}\mathrm e^{\mathrm i\mathbf k_{\parallel}\cdot\mathbf r_{\parallel}} \mathrm e^{\mathrm i k_z\cdot z} u_{\lambda,\mathbf k_{\parallel},k_z}^{s}(\mathbf r_{\parallel},z)
\end{align}
is the Bloch function of a three-dimensional lattice. To obtain a layer-resolved description, we Fourier expand the creation/annihilation operators as follows:
\begin{align}
    \andagtwo{\lambda,\kpar{},k_z}{s} = \frac{L_z}{N_z} \sum_{\ell}\mathrm e^{-\mathrm i k_z \ell} \andagtwo{\lambda,\kpar{}}{\ell,s},
    \label{eq:fourier_expansion_operators_layer}
\end{align}
where $\ell$ is a lattice vector in $z$ direction, $L_z$ is the total lattice length in $z$ direction and $N_z$ is the number of unit cells in $z$ direction. The field operator then becomes:
\begin{align}
    \hat \Psi(\rpar,z,t) = \frac{\sqrt{L_z}}{\sqrt{\mathcal A N_z}}\sum_{\lambda,\kpar{},\ell,s}\chi_s \mathrm e^{\mathrm i\mathbf k_{\parallel}\cdot\mathbf r_{\parallel}} u_{\lambda,\mathbf k_{\parallel}}^{\ell,s}(\mathbf r_{\parallel},z)
    \andagtwo{\lambda,\kpar{}}{\ell,s}(t),
    \label{eq:field_operator_layer_expansion}
\end{align}
where:
\begin{align}
    u_{\lambda,\mathbf k_{\parallel}}^{\ell,s}(\mathbf r_{\parallel},z) = \frac{1}{\sqrt{N_z}}\sum_{k_z}\mathrm e^{\mathrm i k_z (z-\ell)}u_{\lambda,\mathbf k_{\parallel},k_z}^{s}(\mathbf r_{\parallel},z),
\end{align}
with normalization condition:
\begin{align}
    \frac{1}{L_z} \int\mathrm dz\, u^*\vphantom{u}_{\lambda,\mathbf k_{\parallel}}^{\ell,s}(\mathbf r_{\parallel},z) u_{\lambda,\mathbf k_{\parallel}}^{\ell^{\prime},s}(\mathbf r_{\parallel},z) = \delta_{\ell,\ell^{\prime}}.
\end{align}
Hence, the expansion in \eqref{eq:field_operator_layer_expansion} constitutes a Bloch-state representation in the in-plane direction $\rpar$ and Wannier-state representation in the out-of-plane direction $z$. The Coulomb Hamiltonian becomes:
\begin{multline}
    \hat H_{\text{Coul}} = \frac{e^2 L_z^2}{2\mathcal A^2N_z^2 \mathcal A L_z}\sum_{\substack{\lambda_1\dots\lambda_4,\kpar{1}\dots\kpar{4},\\
    \ell_1\dots\ell_4,\qpar,\Gpar,\Gparp,\\
    q_z,G_z,G_z^{\prime},s,s^{\prime}}}\\
    \times \int\mathrm d^2r_{\parallel}^{}\,\mathrm dz\,\mathrm e^{\mathrm i(\kpar{4}-\kpar{1}+\qpar+\Gpar)\cdot\rpar}\\
    \times u^*\vphantom{u}_{\lambda_1,\kpar{1}}^{\ell_1,s}(\rpar,z) u_{\lambda_4,\kpar{4}}^{\ell_4,s}(\rpar,z)\mathrm e^{\mathrm i(q_z+G_z)z}\\
    \times \int\mathrm d^2r_{\parallel}^{\prime}\,\mathrm dz^{\prime}\,
    \mathrm e^{\mathrm i(\kpar{3}-\kpar{2}-\qpar-\Gpar)\cdot\rparp}u^*\vphantom{u}_{\lambda_2,\kpar{2}}^{\ell_2,s^{\prime}}(\rparp,z^{\prime}) \\
    \times u_{\lambda_3,\kpar{3}}^{\ell_3,s^{\prime}}(\rparp,z^{\prime})\mathrm e^{-\mathrm i(q_z+G_z^{\prime})z^{\prime}}\\
    \times G_{\qpar,\Gpar,\Gparp,q_z,G_z^{},G_z^{\prime}} \adagtwo{\lambda_1,\kpar{1}}{\ell_1,s} \adagtwo{\lambda_2,\kpar{2}}{\ell_2,s^{\prime}} \andagtwo{\lambda_3,\kpar{3}}{\ell_3,s^{\prime}} \andagtwo{\lambda_4,\kpar{4}}{\ell_4,s}.
\end{multline}
Evaluate the form factors:
\begin{multline}
    \frac{1}{\mathcal AL_z}\int\mathrm d^2r_{\parallel}^{}\,\mathrm dz\,\mathrm e^{\mathrm i(\kpar{4}-\kpar{1}+\qpar+\Gpar)\cdot\rpar}\\
    \times u^*\vphantom{u}_{\lambda_1,\kpar{1}}^{\ell_1,s}(\rpar,z) u_{\lambda_4,\kpar{4}}^{\ell_4,s}(\rpar,z)\mathrm e^{\mathrm i(q_z+G_z)z} = \\
    \frac{1}{N_z}\sum_{\Gparpp}\delta_{\Gparpp,\kpar{4}+\qpar-\kpar{1}}\mathrm e^{\mathrm iq_z\ell_1}\sum_{q_z^{\prime},\mathbf G_{\parallel,1},G_{z,1}} \mathrm e^{\mathrm iq_z^{\prime}(\ell_1-\ell_4)}\\
    \times u^*\vphantom{u}_{\lambda_1,\kpar{4}+\qpar,q_z^{\prime}+q_z,\mathbf G_{\parallel,1}+\Gpar,G_{z,1}+G_z}^s u_{\lambda_4,\kpar{4},q_z^{\prime},\mathbf G_{\parallel,1},G_{z,1}}^s.
\end{multline}
Now, we assume that the Bloch factors depend only weakly on $q_z$:
\begin{align}
    u_{\lambda,\kpar{},q_z,\Gpar,G_{z}}^s \approx u_{\lambda,\kpar{},0,\Gpar,G_{z}}^s \equiv u_{\lambda,\kpar{},\Gpar,G_{z}}^s,
\end{align}
which is a good approximation, since the allowed $q_z$ values are already small if the supercell length $L_{z,\text{SC}}$ is chosen as correspondingly large, i.e., most of the spatial variation in the out-of-plane direction occurs \textit{within} the supercell. In addition, many DFT codes provide the dominant $q_z=0$ components only. 
We then can perform the $q_z^{\prime}$ sum and obtain:
\begin{multline}
    \frac{1}{\mathcal A L_z}\int\mathrm d^2r_{\parallel}^{}\,\mathrm dz\,\mathrm e^{\mathrm i(\kpar{4}-\kpar{1}+\qpar+\Gpar)\cdot\rpar}\\
    \times u^*\vphantom{u}_{\lambda_1,\kpar{1}}^{\ell_1,s}(\rpar,z) u_{\lambda_4,\kpar{4}}^{\ell_4,s}(\rpar,z)\mathrm e^{\mathrm i(q_z+G_z)z} = \\
    \sum_{\Gparpp}\delta_{\Gparpp,\kpar{4}+\qpar-\kpar{1}}\mathrm e^{\mathrm iq_z\ell_1}
    \delta_{\ell_1,\ell_4}\\
    \times \sum_{\mathbf G_{\parallel,1},G_{z,1}} u^*\vphantom{u}_{\lambda_1,\kpar{4}+\qpar,\mathbf G_{\parallel,1}+\Gpar,G_{z,1}+G_z}^s u_{\lambda_4,\kpar{4},\mathbf G_{\parallel,1},G_{z,1}}^s,
\end{multline}
i.e., now only if $\ell_1=\ell_4$ the form factor is nonzero. The multi-layer Coulomb Hamiltonian then becomes:
\begin{multline}
    \hat H_{\text{Coul}}
    = 
    \frac{1}{2}\frac{L_z^4}{N_z^2}\sum_{\substack{\lambda,\lambda^{\prime},\kpar{},\kparp{},\\
    \ell,\ell^{\prime},\qpar,\Gpar,\Gparp,s,s^{\prime}}}\\
    \times \sum_{G_z^{},G_z^{\prime}} 
    \overline \Upsilon^{\text{sc}}\vphantom{\overline \Upsilon}_{\kpar{}+\qpar+\Gpar,\kpar{},G_z^{},0}^{\lambda_1,\lambda_4,s}
    \overline \Upsilon^{\text{sc}}\vphantom{\overline \Upsilon}_{\kparp{}-\qpar-\Gparp,\kparp{},-G_z^{\prime},0}^{\lambda_2,\lambda_3,s^{\prime}}\\
    \times V_{\qpar,\Gpar,\Gparp,G_z^{},G_z^{\prime}}^{\ell,\ell^{\prime}} \adagtwo{\lambda_1,\kpar{}+\qpar}{\ell,s} \adagtwo{\lambda_2,\kparp{}-\qpar}{\ell^{\prime},s^{\prime}} \andagtwo{\lambda_3,\kparp{}}{\ell^{\prime},s^{\prime}} \andagtwo{\lambda_4,\kpar{}}{\ell,s},
    \label{eq:Coulomb_hamiltonian_layerresolved_final}
\end{multline}
where $\overline \Upsilon^{\text{sc}}\vphantom{\overline \Upsilon}_{\kparp{},\kpar{},G_z^{},0}^{\lambda,\lambda^{\prime},s}$ is the (reduced) supercell form factor:
\begin{multline}
    \overline \Upsilon^{\text{sc}}\vphantom{\overline \Upsilon}_{\kpar{},\kparp{},G_z^{},G_z^{\prime}}^{\lambda,\lambda^{\prime},s} \\
    = \sum_{\mathbf G_{\parallel,1},G_{z,1}} u^*\vphantom{u}_{\lambda,\kpar{},0,\mathbf G_{\parallel,1},G_{z,1}+G_z^{}}^s u_{\lambda^{\prime},\kparp{},0,\mathbf G_{\parallel,1},G_{z,1}+G_z^{\prime}}^s,
\end{multline}
and $V_{\qpar,\Gpar,\Gparp,G_z^{},G_z^{\prime}}^{\ell,\ell^{\prime}} $ is the layer-resolved Coulomb potential:
\begin{multline}
    V_{\qpar,\Gpar,\Gparp,G_z^{},G_z^{\prime}}^{\ell,\ell^{\prime}} = \sum_{q_z}\mathrm e^{\mathrm i q_z(\ell-\ell^{\prime})}
    V_{\qpar,\Gpar,\Gparp,q_z,G_z^{},G_z^{\prime}}\\
    = \sum_{q_z}\mathrm e^{\mathrm i q_z(\ell-\ell^{\prime})}
     \frac{e^2}{\mathcal A L_z} G_{\qpar,\Gpar,\Gparp,q_z,G_z^{},G_z^{\prime}}.
\end{multline}
Note that the prefactor $\frac{L_z^4}{N_z^2}$ in \eqref{eq:Coulomb_hamiltonian_layerresolved_final} arises due to the Fourier expansion in $k_z$, cf.~\eqref{eq:fourier_expansion_operators_layer}, which entails layer-resolved operators in units of one over length, i.e., $[\hat a^{(\dagger)}\vphantom{a}_{\lambda,\kpar{}}^{\ell,s}] = \text{nm}^{-1}$. 
For the Green's function, it holds:
\begin{multline}
    G_{\qpar,\Gpar,\Gparp,q_z,G_z^{},G_z^{\prime}}\\
    = \left(\varepsilon_{\text{mic}}^{-1}\right)_{\qpar,\Gpar,\Gparp,q_z,G_z^{},G_z^{\prime}}
    G_{0,\qpar+\Gparp,q_z+G_z^{\prime}}.
\end{multline}
To obtain the monolayer limit without image charge interactions from the other layers, i.e., supercells, we apply a truncated Coulomb potential:
\begin{align}
    G_{0,\qpar}(z) \rightarrow G_{0,\qpar}^{\text{trunc}}(z) = \frac{\mathrm e^{-|\qpar| |z|}}{2\epsilon_0 |\qpar|}\Theta \mleft(z+\frac{\Omega_z}{2}\mright)\Theta \mleft(\frac{\Omega_z}{2}-z\mright).
\end{align}
In reciprocal space, this transforms to:
\begin{align}
    G_{0,\qpar+\Gpar,q_z+G_z}^{\text{trunc}} = \frac{1 - \mathrm e^{-|\qpar+\Gpar|\frac{\Omega_z}{2}}\cos\mleft((q_z+G_z)\frac{\Omega_z}{2}\mright)}{\epsilon_0 (  |\qpar+\Gpar|^2 + (q_z+G_z)^2)}.
\end{align}
Since $q_z$ is small, we can safely assume that the Green's function does not depend on it (similar argument as above): 
\begin{multline}
    G_{\qpar,\Gpar,\Gparp,q_z,G_z^{},G_z^{\prime}}^{\text{trunc}}  \approx G_{\qpar,\Gpar,\Gparp,0,G_z^{},G_z^{\prime}}^{\text{trunc}}  \\
    \equiv G_{\qpar,\Gpar,\Gparp,G_z^{},G_z^{\prime}}^{\text{trunc}} .
\end{multline}
Hence, the layer-resolved Coulomb potential becomes:
\begin{multline}
    V_{\qpar,\Gpar,\Gparp,G_z^{},G_z^{\prime}}^{\ell,\ell^{\prime}} \rightarrow V_{\qpar,\Gpar,\Gparp,G_z^{},G_z^{\prime}}^{\ell,\ell^{\prime},\text{trunc}}  \\
    \approx N_z\delta_{\ell,\ell^{\prime}} V_{\qpar,\Gpar,\Gparp,G_z^{},G_z^{\prime}}^{\text{trunc}} ,
\end{multline}
where:
\begin{align}
    V_{\qpar,\Gpar,\Gparp,G_z^{},G_z^{\prime}}^{\text{trunc}} = \frac{e^2}{\mathcal A L_z} G_{\qpar,\Gpar,\Gparp,G_z^{},G_z^{\prime}}^{\text{trunc}}.
\end{align}
Now, interlayer, i.e., inter-supercell and, hence, unwanted, Coulomb interactions are removed. We obtain:
\begin{multline}
    \hat H_{\text{Coul}} = 
    \frac{1}{2}\frac{L_z^4}{N_z}\sum_{\substack{\lambda,\lambda^{\prime},\kpar{},\kparp{},\\
    \ell,\qpar,\Gpar,\Gparp,s,s^{\prime}}}
    \sum_{G_z^{},G_z^{\prime}} 
    \overline \Upsilon^{\text{sc}}\vphantom{\overline \Upsilon}_{\kpar{}+\qpar+\Gpar,\kpar{},G_z^{},0}^{\lambda_1,\lambda_4,s}\\
    \times\overline \Upsilon^{\text{sc}}\vphantom{\overline \Upsilon}_{\kparp{}-\qpar-\Gparp,\kparp{},-G_z^{\prime},0}^{\lambda_2,\lambda_3,s^{\prime}} V_{\qpar,\Gpar,\Gparp,G_z^{},G_z^{\prime}}^{\text{trunc}}\\
    \times \adagtwo{\lambda_1,\kpar{}+\qpar}{\ell,s} \adagtwo{\lambda_2,\kparp{}-\qpar}{\ell,s^{\prime}} \andagtwo{\lambda_3,\kparp{}}{\ell,s^{\prime}} \andagtwo{\lambda_4,\kpar{}}{\ell,s}.
\end{multline}
By rescaling $ L_z \hat a^{(\dagger)}\vphantom{a}_{\lambda,\kpar{}}^{\ell,s}\rightarrow \hat a^{(\dagger)}\vphantom{a}_{\lambda,\kpar{}}^{\ell,s}$, which retrieves unitless creation/annihilation operators, and dropping the layer index $\ell$ completely, as there are no interlayer interactions anymore, so that the layer-sum disappears: $\frac{1}{N_z}\sum_{\ell}=1$, we obtain:
\begin{multline}
    \hat H_{\text{Coul}} = 
    \frac{1}{2}\sum_{\substack{\lambda,\lambda^{\prime},\kpar{},\kparp{},\\
    \qpar,\Gpar,\Gparp,s,s^{\prime}}}
    \sum_{G_z^{},G_z^{\prime}} V_{\qpar,\Gpar,\Gparp,G_z^{},G_z^{\prime}}^{\text{trunc}}\\
    \times \overline \Upsilon^{\text{sc}}\vphantom{\overline \Upsilon}_{\kpar{}+\qpar+\Gpar,\kpar{},G_z^{},0}^{\lambda_1,\lambda_4,s}
    \overline \Upsilon^{\text{sc}}\vphantom{\overline \Upsilon}_{\kparp{}-\qpar-\Gparp,\kparp{},-G_z^{\prime},0}^{\lambda_2,\lambda_3,s^{\prime}} \\
    \times \adagtwo{\lambda_1,\kpar{}+\qpar}{s} \adagtwo{\lambda_2,\kparp{}-\qpar}{s^{\prime}} \andagtwo{\lambda_3,\kparp{}}{s^{\prime}} \andagtwo{\lambda_4,\kpar{}}{s}.
    \label{eq:Coulomb_hamiltonian_supercell_final}
\end{multline}
\eqref{eq:Coulomb_hamiltonian_supercell_final} is the final second-quantized many-body Coulomb Hamiltonian, which directly uses Bloch factors and dielectric functions obtained from \textit{ab initio} supercell calculations as input quantities.

\section{Fourier Transformation in Periodic Lattices}
\label{app:fourier_transformation}

In this section, we provide all relevant conventions and relations.

General Fourier transformation of an arbitrary function $f$ with respect to time domain $t$ and frequency domain $\omega$:
\begin{align}
	f(t)=\frac{1}{2\pi}\int\mathrm d\omega\,\efun{-\mathrm i\omega t}f(\omega),\quad f(\omega)=\int\mathrm dt\,\efun{\mathrm i\omega t}f(t).
\end{align}
General Fourier transformation of an arbitrary function $f$ with respect to real space $\mathbf r$ and momentum space $\mathbf k$:
\begin{align}
	f(\mathbf r)=\frac{1}{\left(2\pi\right)^n}\int\mathrm d^nk\,\efun{\mathrm i\mathbf k\cdot\mathbf r}f_{\mathbf k}, \quad  f_{\mathbf k}=\int\mathrm d^nr\,\efun{-\mathrm i\mathbf k\cdot\mathbf r}f(\mathbf r),
    \label{eq:fourier_trafo_continuous}
\end{align}
where $n$ denotes the dimensionality.

Discrete Fourier series within Born-von~K\'arm\'an boundary conditions \cite{vonkarman1912schwingungen}:
\begin{align}
    f(\mathbf R)=\frac{1}{\mathcal V_n}\sum_{\mathbf q}\efun{\mathrm i\mathbf q\cdot\mathbf R}f_{\mathbf k}, \quad f_{\mathbf q}=\frac{\mathcal V_n}{\mathcal N}\sum_{\mathbf R}\efun{-\mathrm i\mathbf q\cdot\mathbf R}f(\mathbf R),
    \label{eq:DiscreteFourierSeries}
\end{align}
where $\mathbf q$ is a crystal momentum in the first Brillouin zone and $\mathbf R$ is a lattice vector, $\mathcal N$ is the total number of unit cells and $\mathcal V_n$ is the $n$-dimensional volume of the total lattice. In the main text, we denote: $\mathcal V_{n=1} \equiv L$, $\mathcal V_{n=2} \equiv \mathcal A$ and $\mathcal V_{n=3} \equiv \mathcal V$. 
From \eqref{eq:DiscreteFourierSeries}, it follows:
\begin{align}
\begin{split}
    \frac{1}{\mathcal N}\sum_{\mathbf R}\mathrm e^{\mathrm i\mathbf R\cdot(\mathbf q-\mathbf q^{\prime})} = &\, 
    \sum_{\mathbf G}\delta_{\mathbf q-\mathbf q^{\prime},\mathbf G},\\
    \frac{1}{\mathcal N}\sum_{\mathbf q}\mathrm e^{\mathrm i\mathbf q\cdot(\mathbf R-\mathbf R^{\prime})} = &\, \delta_{\mathbf R,\mathbf R^{\prime}},
    \end{split}
\end{align}
and:
\begin{align}
\begin{split}
    \frac{1}{\mathcal V_{n,\text{UC}}}\sum_{\mathbf G}\mathrm e^{\mathrm i\mathbf G\cdot \mathbf r} = &\,  \sum_{\mathbf R}\delta(\mathbf r-\mathbf R),\\
    \frac{1}{\mathcal V_{n,\text{UC}}}\int_{\mathcal V_{n,\text{UC}}}\mathrm d^3r\, \mathrm e^{\mathrm i(\mathbf G-\mathbf G^{\prime})\cdot \mathbf r} = &\, \delta_{\mathbf G,\mathbf G^{\prime}}.
    \end{split}
\end{align}
Here, $\mathcal V_{n,\text{UC}}$ is the $n$-dimensional volume of the unit cell.

By expanding the arbitrary momentum $\mathbf k$ in \eqref{eq:fourier_trafo_continuous} via $\mathbf k = \mathbf q + \mathbf G$, where $\mathbf q$ is a momentum within the first Brillouin zone and $\mathbf G$ is a reciprocal lattice vector, we can rewrite \eqref{eq:fourier_trafo_continuous}:
\begin{align}
    f(\mathbf r)=\frac{1}{\left(2\pi\right)^n}\sum_{\mathbf G}\int_{\mathcal V_{n,\text{BZ}}}\mathrm d^nq\,\efun{\mathrm i(\mathbf q+\mathbf G)\cdot\mathbf r}f_{\mathbf q+\mathbf G}.
\end{align}
By using the conversion relation between discrete sums and integrals, which is valid as long as the crystal momenta $\mathbf q$ lie sufficiently dense in the first Brillouin zone \cite{czycholl2007theoretische}:
\begin{align}
	\sum_{\mathbf q}f_{\mathbf q}\leftrightarrow \frac{\mathcal V_n}{\left(2\pi\right)^n}\int_{\mathcal V_{n,\text{BZ}}} \mathrm d^nq\, f_{\mathbf q},
    \label{eq:sum_integral_conversion}
\end{align}
where 
$\mathcal V_{n,\text{BZ}}$ denotes the $n$-dimensional volume of the Brillouin zone (unit cell of the reciprocal lattice), we obtain:
\begin{align}
    f(\mathbf r)=\frac{1}{\mathcal V_n}\sum_{\mathbf q,\mathbf G}\efun{\mathrm i(\mathbf q+\mathbf G)\cdot\mathbf r}f_{\mathbf q+\mathbf G}.
\end{align}
The back transformation can be expressed as:
\begin{align}
    f_{\mathbf q+\mathbf G} = \sum_{\mathbf R}\mathrm e^{-\mathrm i (\mathbf q+\mathbf G)\cdot\mathbf R}\int_{\mathcal V_{n,\text{UC}}}\mathrm d^nr\,\mathrm e^{-\mathrm i (\mathbf q+\mathbf G)\cdot \mathbf r} f(\mathbf r+\mathbf R).
    \label{eq:fourier_backtransformation}
\end{align}
If the function $f(\mathbf r)$ is lattice periodic:
\begin{align}
    f(\mathbf r+\mathbf R) = f(\mathbf r),
\end{align}
the following relations hold:
\begin{align}
    f(\mathbf r) = \sum_{\mathbf G}\mathrm e^{\mathrm i\mathbf G\cdot\mathbf r}f_{\mathbf G}, \quad f_{\mathbf G} =  \frac{1}{\mathcal V_{\text{UC}}}\int_{\mathcal V_{n,\text{UC}}}\mathrm d^nr\,\mathrm e^{-\mathrm i\mathbf G\cdot\mathbf r}f(\mathbf r), 
\end{align}
where we identified: $f_{\mathbf q+\mathbf G} \equiv f_{\mathbf G}\mathcal V_n\delta_{\mathbf q,\mathbf 0}$.

An arbitrary function with two arguments $f(\mathbf r,\mathbf r^{\prime})$ can be Fourier expanded accordingly:
\begin{align}
    f(\mathbf r,\mathbf r^{\prime}) = \frac{1}{\mathcal V_n^2}\sum_{\mathbf q,\mathbf G,\mathbf q^{\prime},\mathbf G^{\prime}} \mathrm e^{\mathrm i (\mathbf q+\mathbf G)\cdot\mathbf r}\mathrm e^{\mathrm i(\mathbf q^{\prime}+\mathbf G^{\prime})\cdot\mathbf r^{\prime}} f_{\mathbf q,\mathbf q^{\prime},\mathbf G,\mathbf G^{\prime}},
\end{align}
with back transformation:
\begin{multline}
    f_{\mathbf q,\mathbf q^{\prime},\mathbf G,\mathbf G^{\prime}} = \sum_{\mathbf R,\mathbf R^{\prime}}
    \mathrm e^{-\mathrm i (\mathbf q+\mathbf G)\cdot\mathbf R}
    \mathrm e^{-\mathrm i (\mathbf q^{\prime}+\mathbf G^{\prime})\cdot\mathbf R^{\prime}}\\
    \times 
    \int_{\mathcal V_{n,\text{UC}}}\mathrm d^nr\,\mathrm d^nr^{\prime}\,\mathrm e^{-\mathrm i (\mathbf q+\mathbf G)\cdot \mathbf r}\mathrm e^{-\mathrm i (\mathbf q^{\prime}+\mathbf G^{\prime})\cdot \mathbf r^{\prime}}
    f(\mathbf r+\mathbf R,\mathbf r^{\prime}+\mathbf R^{\prime}).
\end{multline}
In case that the function $f(\mathbf r,\mathbf r^{\prime})$ is lattice-periodic:
\begin{align}
    f(\mathbf r+\mathbf R,\mathbf r^{\prime}+\mathbf R) = f(\mathbf r,\mathbf r^{\prime}),
\end{align}
the Fourier expansion can be expressed as follows:
\begin{align}
    f(\mathbf r,\mathbf r^{\prime}) = \frac{1}{\mathcal V_n}\sum_{\mathbf q,\mathbf G,\mathbf G^{\prime}} \mathrm e^{\mathrm i (\mathbf q+\mathbf G)\cdot\mathbf r}\mathrm e^{-\mathrm i(\mathbf q+\mathbf G^{\prime})\cdot\mathbf r^{\prime}} f_{\mathbf q,\mathbf G,\mathbf G^{\prime}},
\end{align}
where we identified: $f_{\mathbf q,\mathbf q^{\prime},\mathbf G,\mathbf G^{\prime}} \equiv f_{\mathbf q,\mathbf G,-\mathbf G^{\prime}}\mathcal V_n\delta_{\mathbf q,-\mathbf q^{\prime}}$. Exploiting the lattice-periodicity of $f(\mathbf r,\mathbf r^{\prime})$, the back transformation then reads:
\begin{multline}
    f_{\mathbf q,\mathbf G,\mathbf G^{\prime}} = \frac{1}{\mathcal V_{n,\text{UC}}} \sum_{\mathbf R}\mathrm e^{\mathrm i\mathbf q\cdot\mathbf R} \\
    \times \int_{\mathcal V_{n,\text{UC}}}\mathrm d^nr\,\mathrm d^nr^{\prime}\, \mathrm e^{-\mathrm i(\mathbf q+\mathbf G)\cdot\mathbf r}\mathrm e^{\mathrm i(\mathbf q+\mathbf G^{\prime})\cdot\mathbf r^{\prime}}f(\mathbf r,\mathbf r^{\prime}+\mathbf R).
\end{multline}

The Dirac delta distributions for arbitrary momenta $\mathbf k$ and arbitrary spatial coordinates $\mathbf r$ read:
\begin{align}
    \frac{1}{(2\pi)^n} \int \mathrm d^n r\,\mathrm e^{\mathrm i\mathbf k\cdot\mathbf r} = \delta(\mathbf k), \quad \frac{1}{(2\pi)^n} \int \mathrm d^n k\,\mathrm e^{\mathrm i\mathbf k\cdot\mathbf r} = \delta(\mathbf r).
\end{align}
By expanding the spatial coordinates over the unit cells and the momenta over the Brillouin zones, we obtain:
\begin{align}
    \frac{1}{(2\pi)^n} \int \mathrm d^n r\,\mathrm e^{\mathrm i(\mathbf q+\mathbf G)\cdot \mathbf r} = &\, \frac{\mathcal V_n}{(2\pi)^n}\delta_{\mathbf q,\mathbf 0}\delta_{\mathbf G,\mathbf 0} \quad  \text{($\mathbf q\in \mathcal V_{n,\text{BZ}}$)},\\
    \frac{1}{(2\pi)^n} \int \mathrm d^n k\,\mathrm e^{\mathrm i\mathbf k\cdot(\mathbf r+\mathbf R)} = &\, \delta(\mathbf r)\delta_{\mathbf R,\mathbf 0} \quad \text{($\mathbf r\in \mathcal V_{n,\text{UC}}$)}.
\end{align}
Hence, for a momentum $\mathbf q$ within the first Brillouin zone, the following relation between the continuous Dirac delta function defined under integrals and the discrete Kronecker delta defined under sums can be obtained:
\begin{align}
    \delta (\mathbf q) \leftrightarrow \frac{\mathcal V_n}{(2\pi)^n}\delta_{\mathbf q,\mathbf 0}  \quad \text{($\mathbf q\in \mathcal V_{n,\text{BZ}}$)},
\end{align}
cf.\ also \eqref{eq:sum_integral_conversion}.

\section{Parameters}
\begin{table}[h!]
    \centering
    \begin{tabular}{ll}
$1s$ excitonic energy 	$E_{\text{A}_{1s}}$ & 1.639\,eV \cite{robert2020measurement}\\
Nonradiative linewidth 	$\hbar\gamma_{\text{nrad},\text{A}_{1s}}$ &10\,meV	\\
Nonradiative linewidth 	$\hbar\gamma_{\text{nrad},\text{B}_{1s}}$ &$2\hbar\gamma_{\text{nrad},\text{A}_{1s}}$\\
\hline
Band structure parameters & \\
\hline
\textit{k}$\cdot$\textit{p}-parameter $\gamma_{\text{\textit{k}$\cdot$\textit{p}}}$ & \hspace{-0.05cm}\begin{tabular}{c} $0.5(0.253+0.22)$\\eV\,nm \cite{kormanyos2015k}\end{tabular}\\
Effective electron mass $m_e^{K,\uparrow}$/$m_e^{K,\downarrow}$ & $0.5m_0$/$0.58m_0$ \cite{kormanyos2015k}\\
Effective hole mass $m_h^{K,\uparrow}$/$m_h^{K,\downarrow}$ & $0.6m_0$/$0.7m_0$ \cite{kormanyos2015k}\\
Conduction band splitting & 20\,meV \cite{kormanyos2015k}\\
Valence band splitting & 184\,meV \cite{kormanyos2015k}\\
\hline
Dielectric screening parameters & \\
\hline
Monolayer width $d$ & 0.6527$\,$nm \cite{kylanpaa2015binding}\\
\hspace{-0.16cm}
\begin{tabular}{l}
Static dielectric constant of bulk\\
MoSe$_2$
$\epsilon_{s,0} = \sqrt{\epsilon_{s,\parallel,0}\epsilon_{s,\bot,0}}$
\end{tabular} & 12.0474 \cite{laturia2018dielectric}\\
Interlayer gap $h$ & 0.3\,nm \cite{florian2018dielectric}\\
Thomas-Fermi parameter $\alpha_{\text{TF}}$ & 1.9 (fit to CMR \cite{andersen2015dielectric})\\
Plasmon peak energy $\hbar\omega_{\text{pl}}$ & 22.0$\,$eV \cite{kumar2012tunable}\\
\hspace{-0.16cm}
\begin{tabular}{l}
Static dielectric constant of bulk\\
h-BN $\epsilon_{\text{h-BN},0} = \sqrt{\epsilon_{\text{h-BN},0,\parallel}\epsilon_{\text{h-BN},0,\bot}}$
\end{tabular} & 4.8 \cite{latini2015excitons}\\
\hspace{-0.16cm}
\begin{tabular}{l}
Optical in-plane dielectric constant\\
of h-BN $\epsilon_{\parallel,\infty,\text{h-BN}}$ 
\end{tabular} & 4.87 \cite{cai2007infrared}\\
\hline
Selected calculated parameters & \\
\hline
A$_{1s}$ binding energy & 341.3\,meV\\
B$_{1s}$ binding energy & 375.1\,meV\\
\hspace{-0.16cm}
\begin{tabular}{l}
A$_{1s}$ Coulomb enhancement\\
$\sum_{\mathbf q}\ExWFtwo{1s,\mathbf q}{K,K,\uparrow,\uparrow}$
\end{tabular}& 0.916\\
\hspace{-0.16cm}
\begin{tabular}{l}
B$_{1s}$ Coulomb enhancement\\
$\sum_{\mathbf q}\ExWFtwo{1s,\mathbf q}{K,K,\downarrow,\downarrow}$
\end{tabular} & 1.038
    \end{tabular}
    \caption{Parameters for the calculations of a h-BN-encapsulated MoSe$_2$ monolayer in Fig.~\ref{fig:linear_absorption}.}
    \label{tab:linear_absorption_parameters}
\end{table}

\FloatBarrier

\bibliography{bibliography_full}

\end{document}